\documentclass{article}

\usepackage{adjustbox}
\usepackage{float}
\usepackage[T1]{fontenc}
\usepackage[utf8]{inputenc}
\usepackage{graphicx}
\usepackage{hhline}
\usepackage{multicol}
\usepackage{multirow}
\usepackage{soul}
\usepackage[normalem]{ulem}
\usepackage[svgnames]{xcolor}
\usepackage{xurl}
\usepackage{authblk}
\usepackage[breaklinks]{hyperref}
\usepackage{cleveref}

\title{Open Problems in DAOs}

\author[1,2]{Joshua Tan}
\author[3,1]{Tara Merk}
\author[4]{Sarah Hubbard}
\author[5]{Eliza R. Oak}
\author[6]{Helena Rong}
\author[1]{Joni Pirovich}
\author[7,1]{Ellie Rennie}
\author[8]{Rolf Hoefer}
\author[9,10,1]{Michael Zargham}
\author[7]{Jason Potts}
\author[7]{Chris Berg}
\author[11]{Reuben Youngblom}
\author[3,4]{Primavera De Filippi}
\author[12,13,1]{Seth Frey}
\author[10]{Jeff Strnad}
\author[14]{Morshed Mannan}
\author[7,14]{Kelsie Nabben}
\author[15,11]{Silke Noa Elrifai}
\author[16]{Jake Hartnell}
\author[17]{Benjamin Mako Hill}
\author[18]{Tobin South}
\author[19]{Ryan L. Thomas}
\author[20, 11]{Jonathan Dotan}
\author[20]{Ariana Spring}
\author[21]{Alexia Maddox}
\author[4]{Woojin Lim}
\author[22]{Kevin Owocki}
\author[23,24]{Ari Juels}
\author[10]{Dan Boneh}
\affil[1]{Metagov}
\affil[2]{University of Oxford}
\affil[3]{CNRS}
\affil[4]{Harvard University}
\affil[5]{Yale University}
\affil[6]{Columbia University}
\affil[7]{RMIT University}
\affil[8]{Cultur3}
\affil[9]{BlockScience}
\affil[10]{WU Vienna}
\affil[11]{Stanford University}
\affil[12]{University of California, Davis}
\affil[13]{The Ostrom Workshop at Indiana University}
\affil[14]{European University Institute}
\affil[15]{Paris II (Panthéon-Assas)}
\affil[16]{DAODAO}
\affil[17]{University of Washington}
\affil[18]{MIT}
\affil[19]{Wentworth Institute of Technology}
\affil[20]{EQTY Lab}
\affil[21]{La Trobe University}
\affil[22]{Gitcoin}
\affil[23]{Cornell Tech}
\affil[24]{IC3}

\begin{document}
\maketitle


\begin{abstract}
Decentralized autonomous organizations (DAOs) are a new, rapidly-growing class of organizations governed by smart contracts. Here we describe how researchers can contribute to the emerging science of DAOs and other digitally-constituted organizations. From granular privacy primitives to mechanism designs to model laws, we identify high-impact problems in the DAO ecosystem where existing gaps might be tackled through a new data set or by applying tools and ideas from existing research fields such as political science, computer science, economics, law, and organizational science. Our recommendations encompass exciting research questions as well as promising business opportunities. We call on the wider research community to join the global effort to invent the next generation of organizations.
\end{abstract}

\tableofcontents

\newpage

\section{Introduction}

Decentralized autonomous organizations (DAOs) first began to surface in discussions of the blockchain community around 2013, where people imagined them as digital substitutes for traditional organizations. Technologists argued that DAOs could automate many organizational processes and allow for more broad-based ownership and governance of the digital economy, all on the basis of a cryptographically-secured blockchain \cite{buterin_daos_2014, larimer_bitcoin_2013}. While such DAOs existed mostly in the realm of speculation, with the noteworthy exception of The DAO \cite{dupont_experiments_2017}, the reality of DAOs has drastically changed over the past few years. Driven by the surge of interest in crypto, real world deployment of DAOs surged by as much as 660\% from 2019 to late 2020 \cite{haig_number_2020}, rising to 1.6 million participants collectively managing 16 billion USD across over 13,000 DAOs in 2022 \cite{metisdao_new_2022,jenkinson_remote_2022,deepdao_ventures_wwwdeepdaoio_nodate}.

While there is no single comprehensive definition of DAOs, their core characteristics can be synthesized from various academic uses of the term: DAOs are organizations governed by a smart contract, typically deployed on a blockchain that autonomously enforces rules for interaction among the members \cite{hassan_decentralized_2021}. DAOs also belong to a larger class of digitally-constituted organizations: organizations governed through computational artifacts such as software, hardware, and/or protocols. For example, the Bitcoin, Ethereum, and Tor networks, even though they are not DAOs, can be classified as digitally-constituted organizations.

This article provides an overview of the open problems and research questions relevant to DAOs and related organizations, as well as a number of discrete research and engineering projects across many disciplines. We cover not only problems important to DAO practitioners but also opportunities to address larger research questions through the lens of DAOs. Many of the problems described here will also be relevant to anyone looking at governance issues within online communities.

\subsection{Who is this article written for?}

This article is intended to (1) establish key problems and research opportunities in DAOs, (2) communicate and motivate those problems and opportunities to researchers not already familiar with DAOs, and (3) survey existing research works and projects relevant to DAOs. We hope that our recommendations will prove valuable to several different audiences.

\textbf{Academics and researchers}: Throughout this paper we identify many problems that require conceptual and practical innovations, either within DAOs or within an academic discipline.

\textbf{DAO practitioners}: Many of the problems identified in this paper were sourced directly from our engagement with DAO practitioners, including workshops at Stanford, Harvard, Devcon, and Devconnect. We hope that in reading this paper practitioners can identify areas of research relevant to their problems and begin to think through the implications that existing research has for their practice today.

\textbf{Entrepreneurs:} Many of the problems and questions highlighted throughout this document require the development of various services, tools, and solutions in order to be solved in the long run. Consequently, we believe that this paper presents an interesting starting point for entrepreneurs looking to contribute to the wider DAO ecosystem. 

\textbf{Investors}: Given its growth and potential, the DAO ecosystem presents an interesting area for investors. We believe that highlighting some of the most pressing problems for DAOs today can serve as a useful indicator for allocating impactful investments, be it to research or to support the concrete development of tools and solutions in the space.

\subsection{Call for collaboration}

While this article is divided into disciplinary sections, many of the problems described require collaboration across multiple fields. For example, economic questions around the potential use of DAOs by unions will require input from lawyers in employment law; analysis of DAOs as social computing systems should invoke existing research within organizational science; a study of the cultural evolution of DAOs likely requires significantly more development of off-chain data sets by engineers; and so on. More generally, collaboration across fields is more likely to produce impactful solutions that can handle the complexities of actual DAOs.

We particularly encourage empirical collaborations between academics and builders in the DAO ecosystem. DAOs are not only organizations but constitute a toolbox that one can work with—an empirical advantage that we encourage researchers to take advantage of. And because so much of the DAO ecosystem is being developed on the fly, it is often useful to work with someone aware of current trends in the ecosystem.

To aid readers in finding collaborators, data sets, and other resources, we have built a website to accompany this article at \href{https://daoscience.org}{www.daoscience.org}.

\subsection{The bigger picture}

As an organizational form, DAOs are closely related to earlier forms of online community, especially open-source communities; many of the most successful DAOs are also communities organized around an open-source product. But DAOs also have antecedents in digital cooperatives and platform cooperatives \cite{scholz_ours_2016,mannan_fostering_2018}, multi-organizational networks such as keiretsus \cite{ronfeldt_tribes_1996,kreutler_prehistory_2021}, crowdfunding platforms such as Patreon, the economies of virtual worlds such as World of Warcraft and Second Life \cite{lehdonvirta_virtual_2014}, and networked projects for peer production such as Wikipedia that question the role of the traditional firm \cite{benkler_peer_2015}. 

As a research subject, DAOs occupy an interesting and important intersection between the social sciences and the computer and engineering sciences. As research subjects, DAOs are interesting because of the unique opportunities for intervention afforded by smart contracts, the  native availability of fine-grained data, the vibrancy of the ecosystem, and a fast-moving do-ocratic culture inherited from tech and open-source. Compared to related fields such as platform governance or civic tech, DAO science is more financially sophisticated, more entrepreneurial, and less permissioned by existing institutions. Compared to blockchain governance (with which it substantially overlaps), DAO science is more experimental and human-centered. Compared to older studies of online communities and virtual (game) worlds, DAO science requires significantly more legal, economic, and organizational sophistication; in DAOs, the scope of activities is broader, and the stakes higher. And compared to the many studies of information technology within management science, DAO science is more keyed into questions of governance rather than operations: a digital organization may operate through tools like Slack, Salesforce, or Zendesk without adjudicating rights and power through a digital mechanism; vice versa, a DAO or digitally-constituted organization may govern incentives and roles within a co-op or restaurant that uses very few digital technologies in its day-to-day operations.\footnote{ This is not to say that governance and operations are not linked or correlated—DAOs are still most likely to be used to govern digital artifacts like smart contracts or AI, not hamburger joints—simply that they are not necessarily so.}

Perhaps most importantly, the smart contracts and Ricardian contracts upon which DAOs are based are truly general-purpose instruments. That means that DAOs can express the full range of existing organizational forms—so the choice is not between a corporation and a DAO per se but between a traditional, legally-constituted corporation and a corporation that is digitally-constituted through a smart contract. DAOs are not just curious instances of a certain kind of online community; they have the expressive potential to transport a tremendous amount of institutional infrastructure from the stonebound halls of power onto the open internet; from law and economics into computer science.

DAOs in their current form may or may not become the future of organizations. However, it is already clear that online forms of organization are becoming more and more important in the politics and economies of the world. DAO science is one of the most promising paths forward for tackling and making progress on hard questions of organization, coordination, and governance.

\section{Computer science}

\begin{quote}
    Editor: Joshua Tan. Contributors: Joshua Tan, Michael Zargham, Jake Hartnell, Benjamin Mako Hill, Dan Boneh, Tobin South, Ari Juels, Eliza R. Oak.
\end{quote}

Computer science historically grew out of applied mathematics, but over the past few decades the field has also found remarkable success applying the theory and practice of computation to the challenges of other disciplines such as physics, biology, and economics. DAOs can be understood as a facet of this ``computational lens" \cite{karp_understanding_2011, wigderson_mathematics_2019}: they allow us to convert many long-standing social, organizational, and legal problems into computational ones, opening up interesting new challenges for computer scientists, mathematicians, and engineers.

In this section, we will explore a number of different open problems and research opportunities in DAOs from the perspective of computer science, touching on subjects in cryptography and networking, which have traditionally been the most salient aspects of computer science to blockchains, as well subjects motivated from social computing, online governance, and human-computer interaction (HCI). Perhaps the most important cluster of research opportunities revolves around building up the scientific foundations of research into DAOs, including better benchmarks, data sets, validation criteria, smart contract taxonomies, and research-friendly DAO frameworks that, taken together, would enable faster experimentation, more rigorous statistics, easier formal verification, and even the training of better AI models for organizations.

\subsection{Granular privacy}

\textit{Summary: Full transparency is often disadvantageous in strategic settings and can be counterproductive in social settings when it reduces the incentives for participation. Cryptography can help define new, composable primitives for privacy that allow DAOs and DAO participants to act strategically even while operating on a public ledger.}

Most current DAOs are public and transparent by default, built on pseudonymous and public blockchain protocols. However, there are various use-cases where DAOs require some form of privacy. For example, ConstitutionDAO, which participated in a widely-reported auction for the U.S. Constitution, was affected by the fact that their treasury size was publicly-known, marking a distinct disadvantage for the DAO \cite{patel_meme_2021}. In general, revealing sensitive information about a DAO or DAO participant’s strategies, preferences, or financial positions can lead to manipulation, collusion, and other negative outcomes. Further, full transparency can be socially problematic for many reasons \cite{nguyen_transparency_2022}, e.g. exposing identifying information (even wallet addresses) may deter users from participating in DAOs due to concerns about privacy, security, and liability. On the other hand, too much privacy can lead to co-optation by problematic actors—consider the use of Tornado Cash by North Korean hackers, which provoked governmental responses and the eventual sanctioning of Tornado Cash \cite{us_department_of_the_treasury_us_2022}. More granular privacy primitives enabled by advances in zero-knowledge (ZK) proofs, homomorphic encryption, and secure multiparty computation can solve many of these problems:

\begin{itemize}
	\item \textit{Private donations to a public treasury} allows large, crowdfunded experiments like ConstitutionDAO to participate in auctions while retaining the transparency of a public ledger. Properly executed, the donations would be private before the auction and then become public at a later date.

	\item \textit{Private voting on public proposals} allows the members of DAOs to participate in public governance without sharing public identifiers like wallet addresses or the vote that they cast. This is especially useful for sensitive decisions like member compensation. Several platforms have already begun implementing these features, though scalability and usability questions remain. Depending on its implementation though, private voting may also destabilize the social fabric of a DAO or undermine the legitimacy of an organization’s decisions.

	\item \textit{Private proposals} would allow DAOs to vote securely on proposals whose content is private. These may be useful for sensitive decisions like a merger or acquisition deal that the DAO does not want to make public.

	\item \textit{Private membership} allows DAOs to operate like traditional private member organizations. This is useful for a variety of use-cases, for example in sensitive security situations or for blinded peer review in scientific research.

	\item \textit{Private treasuries} enable DAOs to hide the contents of their treasuries as well as payments to and from those treasuries. Providing for privacy around payments allows for a return to the norms that many businesses and organizations expect. For example, many traditional companies shy away from using DAO tooling as they do not want to allow competitors to see their finances or their employees to see their compensation with respect to their peers.

	\item \textit{Zero-knowledge compliance} allows for DAOs to prove things about themselves without revealing confidential information. For example, proving that no members or payments received are on an OFAC registry or proving the amount of taxable DAO income for a DAO registered as an LLC.

\end{itemize}

More granular privacy primitives could be especially helpful in squaring the circle between the anonymity germane to public peer-to-peer networks and the disclosure requirements of governments and regulators.

\subsection{Private execution}

\textit{Summary: Privacy concerns extend beyond DAO management into the realm of DAO execution, where the actions of the DAO itself (usually through smart-contract execution) are public by default. This could be disadvantageous to some organizations. While some tools for private execution of operations exist, more work in computer science is needed to bridge between the status quo of public DAO execution and the ability of organizations to selectively act discretely for strategic reasons.}

Beyond the scope of integrating privacy primitives into various aspects of DAO management (treasuries, voting, membership), there is a broader issue concerning the privacy of DAO execution. Most existing traditional organizations rely on the privacy of their internal operations to function effectively. Broad operational privacy is crucial for defending against corporate espionage, protecting the confidentiality of individuals and communities served by the organization, enabling surprise product releases, and defending against other vulnerabilities—issues that have been witnessed in DAOs as they have become more sophisticated \cite{nabben_is_2021}.

To fully realize their potential, DAOs must offer greater choice in execution privacy while retaining their ‘autonomous’ character.

In general, this relies upon executing smart contracts privately, of which there has been extensive work. In particular, private contract execution via new chains has spawned both academic work (including theoretical analysis in \cite{kerber_kachina_2021} and in-production public chains. These approaches draw on a wide variety of technology primitives for privacy including secure multi-party computation (e.g., Enigma by Zyskind et al. \cite{zyskind_enigma_2015}, Eagle by Baum et al. \cite{baum_eagle_2022}, hardware enclaves \cite{cheng_ekiden_2019,kaptchuk_giving_2019,woetzel_secret_2022}, and zero-knowledge SNARKS (e.g., Hawk by Kosba et al. \cite{kosba_hawk_2016}, \cite{bonneau_zether_2020}, Bulletproofs by \cite{bunz_bulletproofs_2018}, zkay by \cite{steffen_zkay_2019}, Zexe by \cite{bowe_zexe_2020}, or via other schemes such as including UTXO-based systems \cite{bowe_zexe_2020,vaitiekunas_dialektos_2022} and incentivized agreement of off-chain smart contract execution (e.g. Arbitrum, \cite{kalodner_arbitrum_2018}, if all parties are honest). Other approaches using permissioned blockchains to add privacy to smart contract execution \cite{baliga_performance_2018, brown_corda_2018}.

Many of these systems operate via a fully private on-chain environment, especially those with the strongest privacy guarantees, and necessitate a shift in DAO operations onto a secure chain. Some approaches piggyback and interoperate with existing popular chains, however these privacy solutions are incomplete. For example, zkay and Zether go some way towards addressing this interoperability challenge with Ethereum, but their privacy guarantees are limited to payments (Zether) or specific private values (zkay). At the very least, most interoperable solutions with public chains always leak activity metadata on what contracts are executing. 

Hybrid models, which exchange metadata been private and public execution at a minimum, are possible but need to be integrated with existing DAO tooling and undergo more battle testing and security analysis.

While these technical goals are worth striving for, it exists in tension with the principles of openness set forward by the DAO community. The public nature of DAO execution drives transparency and accountability, a key feature in building competitive advantages based on trust, and considered to many a core ethical advantage. Further, the existence of privately executing DAOs can lead to malicious phenomena such as Dark DAOs \cite{daian_-chain_2018} where a decentralized cartel buys on-chain votes opaquely. Are the benefits of normalizing private DAO usage worth the risks of coordinated large-scale bribery?

While the use of privacy over transparency is not always an ethical cost (such as if privacy enables the protection of vulnerable people), the tension between the existing bias of the traditional business world toward public secrecy and the values of open communities should be examined further, perhaps facilitating the building new models of hybrid transparency enabled through DAOs.

By combining inherently private offline activities with secure DAO operations, the future of DAOs could offer valuable granular privacy choices. This would enable a variety of strategic options that would be unfeasible or disadvantageous in a fully public setting. Ongoing development, testing, and deployment of these tools are necessary as the DAO ecosystem continues to grow.

\subsection{New computational substrates}

\textit{Summary: DAOs and digitally-constituted organizations are built on top of computational substrates such as blockchains. Computer networking, cryptography, and distributed computing can help improve the properties of these substrates, help them interoperate with each other, and even define entirely new substrates.}

Computer science is the material science of digital space. Just as new materials allow us to build taller buildings and longer bridges, new computational primitives allow us to build more accessible spaces, more scalable platforms, and more effective digital organizations. For example, scaling challenges in blockchains have led to new technological developments such as optimistic rollups and zero-knowledge rollups that enable blockchains to securely delegate computation to off-chain substrates, typically a ``layer 2" blockchain. This idea of delegating computation across different computational substrates highlights several open research questions for DAOs: 

\begin{enumerate}
	\item How do we build secure and effective multichain DAOs? Substantial work on cross-chain communication protocols \cite{kwon_cosmos_2019,wood_polkadot_2016}, cryptographically-secure bridges \cite{belchior_survey_2021,xie_zkbridge_2022}, and atomic swaps \cite{meyden_specification_2019} highlights the way in which there is a growing universe of ``ledgers of record". DAOs increasingly operate and manage assets across multiple chains, and they need a secure way of owning and governing themselves across multiple chains.

	\item How do we build secure and effective multi-substrate DAOs, i.e. DAOs whose organizational logic is executed across multiple different computational substrates, e.g. within two blockchains with substantially different properties, within a blockchain and a distributed network such as Holochain \cite{harris-braun_holochain_2017}?

	\item How do we securely and efficiently interoperate between DAOs and the rest of the internet? While DAOs are commonly associated with the smart contract infrastructure of certain blockchains,\footnote{ The ``decentralized" and ``autonomous" in ``DAO" are typically predicated on the guarantees of cryptographic consensus within a widely-distributed peer-to-peer network.} many DAOs already delegate substantial parts of their organizational logic and compute to traditional, off-chain services such as Discord and Discourse as well as decentralized storage services such as IPFS. Substantial research and development has been done on off-chain oracles \cite{al-breiki_trustworthy_2020} as well as various ways of presenting block transactions through subgraphs, but many open questions remain about the architecture (client-server, peer-to-peer, n-tier, etc.) and governance of these off-chain services—which relates to the design of general internet services and applications.

	\item What kind of computational substrate is \textit{sufficient} for DAOs? For example, is it possible to build minimally-viable DAOs on other sorts of peer-to-peer networks beyond the blockchain, or within the internet protocol stack itself? Such a project seems possible: after all, the first usage of the term ``DAO" was part of a proposal for a wireless home network \cite{hassan_decentralized_2021}.

	\item What kind of computational substrate is \textit{optimal} for DAOs? If a blockchain with a smart contract architecture is optimal, what kind of blockchain and what kind of architecture best supports the use-cases of DAOs?

\end{enumerate}

In general, any research that increases the speed, reduces the cost of on-chain storage and compute, or makes smart contract architectures more robust to attack will expand the practical design space of DAOs.

\subsection{Secure voting}
\textit{Summary: Many forms of DAO governance depend on secure and trustworthy voting mechanisms. New research in computer science, especially cryptography, can help DAOs implement voting mechanisms that is more secure along several different dimensions, including integrity, public verifiability, privacy, and coercion-resistance.}

The process of voting on proposals is central to the governance of DAOs. Voting raises many design decisions, such as how to allocate voting power and how voting choices should translate into outcomes. These decisions relate to questions of social choice and tokenomics discussed in this paper. Once a DAO chooses to adopt a desired mechanism for voting, the question arises of how to \textit{implement} it, and in particular how to do so \textit{securely}.

DAO voting differs from conventional voting in political elections in several key ways. First, it is entirely digital. Users do not need to visit polling stations or interact physically with ballots. Second, it is typically pseudonymous: Voters identify themselves by on-chain addresses, rather than real-world identities, as in conventional elections. Finally, while most political elections permit only one vote per eligible voter, DAO voting often weights votes according to token holdings. DAO voting bears some similarity in this sense to corporate proxy voting.

There is a well-established literature on what is often called \textit{verifiable voting} (see, e.g., \cite{benaloh_end--end_2015, chaum_scantegrity_2008, ryan_pret_2009}). While this literature is often concerned with conventional voting, it nonetheless outlines a number of cryptographic concepts and tools relevant to DAO voting. Especially important is an articulation of the desirable security properties for voting systems. These properties vary somewhat across the literature. Informally, though, they typically include:

\begin{itemize}
	\item \textit{Integrity}: Each voter should be able to cast only a single ballot and the choice(s) made by a voter on that ballot should be reflected correctly in the outcome of a vote. 

	\item \textit{Public verifiability}: Anyone should be able to verify (cryptographically) the integrity of a vote.

	\item \textit{Privacy:} No entity, including those tasked with tallying election results, should learn anything about an election other than the outcome. (Privacy in this sense is a special case of private execution, discussed above.)

	\item \textit{Coercion-resistance: }Voters should be unable to prove to anyone how they voted. As a result, \textit{coercion}—which encompasses voter bribery, i.e., vote-buying, as well as voter intimidation—should be infeasible \cite{clarkson_civitas_2008, juels_coercion-resistant_2005}. 

\end{itemize}

Today, the main focus of DAO voting systems has been on integrity—the only one of these properties that is truly indispensable for trustworthy governance—and to some extent on public verifiability. Privacy has been a growing concern, however. For instance, Snapshot, the most popular voting system for DAOs, recently incorporated support for ballot privacy—although it lasts only for the duration of the ballot-casting period \cite{snapshot_labs_wwwsnapshotorg_nodate}. The properties of privacy and integrity are in fact in fundamental tension. For example, one way to achieve public verifiability is to publish all ballots and their tally on chain, but this approach would undermine ballot privacy. Simultaneously achieving integrity and privacy is one of the main focuses of the verifiable-voting literature. Translating results from that literature to the setting of DAOs is an open problem—as is, more generally, a practical realization of the full set of four voting security properties. 

Coercion-resistance is an especially tricky property to enforce. Recognizing the risk it poses in blockchain systems to DAOs and even to staking systems, researchers have considered a range of various countermeasures, such as Buterin’s Minimum Anti-Collusion Infrastructure (MACI) \cite{buterin_minimal_2019}.

Unfortunately, blockchains aren’t just an attractive platform for implementing voting schemes, but also for attacking them through orchestrated coercion or vote-buying. This is the idea behind what is called a Dark DAO \cite{daian_-chain_2018}, a decentralized vote-buying system that uses a secure enclave to conceal the identities of its participants. While not yet deployed in the wild, Dark DAOs are nonetheless a practical threat, realizable today using enclave-based blockchains such as Oasis and capable of attacking DAOs on other chains such as Ethereum \cite{austgen_dao_2023}. Dark DAOs break MACI and other vote-buying prevention schemes.

Effective countermeasures to Dark DAOs—and, in general, practical ways of achieving coercion-resistance in DAO voting—are a critical open subproblem within the larger challenge of realizing secure voting systems for DAOs \cite{kelkar_complete_2023}.

\subsection{Modeling and formal verification for governance}

\textit{Summary: There are many modes of failure within DAOs and many ways in which they might achieve undesirable outcomes. It is hard—but not impossible, given the computational nature of DAOs—to define conditions that provably or statistically constrain such failures. Computer science can help define conditions and build mathematical models for DAOs that produce formal and statistical guarantees about DAOs’ behavior.}

Existing research within social choice theory and computational social choice gives conditions and limitations on the performance of voting and preference aggregation systems, most famously Arrow’s impossibility theorem. While results from social choice and its variants are certainly relevant to DAOs, the organizational logic embedded within a DAO usually extends far beyond the exact voting mechanism. This broadness makes them susceptible to a wide range of other attacks, exploits, and failures beyond those salient in social choice (e.g. 51$\%$ attacks, sybil attacks, free-riding on governance, lack of engagement), but it also sits them downstream of a number of potential improvements.

Formal verification involves the use of mathematical methods to prove the correctness of the implementations of algorithms. Since the hack of The DAO, substantial work has been done to verify the functional correctness and security of smart contracts using formal methods \cite{bhargavan_formal_2016,murray_contracting_2019}, complementing a strong tradition of industry-driven security audits for smart contracts.

There are existing efforts in game theory and computer science to build computer-aided design (CAD) tools for governance \cite{frey_composing_2023,ghani_compositional_2018}, i.e. a modeling language that captures the basic primitives and incentives necessary to design and analyze governance systems. These languages already have applications to contract theory (the economic analysis of legal contracts) and to smart contracts, and could be directly extended to the smart contracts that constitute DAOs.

While agent-based models do not typically produce formal guarantees, they are often quite valuable for exploring and drawing insights into a complex system \cite{abar_agent_2017}. Agent-based models can provide a useful framework for simulating and understanding the dynamics of DAOs, allowing researchers to test various configurations and observe their impact on system behavior \cite{voshmgir_foundations_2019}.

Reachability analysis, a concept borrowed from robotics \cite{althoff_reachability_2010,zargham_generalized_2022}, can be applied to DAOs to determine the feasibility of achieving certain states through unknown sequences of actions. This approach involves defining a "governance surface" analogous to a control surface, which can be used to identify possible outcomes and transitions within the DAO's governance process. By understanding the reachability of different states, we can better evaluate the robustness and adaptability of a DAO's governance structure \cite{zargham_aligning_2022}.

Additionally, computer-aided design methodologies, particularly model-based systems engineering (MBSE) for complex systems \cite{douglass_chapter_2016}, can provide valuable insights into the design and management of DAOs. MBSE focuses on creating, analyzing, and managing complex systems using models, which can help in addressing the unique challenges posed by DAOs. In the context of DAOs, it is crucial to distinguish between verification and validation, as they determine the effectiveness of a software system deployed to administrate a governance process within the DAO. Verification ensures that the system meets its specifications, while validation ensures that it fulfills its intended purpose in the real-world context \cite{cetinkaya_verification_2007}. The extent to which a particular set of governance algorithms entrenched as smart contract logic is suitable for the context in which they deployed is a question of validation, rather than verification \cite{magazzeni_validation_2017}.

In conclusion, the interdisciplinary nature of DAOs opens up opportunities for integrating various computer science concepts and methodologies to analyze and improve their governance structures. By leveraging formal verification, reachability analysis, agent-based models, and computer-aided design methodologies, we can build a more robust understanding of DAOs and develop more secure, efficient, and resilient governance systems.

\subsection{Human-computer interaction for DAO interfaces}

\textit{Summary: DAOs have a usability problem. Human-computer interaction can help design better interfaces and systems through which members and stakeholders organize their decentralized communities.}

Human-computer interaction (HCI) is a subfield of computer science that focuses on the design, implementation, and evaluation of interactive computing systems. In particular, HCI tends to focus on the interfaces and processes through which humans interact with computers and digital systems, as opposed to the internal workings of those systems. Many innovative features of online platforms are first incubated as research projects within the field of HCI \cite{dix_human-computer_2003}, e.g. within content moderation \cite{fan_digital_2020}. While HCI researchers have contemplated and sometimes approached the design questions of blockchain applications \cite{elsden_making_2018}, they have rarely worked on topics specific to DAOs. There are a number of dimensions within which HCI can contribute to DAOs:

\begin{itemize}
	\item \textit{Promoting usability}. Right now, the learning curve to enter a DAO is extremely steep, ruling out use by users with more diverse backgrounds and abilities. Further, the rigid, irreversible nature of smart contracts makes many interactions prohibitively risky for users without a significant financial cushion. HCI can inform the design of inclusive systems that cater to varying levels of expertise, language, and accessibility needs, ensuring that DAOs are available to a wider audience.

	\item \textit{Bring rigorous methodology into the design process for DAOs}. There is a need for more consistent design methodology in this space. For example, while many projects have built dashboards and visualizations for DAOs, a rigorous, HCI-driven evaluation process can help iterate and refine these systems.

	\item \textit{Testing specific design interventions}. A key component of DAOs is to engage participants in governance. While the industry has so far been drawn to cryptoeconomic systems that incentivize participation such as Paladin, design interventions that focus on better interfaces for eliciting and aggregating preferences may be less costly and more effective. HCI is particularly well-suited for evaluating the cognitive and social factors that influence group decision-making in DAOs.

	\item \textit{Exploring bots and technical interventions within DAOs.} Many DAOs make use of bots and automated tools within their communication platforms. These bots facilitate not only blockchain-enabled features such as token-gating but also various forms of governance decision-making. Various studies could explore this dynamic, for example calculating a DAO’s ``sense of virtual community" \cite{koh_sense_2003} and then measuring how bots affect that metric.

\end{itemize}
\subsection{DAOs as social computing systems}

\textit{Summary: DAOs, as novel forms of digital organization, are germane to many core questions in social computing, an interdisciplinary subfield of computer science. Social computing can help provide a design framework for DAOs and contextualize them within a wide range of digital institutions including open-source software, reputation systems, online auctions, and crowdfunding platforms.}

Social computing is a subfield of computer science that studies the interplay between people, technology, and society. As a highly interdisciplinary field, many research questions motivated by social computing overlap with those motivated from political science, sociology, and economics—e.g. how do we understand, model, and visualize the behavior of DAOs, what is novel and not novel about them, how does this particular digital infrastructure change the way that people behave? Unlike social science, social computing research tends to center computing technology as the primary object of study. Although social computing is closely related to HCI, it seeks to understand or design computing systems for social groups and social behavior while HCI research typically focuses on individual users. In this sense, DAOs are classic examples of social computing systems.

Despite clear affinities, there has been relatively little work in social computing on DAOs. DAOs have been discussed as promising sites for future research in social computing work on governance \cite{frey_this_2019} and some early work has treated DAOs as objects of study \cite{arroyo_dao-analyzer_2022,lustig_intersecting_2019}. Social computing can potentially offer a unique and important perspective on DAOs, one that situates DAOs within the larger context of socio-technical systems and emphasizes the role of technology in shaping social processes. Social computing is also often more experimental and interventionist than traditional social science, which matches well with the open and permissionless character of DAOs. Some potential research questions include:

\begin{itemize}
	\item How do specific technological design choices in the context or affordances shape social behavior or social interaction within a DAO? How will specific technological choices made by the creators of DAOs allow, encourage, discourage, or foreclose  certain types of behaviors and outcomes? 

	\item How does social structure (e.g., roles, power structure, norms, expectations about technology, and so on) shape the way that a particular technology is understood or used? For example, how might technically-similar smart contracts ground different communities in very different ways? 

	\item How can we design DAO technology to support certain types of social behaviors or outcomes and to prevent others?

\end{itemize}

\subsection{Measures of decentralization}

\textit{Summary: Research on standardized metrics and definitions of decentralization is ongoing, with approaches ranging from entropy-based indices, SoKs, or empirical analyses of voting concentration. Broader questions about the current state, value, and structures of decentralized systems continue to be explored.}

Research on how to measure DAO decentralization is still emerging, but various works have explored this topic. \cite{feichtinger_hidden_2023} examine the Gini and Nakamoto indices, participation rates, and the monetary aspects of governance. \cite{sharma_unpacking_2023} delve into different entropy concepts and introduce graph-centric decentralization measures. Wright presents a classification system comparing DAOs with other autonomous entities \cite{wright_measuring_2021}. \cite{sun_voter_2023} analyze voting blocs in MakerDAO, using clustering to identify them. Additionally, \cite{austgen_dao_2023} introduce a new, DAO-specific metric called voting-bloc entropy (VBE), merging traditional entropy measures with voting-bloc identification. Another noteworthy concept is "credible neutrality," a community benchmark mentioned in sources like \cite{buterin_credible_2020, buterin_what_2023}.

Moreover, various systematizations of knowledge (SoKs) have sought to assess the landscape as well as propose new measures. \cite{zhang_luyao_sok_2023} note the need for the blockchain community to converge on a standardized definition or metric for decentralization. In their SoK, they first create a taxonomy based on consensus, network, governance, wealth, and transactions. Building on this taxonomy, they apply Shannon entropy to propose a decentralization index for blockchain transactions which they then apply empirically in the context of DeFi token transfers to document that DEXs and DeFi lending protocols appear slightly more decentralized than payment and derivative apps. They also find that EIP-1559 \cite{buterin_eip-1559_2019}, which modified Ethereum's transaction fees, increases decentralization.

\cite{karakostas_dimitris_sok_2022} also present an SoK on the topic of blockchain decentralization, this time taking a stratified approach by first dissecting blockchain systems into multiple strata (hardware; software; network; consensus; economics; API; governance; and geography). They also introduce the ``Minimum Decentralization Test'' as an approach to assessing decentralization.

Finally, another approach to measuring decentralization might focus specifically on governance participation, which \cite{messias_johnnatan_understanding_2023} do by examining decentralized voting to amend DeFi smart contracts. Focusing on decentralized lending protocol Compound, they find a striking concentration of power, with ten voters holding nearly 60\% of voting power. 

In addition to these efforts to develop a systematic measure of decentralization, defining decentralization touches on other fruitful research questions, including:

\begin{itemize}
    \item What is the current landscape of decentralization? 
    \item Why/ when is decentralization valuable? 
    \item To what extent do the unique aspects of web3/ decentralization help or hurt with the challenges of online interactions? 
    \item How do we incorporate the role of foundations and other groups of actors  into our measure of decentralization? 
\end{itemize}
    
One possible angle for mapping the motivations for decentralization could differentiate between the economic; regulatory; technical engineering; and ideological motivations.


\subsection{Democratic infrastructures for governing technology}

\textit{Summary: DAOs are practical examples of an approach to governing technology that foregrounds the role of technology and local communities instead of traditional modalities of regulation. Computer science can help us understand and improve these approaches to the governance of technology.}

The governance of technology and algorithms is more often approached through an institutional or legal lens than a computational one. Studies of past examples tend to foreground regulatory mechanisms such as the law and institutions; for example, research on the governance of internet protocols tends to focus on technical standards bodies such as the IETF or elite forums such as the IGF, assuming that control over the internet protocol was out of the reach of ordinary users \cite{denardis_protocol_2009}. More recent work has highlighted ways to involve ordinary users in the design and governance of digital technologies such as machine learning algorithms \cite{smith_keeping_2020}, but these efforts have not scaled to more widespread use. DAOs are on the forefront of these more community-oriented approaches to the governance of technology. However, DAOs are also quite limited insofar as the technologies they govern are usually protocols and contracts deployed on a blockchain. How can DAOs or related social technologies govern technologies such as websites and domain names, server infrastructure, telecommunications protocols, and even pre-trained AI models—none of which are not implemented in a Web3 context? Perhaps sensitive algorithms in content moderation or policing could be content-addressed, creating the means through which affected parties could participate in algorithmic oversight.

\subsection{Data sets and data standards}

\textit{Summary: DAOs, despite their public nature, are still difficult to study due to lack of standardization, reliance on off-chain services and APIs, and other challenges with data indexing and integration. Computer science can help define data sets, schemas, and taxonomies that would enable easier data collection, faster experimentation, more consistent statistical tests, easier formal verification, and even the training of better AI models. }

DAOs are nominally fully public and transparent, but there are many questions that researchers might like to ask which cannot be readily answered given the current data infrastructure of DAOs. For example:

\begin{itemize}
	\item How many DAOs exist on a given blockchain at any moment?

	\item Who are the actual members of a DAO?

	\item What is the contract space of a DAO, i.e. all the on-chain contracts and multisigs controlled by or associated with a DAO?

	\item What is the level of engagement of members over time?

	\item What is the legal status of a DAO, if any?

	\item What is the organizational structure of a DAO, e.g. in terms of subDAOs?

	\item What is the spread of different voting schemes across all DAOs?

	\item What kinds of activities does a DAO engage in?

	\item What rights are afforded to the members of a DAO? Which rights are code- versus rule-enforced?

\end{itemize}

Again, the issue in these cases is not that the information is private and must be negotiated for (as with typical Web2 platforms) or that the precise definition of a metric is being contested (e.g. of decentralization). Most data relevant to DAOs is publicly-accessible, but interpreting, integrating, and organizing that data is hard. Some of these issues arise from lack of on- and off-chain standardization across DAOs and DAO frameworks, which is being addressed by nascent efforts in the ecosystem to develop standard schemas and taxonomies \cite{daostar_wwwdaostarorg_2023, tan_erc-4824_2022}. Others arise because engaging with a DAO involves interacting with many off-chain platforms, and these platforms either do not report data or do not integrate with early standards efforts. And while some data sets have been built for DAO smart contracts \cite{korpas_governance_2023}, DAO constitutions \cite{tan_constitutions_2022}, and political sentiment \cite{korpas_political_2023}, other data sets simply still need to be built.

Despite the present lack of data infrastructure, DAOs present a unique opportunity to build ``live" data sets, whether through public subgraphs and other automated mechanisms, through technical standards implemented by DAO frameworks and other service providers, or through institutional designs that incentivize reporting on a DAO-to-DAO basis. Having access to such live data is not only a boon to scientific analysis; it is also an important basis for providing real-time feedback to DAOs, drastically reducing the timeline for research to impact the real world. DAOs themselves could substantially benefit from access to such data, whether for operational, reporting, or governance purposes.

\subsection{Automated testing and automated experimentation}

\textit{Summary: DAOs do not have a rigorous testing framework or any infrastructure for experimentation such as which exists in psychology, medicine, or AI. Computer science can help supply the protocols and libraries for automated testing and automated experimentation.}

While most experimentation with DAOs currently takes place outside of academic settings, various quality-of-life improvements to tooling and infrastructure could make it much easier for researchers (and ordinary users) to develop, deploy, and test DAOs.

Automated testing of DAOs could be especially helpful. While there exist platforms for smart contract testing such as Tenderly, none are currently specialized to DAOs. Automated tests or benchmarks could be constructed to query a given DAO’s: 

\begin{itemize}
	\item susceptibility to the loss or inattention of key participants or administrators \cite{patka_exploiting_2022}, 

	\item resilience to 51$\%$ attacks, 

	\item speed and responsiveness to crises, 

	\item ability to handle disagreements, 

	\item susceptibility to forking, 

	\item and likely responses to various adverse financial or technical events.

\end{itemize}

Such tests and benchmarks follow a rich tradition of testing and quality assurance within software engineering and could be used to validate a DAO’s fitness for purpose. Further tests could be devised that could be structured as challenges or games that a living DAO could run through, akin to fire drills, war games, and other tests of capacity.

Another approach to automated experimentation involves working directly with the many already-extant DAO frameworks. A governance framework for DAOs, or just DAO framework, is a smart contract template or set of such templates that can be used to deploy a DAO on a blockchain, typically with an associated management interface. Many of these frameworks are moving toward specialized use-cases, e.g. Safe’s focus on multisig management compared to Aragon’s focus on permissions infrastructure. A modular, open-ended plugin or library that can abstract over these DAO frameworks and connect them to rigorous testing tools—comparable to how libraries like scikit-learn or TensorFlow drastically reduce the hassle of building and testing new learning algorithms in AI—could be extremely useful.

Ultimately, the elements above could be synthesized into an experimental platform for DAOs and DAO frameworks. For example, software such as \href{https://dallinger.readthedocs.io/en/latest/}{Dallinger} \cite{suchow_dallinger_2023}, \href{https://languagelearninglab.gitbook.io/pushkin/}{Pushkin} \cite{hartshorne_thousand_2019}, and \href{https://expfactory.github.io/}{Experiment Factory} \cite{sochat_experiment_2016} automate significant parts of the process for producing classes of psychology experiments via automated statistical design, a domain-specific language for experimental setup, and simple Mechanical Turk and TaskRabbit integrations. They allow a researcher to build and run a psychology experiment in a matter of hours—something that would have taken them months to plan and run previously—and to run ``massive" versions of traditional experiments that were previously impossible. A similar platform for testing governance experiments, including distinct modules for voting, reputation, and privacy as well as integrations for on-chain deployment and data tracking, could vastly accelerate experimentation as well as data collection and synthesis.

\section{Economics}

\begin{quote}
Editors: Jason Potts, Chris Berg, Sarah Hubbard. Contributors: Jeff Strnad, Sarah Hubbard, Jason Potts, Chris Berg, Divya Siddarth, Sarah Horowitz, Michael Zargham.
\end{quote}


Economists who have sought to study blockchains have developed a number of distinct approaches. These include: (1) \textit{cryptoeconomics} (game theory and mechanism design applied to incentive problems in protocol design, also known as \textit{tokenomics}); (2) \textit{microeconomics}, including finance, industrial organization, and public goods, typically using some formulation of Chicago price theory, Berkeley competition theory, and Harvard/MIT welfare economics (e.g. Abadi and Brunnermeier 2018 \cite{abadi_blockchain_2018}, Cong and He 2019 \cite{cong_blockchain_2019}, Buterin et al 2019 \cite{buterin_flexible_2019}, Catilini and Gans 2020 \cite{catalini_simple_2020}), and (3) \textit{institutional cryptoeconomics}, which draws on new institutional economics, public choice theory and Austrian economics (e.g. Davidson et al 2018 \cite{davidson_blockchains_2018}, Berg et al 2019 \cite{berg_understanding_2019}. There is also a distinct school of \textit{crypto monetary economics} (e.g. George Selgin \cite{selgin_bitcoin_2022}, Larry White \cite{white_market_2015}, William Luther, et al \cite{luther_cryptocurrencies_2016}). There is overlap between these mostly complementary approaches, and it is useful to understand these as examining broadly different parts of the crypto-economy at different levels of focus and with somewhat different tools. Nevertheless, the common core is analysis of the crypto-economy through the lens of rational behavior subject to incentives, which are designed mechanisms in which blockchain technologies (including smart contracts) are new technologies that lower specific costs. The purpose of analysis is to trace these changes in costs, through changes in markets, through to new equilibrium outcomes.

The economic approach to DAOs is a new field that sits within the standard economic approach to the theory of the firm as developed by leading economists such as Ronald Coase, Oliver Williamson, Armen Alchian and Harold Demsetz, Jensen and Meckling, and Oliver Hart. In 1939, Ronald Coase asked ‘why do firms exist?" and then answered that they were an organizational technology that economized on the transaction costs of using the market \cite{coase_nature_1937}. This ‘transaction cost’ perspective is the foundation of New Institutional Economics. William and Hart developed this approach further by examining issues of trust and opportunism, and the economic logic of contracting. Jensen and Meckling developed a theory of the firm as a ‘nexus of contracts’ and Alchian and Demsetz developed a model of a firm as a ‘private market’. These approaches sought to understand the existence of firms, as well as their comparative advantages as technologies for coordinating people and capital, through the lens of efficient contracting over situations that involved missing information and uncertainty (i.e. asymmetric information, hazards of opportunism and problems of trust).

A small but growing literature seeks to develop economic analyses of DAOs. Lumineau et al. \cite{lumineau_blockchain_2021} and Santana and Albareda \cite{santana_blockchain_2022} offer useful recent surveys. This literature largely overlaps with the \textit{theory of the firm} and therefore includes topics such as: agency, contracting, entrepreneurship, investment under uncertainty, team production, asset specificity, boundaries of the firm, and corporate governance. However, the economics of DAOs also sits within a broader literature on voluntary or private organizations, integrating the economic theory of \textit{clubs} and the economic theory of \textit{commons} \cite{rozas_when_2021}. Clubs and commons theory is also focused on governance and collective decision-making, as well as the creation and use of common pool resources or local public goods. 

\subsection{Institutional economics}

\textit{Summary: Part of the rationale for DAOs is that they are a technology, much like firms, for minimizing transaction costs and ‘costs of trust’. Institutional economics can help explain present DAO structures and suggest new designs for DAOs based on a rigorous analysis of the DAO’s cost functions.}

Why do DAOs exist? This is the same question that Ronald Coase asked in 1939 that established the concept of transaction cost and the field of New Institutional Economics as a comparative analysis of institutions. Coase’s answer was that firms exist to minimize the ‘transaction cost’ of using a market. These transaction costs are the costs of searching for counterparties, writing contracts, monitoring actions, haggling and bargaining and so on. They include search and information costs, agency and monitoring costs, contracting and decision costs; to a considerable extent they are ‘costs of trust’ that would not exist in a world of perfect costless information with total transparency in a world of counterparties who only ever told the truth and always did what they promised. Such a world would lack uncertainty and opportunism. Firms are a technology for organizing people and resources into coordinated actions. Firms exist when it is cheaper to use that technology than an alternative institutional technology, namely the market. Firms exist because they economize on the costs of using the market. The particular boundaries of firms (e.g. vertically integrated, multidivisional, cooperative, etc) are competitive consequences of those cost-specific cost advantages over a range of margins. The same logic applies to the economic analysis of DAOs.

A DAO is an organization made in part using a new technology: smart contracts. These give DAOs different competitive advantages in relation to transparency, monitoring and auditing, as well assurance and expectation. As such, DAOs have different cost functions with respect to a range of key operational and competitive functions within economic coordination, e.g. some have argued that DAOs exist to economize on the costs of trust compared to firms and markets \cite{berg_understanding_2019}. Note that not all costs are significantly lower. On some dimensions such as regulatory uncertainty and integration with legal and other external systems, DAOs are often more costly than traditional industrial corporations. In other words, DAOs are institutional competitors to other forms of economic organization (also including clubs, coops, trusts, governments, as well as firms and markets). 

The problem, then, is to develop a general economic theory of DAOs that (1) explains their existence in terms of specific costs and the way in which those costs are internalized, (2) grounds those costs in the logic and design of (smart) contracts and other resources within the DAO, and (3) links those resources to a theory of DAO strategy (see ``Dynamics and strategy", below). Within this general framing, we can then organize a range of open questions:

\begin{itemize}
	\item \textbf{Structure of a DAO}. Analogous to the theory of the organizational structure of a firm (pioneered by Alfred Chandler \cite{chandler_strategy_1969}, and the U-form and M-form corporation), what is the governing economic logic determining the structure of DAOs and subDAOs?

\end{itemize}
\begin{itemize}
	\item \textbf{Information economics of DAO coordination}. The economic efficiency of firms and markets (and governments) is in part based on their comparative efficiency in different types of information processing, e.g. in price signals in markets or entrepreneurial coordination (in the work of F.A. Hayek \cite{hayek_kinds_1964}, Mark Casson \cite{casson_information_1997, casson_nature_1996}, Israel Kirzner \cite{kirzner_competition_1973}, George Stigler. An open question is to empirically and theoretically analyze the economic mechanisms and economic efficiency of information flow and processing in a DAO, including through new data sets not present in a typical firm.

	\item \textbf{Entrepreneurship versus management}. Firms are both effective organizational mechanisms for starting and building new projects and engaging in entrepreneurship (e.g. entrepreneurial capitalism, \textit{viz}. Schumpueter \cite{schumpeter_capitalism_1942}, Kirzner \cite{kirzner_competition_1973}), as well as efficient mechanisms for operating large going concerns e.g. managerial capitalism, \textit{viz}. Porter, Bloom and van Reenan). Do DAOs share this joint comparative efficiency, or are they better for starting new projects due to low costs of organizing (entrepreneurship) or for operating existing projects due to low costs of participation in cooperative joint ownership (management)? This is both an empirical and theoretical research question.

	\item \textbf{Management and incentives in DAOs}. What incentive structures (tokenized or otherwise) are evident across DAOs that can be designed and used for effective DAO management? 

\begin{itemize}
	\item How can tendencies towards speculation be identified and minimized mechanistically? What are the impacts of financialized governance on efficacy, ability to fundraise, and internal execution processes? How should token distributions be structured? 

	\item How should incentives be created to get work done from contributors? What is an effective compensation structure for aligning delegate incentives?

	\item How to define and create effective bounties? (Note this issue relates broadly to incentivising community production of local public goods)

\end{itemize}
	\item \textbf{DAO constitutions}. Public choice theory and constitutional economics makes specific predictions about the economic efficiency of different types of constitutions \cite{black_rationale_1948,buchanan_domain_1990}(Black 1948, Buchanan and Tullock 1962 \cite{buchanan_calculus_1962}, Buchanan 1990 \cite{buchanan_domain_1990}). Early data on DAO constitutions suggest a very different form for DAO constitutions, which are designed to complement existing smart contracts \cite{tan_constitutions_2022}. What are the economic principles of DAO constitutions? Does the efficiency principle of supermajority or unanimity hold in internalizing externalities?

	\item \textbf{Mapping rules for DAOs}. Where corporate firms use hierarchical decision-making, DAOs tend to be flatter and egalitarian, like a commons \cite{ostrom_governing_1990}. A large-scale empirical and theoretical research question is to map the governance rules of all DAOs, e.g. through some form of institutional grammar \cite{crawford_grammar_1995,frantz_institutional_2021} or building earlier work on capturing data sets of DAO smart contracts \cite{korpas_governance_2023}.

	\item \textbf{Cultural and institutional economics of DAOs}. A significant literature in development economics focuses on the role of culture in shaping institutional evolution and economic outcomes (e.g. Grube and Storr \cite{grube_culture_2015} 2015, Storr \cite{storr_understanding_2013}, Chamlee-Wright 2002, 2008 \cite{chamlee-wright_cultural_2002,chamlee-wright_structure_2008}). This literature can be usefully re-applied to the study of DAO communities and ecologies.

	\item \textbf{Externalities and social welfare}. What are ‘market failure’ equivalents for DAOs, and what types of goods can be provided by different actors within a given DAO’s ecosystem? How can public goods funding mechanisms piloted in DAOs (e.g. see Gitcoin's quadratic funding rounds) extend existing economic theory, particularly in maximizing positive-sum returns to semi-private, anti-rival goods? How do DAOs bound their club goods? What digital property rights should be enforceable? What public ``bads" can be reduced by DAOs?

	\item \textbf{Infrastructure and public goods}. Due to the relatively low cost of distributed ownership and participation, DAOs have characteristics of clubs, and are therefore effective institutional mechanisms for the private provision of local public goods or quasi-public infrastructure, applying to both the private individual and private collections. Such goods can be provisioned by sub-committees within a DAO or by a trust or foundation under DAO oversight \cite{potts_economic_2021}. A major open question in this arena concerns the design and analysis of incentive mechanisms to provide different types of local public goods, e.g. taxes, bounties, grants, prizes, retroactive funding, and so on \cite{buterin_flexible_2019}. 

\end{itemize}
\subsection{Case studies within institutional economics}

\textit{Summary: Economists do not currently have a good way of measuring and quantifying the transaction and governance costs of DAOs. To make progress on this problem, we can build on top of particular case studies that compare DAOs with comparable firm or market forms of coordination. }

The overarching research project in the institutional economics of DAOs is to map, quantify, and analyze the transaction costs and governance costs within DAOs versus comparable firm or market forms of coordination. In particular, quantifying these costs forms the basis for comparative static institutional analysis. From an empirical perspective, such an examination must begin with case studies of particular application domains before building to sectoral data sets. For example:

\begin{itemize}
	\item \textbf{Exchanges}. What are the costs within centralized exchanges such as Coinbase versus those in decentralized (and DAO-governed) exchanges such as Uniswap? How do costs compare between even different decentralized exchanges?

	\item \textbf{Unions}. What are the collective action costs within traditional labor unions versus in those in (theorized) DAO-based unions \cite{allen_cryptodemocracy_2019}?

	\item \textbf{Courts}. What are the governance costs within traditional governance institutions such as courts as compared with those in blockchain-based dispute resolution systems such as Kleros, Celeste, and Aragon Court?

	\item \textbf{Open source}. Open-source software is a good example of a public good that is easier to build than to maintain \cite{eghbal_working_2020}, and DAOs have previously been theorized as an evolution of open-source communities \cite{korpas_political_2023}. In a DAO, what is the balance between costs associated with developing such goods and costs associated with maintenance and operations; how do these costs compare with costs in typical open source communities?

\end{itemize}

Note: while institutional comparative statics is an empirical research exercise, it is also a \textit{market test} \cite{alchian_uncertainty_1950}, as we expect that DAOs with lower relative costs will survive and grow, whereas DAOs with higher costs will be outcompeted by alternative forms of organization. Of course, that claim only holds in the long run and under the selection pressure of fair and robust market competition, so there are many reasons that short or medium run results may depart from economic optimal conditions and predictions. 

\subsection{Corporate governance and principal agent problems}

\textit{Summary: Research and best-practices in corporate governance can help inform existing DAO practices, especially with respect to principal agent problems between managers and owners. On the other hand, there are many open questions about whether and how DAOs can imitate various aspects of corporate structure.}

Separation of ownership and control is an important source of value in the modern corporation, but it is costly and requires governance. The field of corporate governance studies this trade-off and its economic consequences, though its results apply far beyond traditional corporations, e.g. principal agent problems between labor unions and union bosses. For example, Davidson and Potts 2022 \cite{davidson_corporate_2022} argue that corporate governance, not democratic voting, is the correct model for understanding decision-making in DAOs. DAOs offer the prospect of significantly reducing agency problems in organizations through lowered costs of governance and collective-decision making, e.g. through new forms of liquid democracy. But aside from the typical social choice critiques of mechanisms such as liquid democracy, many of these mechanisms expose agents to higher information and attention costs. Some of these costs may be offset with greater potential for automation.

A central line of open questions relates to what we can learn from from models and structures explored in the corporate governance literature, covering decision-making hierarchy, accountability (to boards or other entities), and measures of success. What best practices should DAOs learn from traditional corporate governance, and what best practices should they reject? If a DAO chooses to, how can it appoint board(s) of directors, advisors, and C-suite executives with roles and powers as expected in modern corporations? Is there room for an on-chain CEO? More generally, can DAOs and smart contracts change the parameters of principal agent problems? On the economics of corporate governance, see Jensen and Meckling \cite{jensen_theory_1976}, Fama and Jensen \cite{fama_separation_1983}, Shleifer and Vishny  \cite{shleifer_survey_1997}. On applications to blockchains see Yermack  \cite{yermack_corporate_2017}.

\subsection{Dynamics and strategy}

\textit{Summary: Like traditional firms, DAOs need to make strategic decisions based on an evaluation of their internal constraints as well as of the external, macroeconomic environment. Microeconomic and macroeconomic theories can help provide the analytic foundations for DAO strategy, especially for competition and cooperation within an ecology of other DAOs, L1s, and traditional institutions such as governments.}

A range of questions also concern the dynamics of DAOs. There are the internal dynamics i.e. how we expect a particular DAO to evolve and change over time; these dynamics can be measured through a range of variables including organizational structure, code, revenue, activities, and culture. There are external dynamics, which relate to the behavior and evolution of populations of DAOs within an ecology of other types of economic organization. Between the internal and external dynamics is the emerging field of DAO strategy (and DAO policy), which is effectively how to navigate the microdynamics and macrodynamics in a world of both other DAOs, other competing organizations and institutions, and other complementary organizations and institutions in the network ecology. DAO strategy will thus include study of phenomena including DAO-to-DAO partnerships \cite{dagdelen_exploring_2021}, mergers and acquisitions (ref. Fei and Rari \cite{peterson_match_2021}, Gnosis and xDAI, Hermes and Polygon), and DAO-to-government relations. These phenomena give rise to a range of specific questions, including:

\begin{itemize}
	\item What are the conditions and best-practices for strategic cooperation between DAOs? How do smart contracts, token swaps, and other technical interfaces change the costs and incentives for such relationships?

	\item How can token swaps between DAOs support coalitional dynamics within DAO markets?

	\item What determines coalition formation within DAO ecosystems? What are the dynamics between economic and political coalitions? 

	\item What is the role of inter-DAO governance in structuring the markets that DAOs participate in? E.g. the role of competition governance, antitrust authority, and industry associations or consortia.

	\item What are the costs and benefits of governmental lobbying by DAOs? Empirically, which DAOs are already contributing to such lobbying, and for what reasons?

\end{itemize}

Many problems that arise in the arena of DAO strategy are specific to the layer 1 (L1) blockchain that hosts the DAOs. For example, the Inter-Blockchain Communication (IBC) protocol in Cosmos allows communication between chains; each chain in Cosmos effectively functions as a little DAO connected to other DAOs by bridges.\footnote{ Note that IBC-style communication requires ``fast finality" but the current architecture of Ethereum makes this hard. Figuring out how to extend additional bridge functionality to EVM-based architectures is a problem for computer scientists.} There is competition between bridges, and the bridges are not fungible with each other, which is a problem.

\begin{itemize}
	\item How does bridge competition or strategic partnerships interact with L1 chain competition?

	\item What is the role of L1 foundations and protocols with ecosystems? How do we balance the power of these foundations and ecosystems with the DAOs that operate within them? 

	\item How should L1s fund different DAOs and organizations? What’s the right distribution? 

	\item How can and should L1 foundations support (and possibly regulate) DAOs and other initiatives working on top of their infrastructure?

	\item When the costs of bargaining are too high, public regulation is more efficient; this is the basis of the institutional possibility frontier analysis, see Djankov et al. \cite{djankov_new_2003}. Insofar as L1s represent a form of public regulation on DAOs and other players in the ecosystem, in what circumstances does public provision or standardization at an infrastructural level make more sense than having it be organized at the level of governance, through a DAO?

\end{itemize}
\subsection{Tokenomics and platform economics}

\textit{Summary: Many DAOs are governed through tradeable tokens. ‘Token economics’ or ‘tokenomics’ is a term given to the economic incentives entrenched within a token’s smart contract or protocol, and the particular fiscal and monetary policies DAOs may implement for these tokens through governance. Additional research on tokenomics can help expose the limitations and tradeoffs of tokens and coin-voting as a mechanism for governance.}

Tokenomics relies on the same first principles as platform economics, as designers aim to imbue their tokens with ‘utility’ to satisfy regulatory constraints. Utility most commonly takes the form of facilitating transactions within a platform economy; the supply side provides a service, the demand side provides the service, and the token is incorporated as a tool to facilitate transactions between actors who are not assumed to trust each other \cite{evans_rise_2016,swartz_theorizing_2022}.

Another common consideration is how to initialize token supply across actors contributing labor and capital prior to token release; the initial coin offering (ICO) craze led to large distributions of tokens to founders and investors. A second wave of projects followed the Fair Launch principle \cite{russo_fair_2020} which emphasized all stakeholders starting on an equal playing field. When considering systems where tokens are used for voting, it is especially important to consider the initial token distribution.

One of the most exciting but also most challenging intersections between DAOs and tokens is the concept of a dynamic supply token: tokens which do not have a fixed total supply but rather have rule systems and parameterized smart contracts which determine how they are created and destroyed. The term ``governance surface" was first used to describe the set of rules governing the parameters of the RAI token \cite{zargham_aligning_2022}. Other large DAOs, such as MakerDAO, actually govern parameters which directly and indirectly determine the supply of tokens and the risk borne by token holders.

Tokens are commonly traded on secondary markets such as decentralized exchanges (DEXes), and the prices on these exchanges impact those tokens effectiveness both in their roles as platform economy enablers and as skin-in-the-game for governance. Mechanisms such as constant function market makers, DEX aggregators and bonding curves are used to provide liquidity, and in ideal settings reduce volatility \cite{zargham_economic_2020,zargham_curved_2020}.

Additional economic incentives associated with tokens are subsidies and grants. Subsidies generally target increased adoption and involve distributing tokens to server providers and/service consumers. The concept of yield farming involves users jumping from token to token engaging precisely to capture these subsidies. Curve pioneered the concept of gauge weights where their DAO votes on how to allocate their subsidies across various pools. This kicked off a process wherein a DAO emerged to capture controlling interest in another DAO \cite{thurman_curve_2021}. In the Balancer DAO, users faced down and eventually came to a compromise with a whale who insisted on placing high gauge weights on a pool they controlled in order to capture subsidies \cite{blockscience_applying_2022}. While grants are notably less complex than programmatic subsidies, they most directly surface the relationship between DAOs and traditional challenges in treasury management and public finance \cite{hasu_new_2021}. Unsurprisingly, a class of ‘protocol politicians’ has emerged to attempt to steer decision-making, especially financially impactful decision making in DAOs.

Due to the prevalence of DAOs with a tradeable governance token, token design cannot be cleanly separated from governance. Indeed, governance tokens are often used (perhaps more often used) for speculation rather than for their intended governance functions. Many authors have raised various critiques of token-based governance \cite{schneider_cryptoeconomics_2021,buterin_nathan_2021}. Thus, a key question of tokenomics as applied to DAOs is \textit{how to prevent governance tokens from becoming an end in themselves}. How can DAOs distinguish between voting rights and other considerations?

But for the foreseeable future, many DAOs will still be funded and governed by a form of token-based governance; DAOs with tokens have proven to be an effective mechanism to raise capital for enterprise operations. Given this, a number of questions arise:

\begin{itemize}
	\item What’s the right way to do token distributions, both the initial distribution as well as ongoing distributions? Tokens tend to be heavily concentrated in the hands of initial founders, which leads to centralization and gives contributors less buy-in.

	\item What is the right way to distribute / manage tokens in a given treasury? a lot of DAOs seem to have just one token, so one sees huge fluctuations in the value of these treasuries, which is problematic. Can hybrid models, which use a mix of transferrable and soul-bound tokens, provide the benefits of marketable tokens while minimizing the downsides?

	\item DAOs hold financial assets (digital or real) in collective ownership for future use and disbursement. How can principles of optimal capital management and portfolio diversification developed from principles of modern finance (e.g. Markowitz's modern portoflio theory, Miller's work on corporate finance, Fama and French on risk factors \cite{fama_common_1993}) apply to DAOs? Note that this question is tightly coupled with legal questions of liability, ownership, and legal / fiduciary responsibilities that vary depending on the DAO’s particular legal structure \cite{wright_rise_2021}.

\end{itemize}

\subsection{Labor economics} 

\textit{Summary: Many DAOs, especially protocol DAOs, constitute a new way of organizing production and labor that has evolved from the open source movement and various modes of peer production. Research in labor economics can help us understand this ``future of work" and identify more productive and equitable arrangements. Further, organized labor has many structures and institutions that could benefit from DAO-based infrastructure, but there are many practical and theoretical difficulties standing in the way of adoption that additional research could help us bridge.}

A large range of standard analysis in labor economics that has previously focused on firms and markets carries over to the analysis of employment and work in DAOs. See Ilyushina and Macdonald \cite{ilyushina_decentralised_2022} for a review of this emerging field with a focus on new types of employment. Open questions include:

\begin{itemize}
	\item How many people work for DAOs, and what do they do? Some preliminary statistics exist in industry aggregators such as DeepDAO, Messari, and Dune, but a rigorous global survey would be extremely relevant here.

	\item What labor and employment relationships are being set up in the DAO context, and how do they interact with existing labor and employment rights, gig work paradigms, and ownership structures? Do DAOs form a structurally new mode of joint employee ownership? How does labor contract automation interact with labor rights and current practices?

 \item How might DAOs be used by unions? For example, unions run elections that could benefit from some of the privacy and security properties DAOs. 

	\item How does DAO infrastructure shift the operation of labor markets? What new types of organizations are possible? 

	\item How do bounties incentivise, and what particular tasks are best done through bounties? Note that bounty-based compensation is often inappropriate for ongoing software contributors, who heavily front-load learning.

	\item What are the leading/main DAO compensation/reward primitives?

\end{itemize}
\subsection{Social choice}

\textit{Summary: Use of newly-theorized mechanisms in social choice such as quadratic voting and variants of liquid democracy by DAOs have been the subject of considerable discussion as well as actual implementation in some cases. As such, DAOs represent an interesting testbed for experimentation within social choice and an exciting new source of data; social choice can also suggest improvements to existing governance within DAOs.}

Social choice theory is an important body of work that is relevant to DAO mechanism design. At the same time, DAOs provide a testing ground for how social choice mechanisms perform in a real-world setting. Some of the social choice and DAO design questions are: 

\begin{enumerate}
	\item \textbf{Applicability and effectiveness of quadratic voting}. Quadratic voting has clear applicability to the allocation of public goods, but its creators themselves have stated that ``applications to politics $\ldots$ or corporate governance $\ldots$ remain far more speculative and are not advisable without further experimentation at smaller scales," despite citing their own work suggesting possible application in both domains \cite{lalley_quadratic_2018}. Although there has been a great deal of theoretical interest, quadratic voting has not found many applications in DAO governance to date. It has been ``much discussed but rarely implemented" \cite{moore_what_2023}. What are possible ways in which quadratic voting can contribute to DAO governance?  What are the practical limitations such as evasion through creating multiple identities akin to a Sybil attack? How would quadratic voting (and quadratic funding) in DAOs affect minority rights and the influence of majority stakeholders? What would be the real-world implications of quadratic voting for the efficient allocation of resources and decision-making within DAOs?

	\item \textbf{Long-term vs. short-term decision-making}. What are the observable effects of short-term incentives on DAO decision-making? How can DAOs be structured to prioritize long-term sustainability over short-term gains?

	\item \textbf{Iterative decision-making.} How can DAOs implement iterative decision-making processes that adapt to real-time feedback? What are the advantages and potential pitfalls of rapid iteration in DAO governance? How can existing ideas from cybernetics and control theory provide a lens for thinking about these questions?

	\item \textbf{Hybrid voting systems}. Some DAOs implement parallel or hierarchical voting systems that may employ different voting methods in stages of a single proposal process or within parallel tracks and grant programs. How can diverse voting methods (like ranked-choice voting, quadratic voting, and liquid democracy) be cohesively integrated within a DAO? What are the emergent properties of DAOs that adopt hybrid voting systems?

	\item \textbf{Vote buying.} Systems such as Paladin and Bribe Protocol facilitate vote buying of governance tokens. Can vote buying enhance DAO governance? Vote buying can capture intensity of preferences, but it can also be a vehicle for malicious actors to secure control. What is the balance of positive and negative on that front, and how can that balance be addressed by structuring the vote buying system with appropriate incentives and limitations?  

	\item \textbf{Applicability and effectiveness of liquid democracy}. Blockchain governance includes multiple instances of the use of liquid democracy, and scholars have examined the performance and characteristics of some prominent cases. \cite{li_liquid_2023} provides an excellent example of one such study. The study captures the actual behavioral patterns of participating voters, raising questions such as what to make of the apparent tendency of the studied instances of liquid democracy to concentrate voting among a few delegatees. There also is a substantial theoretical literature on the strengths and limitations of various liquid democracy implementations both from the vantage point of the possibility or impossibility of embodying an effective decision process and from the vantage point of computational feasibility. A prominent example is \cite{becker_unveiling_2021}. These theoretical studies are of obvious relevance to DAOs. A combination of theoretical and observational work can provide answers to the question of whether liquid democracy has a role to play in DAO governance, and, if so, which implementations are likely to be most effective and desirable both with respect to decision outcomes and power dynamics.

\end{enumerate}

Many of these methods have potential in the context of DAO governance, but, at present, their effectiveness and desirability is unclear. We are only in the early stages of gaining understanding on that front. That makes studying the methods both theoretically and observationally all the more important. 

\section{Ethics}

\begin{quote}
    Editor: Reuben Youngblom. Contributors: Reuben Youngblom, Joshua Tan, Tara Merk.
\end{quote}

DAOs are a powerful tool for organizing and governing digitally-constituted entities, but little attention has been paid to the potential ethical implications of housing jurisdictionless, corporate-like structures on an immutable substrate. Despite this, or perhaps because of it, ethical questions abound, ranging from the practical (e.g. are there moral obligations when architecting a DAO?) to the meta (e.g. what kind of ethics are embodied within the DAO structure?) to the ontological (e.g. who is morally culpable if a DAO performs an unethical action? Are DAOs moral agents?). Finding universally-accepted answers to all of these questions may prove to be an intractable problem, but hopefully just being \textit{aware} of some of the open questions will give us an edge as we think through the answers or, more pessimistically, arm us with the requisite knowledge to make good decisions as unforeseen ethical issues inevitably crop up. Early work on DAO ethics has focused largely on DAOs as tools for ethical enforcement \cite{sulkowski_tao_2019}, or blockchain ethics within organizations \cite{sharif_ethics_2022}, or, even more generally, on financial ethics, but this space can be significantly expanded. Useful fields that may offer insights could be moral theory \cite{driver_moral_2022}, tech ethics \cite{brey_ethics_2017}, or organizational ethics \cite{kaptein_corporations_2017}, among others.

In exploring DAO ethics, it is critical to keep in mind that if DAOs are, for example, capable of being leveraged for unethical ends, this does not automatically equate to ``bad." The crux of this exercise is, in part, to examine the extent to which DAOs might be used unethically (but also explore other questions like their moral agency, etc.), primarily so that this line of thinking simply gets surfaced—it’s very possible that, once the potential negative externalities of DAOs are collected and analyzed, the community still sees DAOs as representing a significant net positive. Further complicating the issue is the idea that the very definition of DAOs can be a bit squishy, and as affordances change, so too does the ethical landscape. For instance, a true smart-contract-based DAO, running on a decentralized virtual machine, that is entirely governed by on-chain code, should perhaps be considered differently than a ``DAO In Name Only" (DINO), such as a loosely-organized Discord group or similar. To take this a step further, many of the issues raised below won’t apply to most DAOs, and this is by design. The important question is not whether most DAOs are reasonable, but rather, what is possible at the edges?

As an important note, it is also good to keep in mind that, while devolving into full ethical relativism is likely to be unhelpful, ethics tend to be (or, at least, are arguably) subjective or relative. When statements about ethics are made in relation to DAOs, it’s useful to understand any claims about ``ethical" or ``unethical" as ``whatever an individual understands to be ethical or unethical in this context."

\subsection{Can DAOs be unethical?}

\textit{Summary: What capacities do DAOs have to cause harm in the world? Understanding the ways in which DAOs can be leveraged to create positive change is important, but so is understanding how they might be used to further immoral ends. What kind of undesirable (unethical, net negative, etc.) actions are enabled by DAOs?}

As noted elsewhere in this document, there are over 4000 active DAOs that collectively manage over $\$$20 billion USD. By any metric, this is a non-negligible industry that can affect similarly non-negligible change on the world. Given this potential, it’s important to consider the full scope of possible positive impacts that DAOs may have on organizations \cite{boder_how_2022}, labor \cite{glaveski_how_2022}, coordination \cite{hubbard_beyond_2022}, computation, and the like, but equally as important to explicitly illuminate any potential negative externalities. This is particularly dangerous if we believe that DAOs have the potential to be a tool (or, hopefully, one of many tools) that can be used to reshape society. Then, if DAOs are able to have an unknown but significant impact on the world, it seems shortsighted to assume that the only outcome(s) are likely to be positive. Given this, it is worth questioning the extent to which DAOs are capable of evil. This is a two-part question or, at least, a question that can be interpreted in one of two ways. One interpretation asks whether or not DAOs, as a structure, are able to cause negative consequences that we might consider to be unethical (above and beyond what might be possible as individuals, or under existing coordination architectures). The second is whether DAOs, as a structure, ought to be held morally responsible for these negative consequences—in other words, whether or not they are true moral agents (\textit{see below}).

Of course, the answer to whether or not DAOs are able to cause negative consequences is almost certainly ``yes," in the same way most things have ``the potential" to make the world worse if they are bent into service in such a way by the wrong individuals. Still outstanding, however, are targeted conversations around scenarios that are among the most likely to occur, and what these scenarios mean in practice. This requires community input, as no one small group of academics has the perspective or the collective intelligence to do this well. To that end, it is a worthwhile exercise to think through the ways in which DAOs could be leveraged as a tool for evil. For example:

\begin{itemize}
	\item DAOs could be leveraged to defraud retail investors

	\item DAOs can encode business ethics, but then can also encode rules/outputs that we would consider unethical, e.g. price gouging

	\item DAOs can enable markets for coordination around goods and services that serve antisocial ends, e.g. organ markets or assassination markets

	\item DAOs might replace traditional corporate structures with something that is ultimately worse and less equitable, either from a financial perspective or beyond

\end{itemize}

Further exercises might include (a) placing these within the larger context along a spectrum of good vs. evil; and (b) considering mitigating solutions.

\subsection{Are DAOs moral agents?}

\textit{Summary: It’s not clear whether or not DAOs, as ``autonomous" organizations, ought to be considered full (or part) moral agents, and thus whether or not they should be held ethically liable for their actions. Or, if DAOs themselves are not held ethically responsible, who—if anyone—should be? }

A ``moral agent" is an entity capable of being held ethically accountable for their actions, as they have the ability to do things like conceptualize distinctions between right and wrong, to formulate plans of action that can be against one’s better judgment or interests, and to put these plans of action into motion \cite{parthemore_what_2013}. Whether or not DAOs (or, indeed, any organization) rises to this level is a sticky question.

As an imperfect starting point, it may be helpful to think about organizational structures we are more familiar with: say, a U.S. corporation. On some level, we believe that the actual entity that is a corporation can be held responsible for the unethical actions performed by said entity—not only are corporations recognized as ``legal people" in a number of jurisdictions \cite{werhane_corporate_2016}, but most jurisdictions also levy punishments for misdeeds committed by the corporation primarily at the ``corporate entity" level \cite{cohen_punishing_2019}, and they (as an entity) can arguably even engage in moral interactions \cite{lebar_corporations_2019}. In other ways, this does not fully track with our intuition of a corporation, which is, realistically, an inanimate shell that is incapable of acting without human intervention. To reconcile this, philosophers have proposed the idea of ``group agency," underpinned by the idea that groups can often ``intend" to act (in the same way even a humanless corporation ``intends" to act by virtue of its existence), and can even intentionally act in ways that none of the decision-making, input-providing humans expect or want the group to act \cite{bjornsson_corporate_2017}. Interestingly, this is also often formulated In terms of corporate culture as an emergent property that is unable to be reduced to any individual beyond macro intentions and coincidence \cite{roth_shared_2017}.

The overarching ethical question here is whether DAOs, specifically, have this group agency and/or can ``intend" to act. This is, in part, a question about the implications of autonomy as defined within a DAO context. Shepherd \cite{shepherd_are_2015} has likened corporations to ``extremely dangerous psychopaths [...] capable of manipulating their own responses to achieve the ends they truly value"—a sentiment worth exploring further by extending the thought process to DAOs. The psychopathic tendencies of corporations is due less to the individuals existing within/around the corporate sphere, and more to the environmental factors (up to and including collective culture) that enable a corporation to act in certain ways. Is this notion of group agency, possibly catalyzed by environmental factors, a reasonable framework to apply to DAOs, given that they are a similar type of organization, but \textit{without} many of the controls found in traditional corporations (size, scope, jurisdictional pressures, permissions, centralized control, etc.)?

Other open questions include:

\begin{itemize}
	\item Do DAOs have moral agency?

	\item What might be determinative of moral agency for a DAO?

	\item Who is responsible if a DAO does something anti-social (or evil, etc.)?

	\item What exactly do we mean when we say DAOs are ``autonomous"?

	\item What external factors, if any, can influence the behavior of a DAO?

\end{itemize}

\subsection{Running an ethical DAO}

\textit{Summary: Are there ethical obligations to consider when constructing a DAO, and if so, how ought a DAO be architected such that it follows ethical best practices?}

How do we run a DAO ethically; alternately, how do we code an ethical DAO?This, too, is a multi-part question. DAOs are capable of being used as a tool for evil.However, most nouns can be used for evil, though we don’t consider most things inherently immoral. The more interesting question is the extent to which we (as a community) have an obligation to ensure that DAOs are not used unethically and the steps we ought to take to achieve this. How does one run a DAO ethically? Is there a framework or a set of best practices that can be followed such that we strike a balance between ethical norms that vary between cultures?

Perhaps a good reference can be found in the field of bioethics. Likely as a way of recognizing that overly-prescriptive rules don’t allow for the flexibility to make nuanced ethical decisions in real time, the best-practices framework in (western) bioethics is composed of four guiding principles: Justice, Beneficence, Non-malevolence, and Respect for Autonomy \cite{beauchamp_principles_2009}. One open question is whether constructing a similar framework would be helpful for DAOs, as well as what should be included in this framework. Future work will include fleshing out and defining the components that the community finds most relevant, but a potential framework can be found below.

Potential framework components:

\begin{itemize}
	\item Transparency

	\item Equity

	\item Immutability

	\item Prosocial purpose

	\item Respect for autonomy

\end{itemize}

\section{Law}

\begin{quote}
    Editors: Joni Pirovich, Primavera De Filippi. Contributors: Joni Pirovich, Chris Wray, Morshed Mannan, Primavera de Filippi, Silke Noa Elrifai, Tara Merk.
\end{quote}


As novel mechanisms for organizing and governing, DAOs raise a host of legal issues and risks. Given the scope of existing legal rules, DAO practitioners have repeatedly highlighted the need for greater legal certainty \cite{ghavi_primer_2022}, particularly concerning the sharing of ownership or rights to govern in DAOs with tokens \cite{wigginton_cooperatives_2023}. This includes legal certainty on liability issues for DAO members and contributors \cite{farmer_daos_2022}, as well as potential liability for the organization itself, including when the DAO has been incorporated or formed as a legal entity distinct from its members \cite{jennings_how_2022}. More legal certainty has also been demanded for structuring the interaction between DAOs and traditional entities, particularly non-Web3 native organizations, including in relation to the ownership of traditional physical assets (e.g. real estate or land) \cite{ruane_what_2022} and the engagement of contributors \cite{ilyushina_decentralised_2022}. Furthermore, DAO practitioners are curious to better understand the regulatory landscape and where the application of existing law should be clarified versus reformed in order to formulate clearer policy demands. Lastly, many practitioners are interested in exploring whether and how legal tools can improve governance structures and accountability within DAOs \cite{murray_contracting_2021}.

\subsection{Legal definition}

\textit{Summary: Legal jurisdictions have a hard time classifying DAOs using existing entity types. Legal research can contribute insights as to where existing legal frameworks can be applied to DAOs and when new regulatory approaches are required. }

DAOs currently face many open problems that could benefit from increased attention from the legal community. On the one hand, some legal scholars question the feasibility of DAOs being alternatives to the traditional corporation \cite{low_company_2022}. On the other hand, other scholars observe that while existing laws are well-equipped to address certain challenges presented by DAOs, there are socio-technical features of DAOs (e.g., fluid transnational membership, pseudonymous contributions) and the open-source protocols they govern (e.g. what smart contract parameters are subject to governance) that ‘test’ the boundaries of legal orders—requiring amendment of the law, a reassertion of existing laws, or a complete reappraisal of how the law perceives certain activities and transactions \cite{de_filippi_alegality_2022}. These unique features present a challenge to adequately classifying DAOs as existing default entity types (e.g., general partnerships, unincorporated associations) across jurisdictions \cite{hitchens_decentralised_2023,metjahic_deconstructing_2017,santana_blockchain_2022,schillig_decentralized_2022,wang_decentralized_2019} or ‘wrapping’ DAOs (or parts of DAOs) into existing legal entity forms, such as corporations, cooperatives or other for-profit and nonprofit entities \cite{brummer_legal_2022}.

\subsection{Legal liability}

\textit{Summary: Lack of limited liability is a major concern for DAO participants, as is the potential liability arising from governance participation. Legal researchers can analyze and create frameworks to help address core concerns both for the community DAO practitioners and regulators. }

One major overarching issue that legal scholarship can contribute to is helping us better understand the specific liability risks for members and contributors when operating through a DAO that is not wrapped in a legal entity (whether registered or unregistered, noting that general partnerships and unincorporated associations can exist as legal entities without registration), thereby not benefiting from separate legal personality or limited liability \cite{chiu_regulating_2021}. While the risk of joint and several liability has often been flagged, certain issues can be explored further: 

\begin{itemize}
	\item What are the grounds on which a claim could be brought against a DAO, its members and/or contributors? 

	\item What are the duties of members or contributors (including core developers, administrators or delegates), the fulfillment of which would help avoid such liability? 

	\item How will claims be brought against DAOs, contributors, or its members, individually or collectively? Who will bring such claims? 

	\item When are DAOs most vulnerable to such claims (e.g., at times of financial distress, or following a hack, or strategic forks in the road)? 

	\item What do theories on corporate attribution and rules on attribution tell us about the attribution and allocation of responsibilities and liabilities in the context of DAOs? 

	\item What legal protections and tools are available to DAOs and their members to avoid liability? 

	\item How will such liability be imposed if a judgment is delivered against a DAO or its members? If a judgment is delivered against a DAO or its members, how would the remedy be effective and timely if it must be formally recognised in each jurisdiction that the remedy is to be enforced? 

	\item What civil and criminal remedies are available against DAOs and their members?

\end{itemize}

Ongoing legal proceedings (e.g., Ooki, bZx, Tulip Trading) continue to shed light on how courts view such issues. 

There are also foreseeable liability risks that have yet to materialize in practice, such as the insolvency of a DAO. While a ‘wrapped’ DAO may be subject to the insolvency procedures of a traditional entity in the jurisdiction of registration or seat, it remains to be seen how creditors and bankruptcy courts treat unwrapped DAOs that are unable to pay their debts and how the technical and governance features of such DAOs figure in procedural arrangements and substantive considerations. For example, Oasis.app was ordered by the High Court of England and Wales to use their accidental multisignature signing permissions to return stolen tokens from a vault to the rightful owner \cite{gilbert_counter-exploit_2023}, which offers insights into how the legacy justice system relies on such permissions existing whereas DAO practitioners are designing systems to optimize for security through decentralization so that such permissions do not exist. Research on the insolvency of general partnerships and unincorporated associations, among other things, may shed light on this issue.

\subsection{Financial regulation}

\textit{Summary: As DAOs often issue and rely on cryptographic tokens in their governance and operations, the relation of these financial technologies to existing financial regulation such as securities laws, financial services, and anti-money laundering regulation and taxation. Legal research can help to discern when and where existing regulation can apply, inform new regulatory frameworks and innovate by proposing Web3 enabled mechanisms to achieve regulatory equivalence in these areas.}

Other important open issues for research include the relationship of DAOs with securities, financial services, and anti-money laundering and counter-terrorism financing (AML/CTF) frameworks designed to prevent, detect and prosecute financial crime, which were developed with legacy system entities with centralized management and more static membership in mind.

Within securities law, the key question is: should a DAO’s governance tokens be treated as securities? Inherently related to the processes around legal recognition of DAOs is the question of whether the crypto-tokens required for governance interactions (i.e. governance tokens) are or should be securities or otherwise regulated as financial products. A number of governance tokens have been alleged by the U.S. Securities Exchange Commission (SEC) as securities, in legal actions by the SEC against Coinbase, Binance, and others. These lawsuits will proceed over the course of 2022 and 2023, alongside policy efforts in the US and worldwide to establish the regulatory purview over crypto-tokens as commodities, securities, or new things for which new regulation is inspired by financial regulation but not constrained by it.

There are a range of legal questions within AML/CTF. For example, how can DAOs comply with the objectives and spirit of laws and regulations intended to combat illicit activities, while still retaining their distinctive features (e.g. pseudonymous or anonymous contributors)? More fundamentally, given the shortcomings of existing AML/CTF frameworks \cite{pol_anti-money_2020}, can DAOs achieve these objectives more effectively and in a more privacy-preserving manner than these frameworks? Even more broadly, how can DAOs contribute to the regulatory discourse on the activities and transactions that are deemed to be (prohibited) money laundering? DAOs after all provide a fertile ground for experimentation, enabling novel approaches for structuring the interplay between the technical and legal codification of rules \cite{de_filippi_blockchain_2018}.

Finally, another major open issue for research is taxation: whether and how should DAOs, their members, and contributors be taxed? How are they already being taxed? How may double-taxation and penalties be avoided?

\subsection{Incorporation and legal recognition}

\textit{Summary: Various jurisdictions have attempted to better regulate DAOs by creating bespoke legal entity forms for them. More comparative legal research, especially from the law and society perspective, can identify the benefits and tradeoffs of different approaches, contrast how each behaves in the face of exogenous shocks or changes, and analyze potential convergence of high-level approaches. }

A number of jurisdictions have begun introducing legislation that creates bespoke entity forms for DAOs, starting with the US state of Vermont with its Blockchain-Based Limited Liability Company in 2018 \cite{wright_measuring_2021}, and followed soon by the US state of Wyoming amending its corporate law statutes to enable the registration of DAO LLCs in 2021 \cite{bannermanquist_marshall_2022} and the Marshall Islands also allowing for such registration (with different conditions) in 2022 \cite{bannermanquist_marshall_2022}. While each of the DAO legislations are different, the recognition of legal personality and limited liability of DAOs in all of these jurisdictions are predicated on the DAO registering with a registrar in their jurisdiction. Most recently, however, in March 2023 the US state of Utah adopted a DAO Act that will come into force in 2024, which amends the Utah Revised Uniform Limited Liability Company Act to ‘recognize’ a DAO as a limited liability DAO (an LLD) as equivalent to a Utah LLC so long as it meets certain requirements, such as appointing a registered agent in the state of Utah \cite{fannizadeh_lawmakers_2023}. The Zone Authority of the Catawba Digital Economic Zone also passed in February 2023 specific legislation to recognise a DAO as a limited liability company or unincorporated not-for-profit association \cite{mckinney_pr008_2023}. There are several jurisdictions across the globe that are exploring DAO legislation, including Australia \cite{senate_of_the_parliament_of_australia_select_2021}, New Hampshire \cite{fannizadeh_lawmakers_2023}, Malta \cite{ganado_mapping_2020}, the United Kingdom \cite{uk_law_commission_decentralised_2022}, and St. Helena \cite{brooks_st_2022}, while there are others in which academics have called for their introduction (e.g., North Carolina) \cite{conway_blockchain_2022}. In addition, there have been jurisdictions like Malta that provide certain legal assurances to DAOs without recognizing their legal personality yet \cite{chiu_regulating_2021,magnuson_blockchain_2020}.

Legal recognition of DAOs, and the ensuing questions around regulation of DAOs as another form of distinct legal person or if that is not appropriate then the responsibility and liability attribution frameworks that should apply, is still in its infancy, but there is a proliferation of views on whether and how each should be best achieved. The LAO had a strong influence on the Wyoming DAO LLC legislation \cite{state_of_wyoming_legislature_wyoming_2021}, while MIDAO has championed the incorporation-based approach of registering DAOs by the Marshall Islands \cite{midao_midao_2022}. The Utah DAO Act mirrors many provisions of the COALA Model Law on DAOs, while departing from its text in important respects \cite{fannizadeh_lawmakers_2023}. And the ongoing discussion on DAO legislation in Malta reveals that their approach to create a DAITO (Decentralized and Autonomous Innovative Technology Organizations) will also depart from the approaches mentioned above \cite{ganado_mapping_2020}. Future research could compare these different legislative approaches to regulating DAOs, particularly from a law $\&$ society scholarship perspective, as well as assess their resilience in the face of changes brought on by market developments, lawsuits, and other regulatory actions. It remains to be seen whether there will be a convergence of approaches on how DAOs are regulated, particularly as the questions outlined above are addressed, and if one ‘model’ ultimately prevails. Research on private international, stateless, marine law, space law, and transnational law approaches as applied to DAOs is also of great interest and may inform new innovative mechanisms to place DAOs within regulatory frameworks.

\subsection{Dispute resolution systems}

\textit{Summary: Smart contracts cannot cover disputes that have not or cannot be codified in software, forcing DAOs and DAO members to settle disputes and seek recourse through the legacy legal system. Legal research can contribute to studies of and solutions for on-chain versions of dispute resolution.}

Plain language laws have a number of use-cases that deterministic code cannot cover, and vice versa, meaning that DAO members often have to resort to interacting with the legacy legal system in order to seek redress for grievances, to affect assets not governed by a smart contract, or to go around the contract logic by compelling off-chain enforcement. This suggests a number of different research questions:

\begin{itemize}
	\item What has been the result of experiments with existing on-chain dispute resolution systems such as Kleros, Aragon Court, and Celeste \cite{ast_when_2021}?

	\item How do such on-chain dispute resolution systems compare with other forms of alternative dispute resolution? How have they fared? Are there fundamental limits to code-is-law-style dispute resolution systems?

	\item How do we incentivize more efficient participation and fairer judgments in these dispute resolution systems?

	\item Are there ways of integrating smart contract enforcement with other forms of alternative dispute resolution?

\end{itemize}

While DAOs may want to interact with legacy legal systems for many reasons (and their ability to interoperate with the existing legal system may even be a competitive advantage relative to other forms of online organization), more research is needed to understand how to allow DAOs to become more autonomous from legacy legal systems.

\section{Organizational science}

\begin{quote}
    Editors: Rolf Hoefer, Ellie Rennie. Contributors: Rolf Hoefer, Seth Frey, Sarah Hubbard, Tony Douglas, Scott Moore, Michael Muthukrishna, Mason Youngblood, Michael Price, Ellie Rennie, Alexia Maddox, Anna Weichselbraun, Kelsie Nabben, Primavera de Filippi, Tara Merk.
\end{quote}


DAOs constitute a fascinating, rapidly emerging organizational form in the wild. In a prototypical DAO, on-chain software is central to the organizing process and every member is able to directly influence the execution of key organizational decisions. Blockchain infrastructure also offers systematically rich, granular, and longitudinal data.

DAOs represent one of the most exciting empirical phenomena in organizational science. As of 2022, DAO participants in over 4000 active DAOs collectively managed over $\$$20 billion USD worth of assets in their treasuries. DAOs are organizations operating in all kinds of industries, organizing around common goals ranging from everything in finance to gaming, politics, culture, arts, and civil society. One illustration of the potential of DAOs and other Web3 developments comes from reflecting on Web2’s impact on organizational scholarship. The Internet’s second wave made new, unprecedented organizational forms possible. For example, Wikipedia demonstrated the scalability of flat organizations and the ability of well-designed technological mediation to organize the incremental contributions of millions of people \cite{puranam_whats_2014}. Poorly explained by existing organizational thought, Wikipedia and other examples motivated the development of Benkler’s theory of peer production, which provided an intellectual foundation for a new generation of organizational scholars \cite{benkler_peer_2015}.

What new organizational forms will DAOs permit, and what flaws will they reveal in how we conceptualize organizational possibilities? New mechanistic corrections for the shortcomings of decentralized organization will further advance the frontier that was opened by Web2. New forms of collective ownership will blur the line between ownership and management regardless of whether they embrace decentralization. Instantaneous algorithmic design and incorporation of ephemeral businesses will blur the lines of Williamson’s transaction cost rationale for firm versus market exchange \cite{williamson_markets_1975}. For their potential to both build on and extend a century of organizational thought, DAOs merit close attention from organizational scholars.

In the following we neither claim nor attempt to depict the breadth and depth of overlap between organizational science and DAOs. Instead, we pick a few topics we believe are interesting and fruitful explorations for those interested in DAOs and organizational science.

\subsection{Organizational imprinting}

\textit{Summary: We still do not fully understand why and how emerging organizations come to adopt their particular social structures and strategies. DAOs, which generate a significant amount of granular, longitudinal data through on-chain processes, can provide valuable insights into the evolution of social structures and practices within organizations, offering opportunities to explore dynamics, histories, and the impact of elements like smart contracts, incentive schemes, and forks.}

A seminal piece in organizational theory is Stinchcombe’s 1965 chapter on \cite{stinchcombe_social_2000}. Over time, this chapter has come to be known for highlighting an observation unique to organizational theory: the phenomenon of organizational imprinting. Organizational imprinting describes the observation that organizations founded at one time tend to have a different social structure from those founded at another time. Once an organization forms its social structure, its social structure tends to persist for an extended period.

While Stinchcombe’s observation is widely taken-for-granted, why and how emerging organizations come to adopt their social structures and strategies has been underexplored \cite{hannan_inertia_1996,johnson_what_2007,johnson_backstage_2009}. In this context, DAOs are interesting because they represent a new, fledgling organizational form. As a novel type of organization, DAOs are in the process of forming and adopting the social structures and strategies that will, according to past organizational imprinting research, persist for a long period of time. This suggests that DAOs represent a fertile ground for researchers to understand the formation phase of organizational imprinting, including when, why, and how an organization does or does not become imprinted.

\subsection{Evolutionary social science}

\textit{Summary: Evolutionary social science applies ideas from biology and evolution to questions in the social domain. DAOs represent a fertile new domain for developing and testing some of these ideas due to the availability of data, their use of technical infrastructure that can be precisely characterized, and the speed at which they evolve.}

Historically, much research on organizational imprinting has drawn inspiration from biology and evolution \cite{lorenz_kumpan_1935} with concepts such as an organization’s ``DNA"; organizations evolving through adaptation and selection pressures \cite{zhong_quantifying_2022}; and complex patterns of mutualism \cite{teblunthuis_identifying_2022}. More broadly, the field of evolutionary social science has a long history of applying evolutionary ideas to questions of social evolution, cultural evolution, and organizational evolution. DAOs represent a fertile new domain for developing and testing some of these ideas.

A research program applying evolutionary ideas to DAOs would need to include at least three components: conceptual mapping, data and tooling development, and comparative case studies.

\begin{enumerate}
	\item \textbf{Conceptual mapping}. To apply evolutionary methods to a field, we need to define a set of evolutionary primitives, in particular mechanisms for variation, transmission, and variation reductive (which, if adaptive, should be selective). What is the mechanism by which variation is created within DAOs? Variation is produced sometimes in the act of transmission—is that true for DAOs? For any evolutionary system you want high fidelity in transmission, but how much actual transmission is there? Finally, what are the mechanisms for selection? Are there informal mechanisms, ways that communities can change, or fork to change? Note: having ``genes" or ``building blocks"—e.g. granular privacy primitives—can be useful but it is not essential for this modeling.

	\item \textbf{Data and tooling development}. How does the infrastructure of a DAO allow a working evolutionary social scientist to engage with either individual DAOs or the broader ecosystem? Given the issues with off-chain data, what affordances and tooling exist to allow researchers to understand and parse what is going on within DAOs? What particular aspects of DAO data are evolutionary social scientists particularly interested in?

	\item \textbf{Comparative case studies}. We want to compare DAO evolution to existing examples within social evolution, for example the evolution of churches \cite{finke_churching_2005}, of constitutions \cite{rockmore_cultural_2018}, within coding competitions \cite{miu_innovation_2018}, and of other online communities such as subreddits \cite{tan_tracing_2018}.

\end{enumerate}

Within the scope of such a research program, there are a number of different open questions:

\begin{itemize}
	\item \textbf{Forks}. Forks are phenomena where an organization splits from another organization yet retains the exact same history as the previous organization \cite{berg_exit_2017}. Due to its public nature, the data and contracts that define a DAO can be easily forked as a technical matter. Indeed forks have happened, e.g. SushiSwap from Uniswap or during the attempted takeover of Steemit, and the threat of forks represent a significant check on bad behaviors by founders and other powerful members of a DAO or blockchain.

	\item \textbf{Infrastructure for merging}. The US federal government is a way for states to merge. The EU is another model for merging. Australia is a very different model. In the infrastructure of a DAO (or a blockchain), what are the mechanisms that would allow you to explore new possible configurations and to possibly incentivize merging of communities?

	\item \textbf{Cultural evolution}. How does the new technical infrastructure of a DAO enable and constrain the variation, transmission, and variation reduction of \textit{cultural} practices within the DAO? How can these infrastructures translate into non-Web3 contexts? 

	\item \textbf{Simulating cultural evolution}. Ideas from cultural evolution could be particularly useful in simulating DAO behavior. There is a suite of methods in cultural evolution for simulating how conformity bias, prestige bias, and payoff bias interact with one another in extremely complex ways, with norms co-evolving with these biases. Considerable time and expense has been applied to collect data from social networks that can then be applied to these models, whereas these methods could be applied directly to DAOs based on already-extant data. Such simulations could eventually be incorporated into DAO design.

	\item \textbf{Theory of cultural evolution}. Insofar as they represent interesting examples of self-governing communities, DAOs could inspire new mathematical treatments of agents’ ability to self-determine and create their own games, as opposed to traditional treatments that operate within a tight framework of phenotypes and behavior.

	\item \textbf{Cultural traditions as cognitive tools}. Cognitive tools are artifacts or processes that facilitate cognition. These are not just physical things but also ``ways of seeing the world", i.e. cultural traditions. For example, ``mountain calendars" from Mexico \cite{ezcurra_ancient_2022} or ``knuckle mnemonics" for months of the year. What kinds of cultural traditions operate as cognitive tools within DAO governance and DAO operations? How can existing infrastructure support such cultural traditions?

\end{itemize}
\subsection{Neo-institutional theory}

\textit{Summary: Building and maintaining legitimacy is a key concern for traditional institutions but especially in DAOs, which often govern through flat hierarchies and informal cultural norms. Neo-institutional theory shifts the focus of research on legitimacy to its communicative aspects and allows for a deeper understanding of how organizations, including DAOs, achieve legitimacy through language and communication rather than mere diffusion of practices. Thus, neo-institutional theory can help practitioners and DAO managers navigate the emergence, intercultural participation, chaos, and complexity that characterize DAO environments.}

For decades, institutional theory and its variances have been a dominant research stream within organizational science. A core concept in institutional theory is the concept of legitimacy \cite{berger_social_2011,dimaggio_iron_1983,friedland_bringing_1991,scott_institutional_1994,suchman_managing_1995,zucker_role_1977}. While much work has measured legitimacy as a function of the successful diffusion of organizational practices, the process by which organizations such as DAOs achieve legitimacy is fundamentally communicative. Consistent with the phenomenological tradition in institutional theory \cite{scott_institutions_2014}, legitimacy is ``built upon language and uses language as its principal instrumentality" \cite{berger_social_2011}. 

A focus on legitimacy as a communicative phenomenon is interesting for two reasons. First, from a theoretical perspective, adopting the view of legitimacy as a communicative concept avoids the suggestion that the diffusion and adoption of organizational structures, practices, and strategies is evidence of legitimacy. Organizational structures, practices, and strategies can be sparsely adopted yet legitimate, and they can be widely adopted yet not legitimate \cite{zilber_institutionalization_2002,green_rhetorical_2004}. Focusing on language also resonates with the linguistic turn in institutional and organizational science \cite{alvesson_taking_2000}. Recent research efforts at the intersection of entrepreneurship and institutional theory, such as the work on cultural entrepreneurship \cite{johnson_what_2007,lounsbury_cultural_2001}, have started to examine the ways that the legitimacy of organizations is a function of language, suggesting an emerging body of work that scholars can build on and extend. Second, a focus on legitimation as rooted in communication is interesting from practitioners’ perspectives \cite{buterin_most_2021,de_filippi_report_2022}. DAOs are internet-native organizations. Coordination is managed digitally in a fast-paced world. When environmental change is high, organizational systems need to adapt quickly, and this work is typically facilitated by people who focus almost exclusively on coordination as opposed to execution —and that is the role of management, or, put into the language of DAOs, that is the role of community managers and delegates \cite{burton_github_2017}.

Persuasion in DAOs often happens more with stories rather than tables, words rather than numbers, beliefs rather than facts. Managers must become skilled at persuasion with stories, words, and beliefs. Persuasion increases social stickiness, and social stickiness is what keeps organizations as communities and social networks together in the absence of tight, hierarchical structures and vague incentive structures. The current mintage of DAOs is characterized by emergence, intercultural participation, chaos, and complexity. Managing these DAOs requires managers to become skilled communicators. Those who lean on the knowledge gained from research efforts in institutional theory such as those on cultural entrepreneurship may benefit tremendously.

\subsection{Organizations as complex adaptive systems}

\textit{Summary: Understanding DAOs as complex adaptive systems (CAS) presents opportunities to study properties such as path-dependency, sensitivity to initial conditions, emergent changes, and episodic shifts. Researchers can employ grounded theory and comparative case study approaches to advance empirical understanding and generate theories around organizational evolution. Additionally, the unique data properties of DAOs offer avenues for fascinating studies on social networks, dynamic tie evolution, and the use of agent-based models to project and understand organizational dynamics over time. This stream of organizational research can further help to address questions of how and why productive self-organization emerges in organizations and DAOs.}

Many DAO participants believe DAOs offer new opportunities for organizational systems to self-organize. There is a real opportunity for academic work that may help practitioners understand how, when, and why productive self-organization emerges. Much existing work is rooted in views of organizations as complex adaptive systems evolving at the edge of chaos. Indeed, the contributions of scholars such as Herbert Simon and James March were as fundamental for the development of complex systems as organization science \cite{simon_sciences_2019}, while Elinor Ostrom explicitly bases her foundational concept of institutional diversity on the proto-CAS thought of cybernetics \cite{ostrom_designing_1995}. Interestingly, empirical research in this domain has been slowed down by the lack of appropriate empirically-grounded data and methods to test researchers’ theoretical hypotheses. DAOs present an opportunity to not only test these hypotheses but also to inform and generate new ones. 

Complexity, in brief, views systems such as organizations as wholes that are more than the sum of their parts and even as archetypes for the multi-scale structure that characterizes emergent complexity \cite{ostrom_designing_1995}. The literature on complexity highlights properties such as path-dependency, sensitivity to initial conditions, emergent (uncertain but not random), and episodic changes \cite{beinhocker_origin_2007}. Of particular interest to practitioners in the DAO space might be how to manage complex organizations while staying true to emergent self-organization \cite{ostrom_beyond_2009}.

DAOs are a fertile ground with plenty of exceptional data to study self-organization. On-chain data does not suffer from left-censoring; however note the learnings from the Curve wars regarding manipulation of the governance that occurs on-chain. Certain datasets have perfect data, obviating the need for sophisticated statistics to address gaps. The recent explosion of DAOs also offers sufficient freedoms of observation. This presents rich opportunities. On one hand, grounded theory and comparative case study approaches would advance our empirical understanding of DAOs. It may also generate theory around organizational evolution. Models of organizational change as punctuated equilibrium traditionally invoke change as developing from a growing discrepancy between an organization and its environment. The causal mechanism typically invoked is organizational inertia. Yet considering organizations as complex adaptive systems suggests alternative mechanisms. For example, organizations may change via punctuated equilibrium because patterns of both small and large changes over time can often naturally lead to a pattern of change in the form of a punctuated equilibrium.

Moreover, a unique research opportunity rests in the data properties around DAOs. Empirically inclined researchers may create fascinating new studies that leverage this data:

\begin{itemize}
	\item Researchers may analyze social networks of agents not only at a given point in time, but dynamically as these social networks evolve over time, and agents and their organizations begin to co-create each other \cite{zhong_institutional_2022}. Much research in social networks is based on the presence or absence of connections between some or the other actor. Data related to DAOs allows researchers to see how ties dynamically evolve. This can be similarly studied at the dyadic, triadic, cluster, or network levels. 

	\item Researchers viewing organizations as complex adaptive systems can easily measure, more easily than ever, how a focal actor’s behavior in the current time period affects the behavior of other actors in the next time period. 

	\item An even more fascinating approach would be to train or condition agent-based models on real-life data points. DAOs may permit faithful implementations and rigorous tests of classic computational models of organizations, from the anarchic garbage can model \cite{cohen_garbage_1972} to Carley’s ORGAHEAD of organizational adaptation \cite{ilgen_organizational_2000}. For example, an agent-based model’s first iterations modeled using real data and the revealed dynamics could continue projecting an organization’s evolution in the next few iterations. Such models would approach scientific work with practitioner work, both tremendously benefiting from each other via superior, shared understanding and greater empirical and theoretical groundedness.

\end{itemize}
\subsection{Organizational methodology in the era of complete data}

\textit{Summary: The questions scholars ask about organizations tend to be a function of the limited data they have access to. As DAOs provide complete data from the level of individuals to the level of populations of organizations, the completeness and transparency of this data promises to advance organizational scholarship, particularly as organizational scholars increasingly work to reconcile and integrate disparate research frameworks.}

Organizations are difficult to study. They are typically too big, dynamic, and unobservable to provide the elements necessary for controlled scientific inquiry. DAOs correct for this by being highly instrumented, entirely transparent organizations that offer systematically rich, granular, and longitudinal data. The comprehensiveness of DAO data is atypical of what academics have encountered to date. As a result, DAOs represent not just a novel empirical setting but an opportunity to develop and test novel empirical methods that were previously impossible to apply.

As DAOs rely on on-chain processes for membership, governance, and operations, they generate a wealth of granular, longitudinal data about the evolution of their social structures and practices. Often this on-chain data does not suffer from typical left-censoring or sampling issues, and, apart from community moderation, other public data from governance and operational interactions is uncensored. We have already discussed the longitudinal tracing of how social structures evolve via on-chain elements, dynamics, and histories such as smart contracts, incentive schemes, and forks, but many other opportunities remain.

Looking ahead, this data could be used to power multiscale studies that bridge (or challenge) Hannan and Freeman’s five levels of organizational analysis: members, subunits, individual organizations, populations of organizations, and communities of populations of organizations \cite{hannan_inertia_1996}. This is important because the major paradigmatic differences driving contemporary organizational scholarship are more a function of data availability than any epistemological differences. An ideal example lies in firm- versus population-level theories of organizational survival. Frameworks like the firm-level ``resource-based view" rely on behavioral data of members of organizations to discover internal determinants of organizational success. By contrast, frameworks like the population-level organizational ecology view rely on data about populations of interacting organizations to reveal environmental determinants of success. While scholars acknowledge that both must be true \cite{goll_relationships_2005,leiblein_empirical_2003,volberda_contingency_2012}, the difficulty of collecting large, complete, multi-scale datasets has made it impossible to understand how internal and environmental theories of organizational performance interact and complement each other. From this perspective, DAOs promise to be the substrate upon which organizational scholars develop a unified multi-scale conception of firm success and even emergent social order.

Looking even further ahead, it is important to recognize that the people who start and operate organizations are themselves organizational scholars engaged in the challenge of building a social system capable of learning and adapting \cite{senge_fifth_2006}. Classic work by March emphasizes the spotty and imperfect nature of trying to theorize about a single organization while situated within it \cite{march_learning_1991}. With their automation, instrumentation, standardization, and transparency, DAOs promise to transform organizational learning by making it easier for members to automatically register and learn from each others’ mistakes both within organizations and across many organizations. Although it has long been recognized that organizational learning can happen at the level of communities of populations of organizations \cite{baum_survival-enhancing_1998}, even lessons at that scale tend to be anecdotal, and lack the automatic, systematic comprehensiveness of potential future DAO-to-DAO organizational learning systems.

\subsection{Organizational ethnography}

Beyond analyzing DAOs through their on-chain and quantitative data, qualitative research methods are widely employed throughout organization studies. In this section, we make a case for ethnography, a specific type of qualitative research method, as a means to surface DAO-based social phenomena and practices that may push the boundaries of existing disciplines.

Ethnography is a qualitative approach in which the  researcher is immersed in the phenomena and community they are studying. It is typically used to make visible the invisible, surface new and recurring questions and themes, and challenge normative assumptions with an eye to identifying discrepancies that exist between what communities say they do versus what they actually do. The process involves producing ``thick description" \cite{geertz_interpretation_1973}, meaning an interpretative account not just of actions, but the meanings, symbols and structures that inform behaviors. As ethnography looks to ``first-hand" knowledge rather than relying exclusively on second-hand accounts \cite{hine_ethnography_2015}, researchers will spend significant time with the people, communities, and contexts they are researching to understand their language, design, rules, and customs. In addition to observation field notes, ethnographic data may include interviews, focus groups, collecting data across multiple media sites \cite{atkinson_ethnography_1998} or even computational data \cite{maddox_netnography_2020}. 

In the case of DAO research, the ethnographer’s task lies in documenting and understanding the social dynamics underpinning a particular context and the interactions among both human and non-human actors constituting a field site. A DAO community is not always the thing being studied; the ethnographer may be tracing various phenomena such as values, standards, infrastructures, and automated policies by observing how they occur in and through DAOs. 

The ethnographer may choose to extend upon traditional ethnographic research methods by including the study of computer mediated social interactions, the study of online communities and the use of digital technologies in data collection and analysis \cite{abidin_private_2020,pink_digital_2015}. For instance, ethnographers studying in DAOs may need to follow both human and non-human actors \cite{seaver_algorithms_2017}, and make sense of their interaction with each other and the community’s wider social and material context \cite{latour_reassembling_2007}. Specific techniques such as situational mapping, arena mapping, and positional mapping \cite{clarke_situational_2003,clarke__2016} may be used to synthesize empirical observation without excluding technical actors, infrastructural components, discursive dynamics or the broader context. 

Conducting ethnographic research in DAOs and Web3 comes with specific challenges. Firstly, ethnographers are required to define, access and make sense of the deluge of data gathered through their field site, including online forums such as Discord servers. In the context at hand, this requires us to ask questions such as: Where is the DAO? Where, when and how can we see it, and where not? Secondly, while ethnographers are relatively free to choose the specific techniques most appropriate to making sense of their data, it is not always easy to get informed consent from all actors being observed in a DAO. One example of an attempt to overcome this challenge  is the Telescope consent bot, developed by Rennie et al. \cite{rennie_toward_2022} to automate parts of the consent process for data collection within Discord chats. 

To date, researchers have used ethnographic methods to explore the informal aspects DAO governance \cite{dupont_experiments_2017,shorin_uniswap_2021}, cultural dynamics and beliefs in cryptocurrency communities \cite{brody_ideologies_2021,faustino_myths_2022}, activism in darknet marketplaces \cite{maddox_constructive_2016} and the use of Web3 technologies as a form of ‘self-infrastructuring’ \cite{nabben_web3_2023}. Researchers employing ethnographic methods have made valuable contributions to DAO science by revealing unformalized and somewhat messy off-chain governance dynamics, anchor on-chain components within specific context and meaning to understand their mediating capabilities, and by developing new participatory methods to involve DAO practitioners in the research process. Ethnography is therefore a promising approach for identifying emergent ‘open problems’ in DAO science.   

\section{Political science and philosophy}

\begin{quote}
Editor: Eliza R. Oak. Contributors: Eliza R. Oak, Woojin Lim, Hélène Landemore, Danielle Allen.
\end{quote}

Political science is the systematic study of governance (Goodin and Klingemann, 1998; Roskin, 2005). At the heart of political science is the study of the transfer and allocation of power in decision-making processes, and of the emergence and consequences of different governance systems. Studying the governance of DAOs and other digitally-constituted organizations is interesting for political scientists because (1) DAOs offer a new platform to empirically test theories from the political science canon; (2) DAOs are actively seeking expert knowledge to consult their governance design decisions so there is real-world demand for scholarship on this topic; and (3) as all aspects of society become increasingly digitized, understanding best-practices for DAO governance paves the way for re-imagining and potentially re-designing current political processes \cite{bernholz_digital_2021}.

Societies and organizations have puzzled over governance questions for millennia, and the contested idea of democracy—a system of governing which depends on the will of the people—is central to these debates (e.g. \cite{plato_republic_1892, aristotle_nicomachean_353bce, locke_two_1690,rousseau_social_1762,dahl_democracy_1989,przeworski_democracy_2012,landemore_democratic_2017}). In theory, DAOs present innovative opportunities for collective decision-making, though in practice there are many possibilities for the anti-democratic seizure of DAOs. Insofar as DAOs rely on blockchain-based automated data storage mechanisms and smart contracts, how can challenges related to technocracy be avoided \cite{buterin_flexible_2019}—and are such automated decision-execution protocols even desirable? How can plutocracy or sybil attacks be avoided, when common ``one-token-one-vote" mechanisms  mean that wealthy users can buy a disproportionate number of tokens and subsequently gain a disproportionate amount of voting power? Because of these and other problems in practice, more thought should be dedicated to questions of whether financializing governance actually make sense—and to formalizing the conditions under which certain voting models are appropriate.

\subsection{Institutions}

\textit{Summary: As DAOs create political institutions, they confront classic coordination puzzles such as preference aggregation, credible commitments and audience costs, information asymmetry, or representation and accountability. As a unique empirical setting, DAOs can help political scientists (1) study foundational theories about political institutions and (2) generate novel theories to be tested in digital governance settings.  }

The potential problems of tokenized governance raise interesting questions concerning the design of durable and representative political institutions. Political institutions tend to refer to ``the formal and informal rules, procedures, and organizations that govern the behavior of individuals and groups in a political system" \cite{hall_political_1996}—that is, the ``rules of the game in a society, or more formally, the humanly devised constraints that shape human interaction" \cite{north_institutions_1990}. Scholars have debated the implications of different types of institutional design, including separation of powers, federalism, the strategic design of non-democratic institutions, or institutional change. How might these theories map onto the design of digitally-native governance institutions? For example, does separation of powers in DAOs prevent any one entity from becoming too powerful \cite{montesquieu_spirit_1748}, increasing transparency and accountability \cite{lijphart_patterns_2012}, or does this instead result in gridlock and indecisive government action \cite{tsebelis_veto_2002}? Does dividing power between central and ``regional" entities in DAOs lead to greater democratic participation, as the threat of regional secession (forking?) can create incentives for the central government to compromise and increase responsiveness \cite{riker_federalism_1964}? In order for institutional change to occur in DAOs—where change can occur much more rapidly—must there still be significant buy-in from political elites and a credible commitment by political actors to respect the new institutional rules \cite{weingast_economic_1995}? 

How can DAOs avoid elite interest capture? For example, non-democratic regimes will sometimes create electoral institutions in order to gain information about who to trust among the ruling elites, meaning these institutions are little more than performative ``window-dressings" of democracy designed to promote non-democratic regime survival \cite{gandhi_political_2008}.

Other research has examined how despite the same institutional composition, variation in the levels of ``civicness" explains why some regional governments are more efficient and more responsive \cite{putnam_making_1993}. Recent scholarship has also argued that the disruptive Schumpeterian consequences of blockchains make it an institutional technology rather than a general purpose technology, and that blockchains themselves are an instance of institutional evolution \cite{davidson_blockchains_2018}. 

\subsection{Turnout}

\textit{Summary: The coordination challenge in collective choice, which involves aggregating individual preferences, also applies to DAOs, prompting questions about voter turnout, representation, and accountability. Political science can help DAOs strike a balance between more participatory decision-making and the inefficiencies caused by uninformed voting, e.g. through new (and old) incentivization methods to address issues of accessibility and quality of participation. The formal voting mechanisms and informal mobilization strategies that influence online, token-incentivized civic participation in DAOs remain open questions in this context.}

The coordination puzzle of aggregating individual preferences by some method such as voting to produce a social outcome (i.e., collective choice) underpins questions about governance and institutions. Collective choice is shown to involve ``surprisingly intransigent paradoxes that seem to challenge the possibility of fair democratic decision-making'' \cite{schwartz_collective_2011}. DAOs similarly confront classic questions of turnout, representation, and accountability as they design their governance institutions. DAOs suffer from similarly low levels of voter turnout to those documented in the real world and therefore aim to balance the desire for more participatory and inclusive decision-making processes, without inducing low levels of participation and inefficiencies from uninformed decision-making \cite{bueno_de_mesquita_paying_2022}. As DAOs experiment with different voting mechanisms, important questions related to participation arise, for example: 

\begin{itemize}
	\item What are the most compelling ways to incentivize voting participation — without building-in perverse incentives? 

	\item Is participation due more to lack of interest or lack of accessibility? 

	\item How should we think about the quality of participation? Quality of participation is especially relevant in DAOs where users can earn tokens based on participation, as this opens up subjective questions about who gets to decide what type of participation is considered deserving of more tokens.

	\item How do informal concepts like ``do-ocracy" and ``lazy consensus" become relevant and operationalized in these digital contexts?

\end{itemize}

Open questions remain regarding which individual-level characteristics, institutional factors, and mobilization strategies influence online, token-incentivized civic participation in DAOs. Do DAO voters, for example, also tend to be individuals with higher levels of income, education, and social status are more likely to have access to political information and feel a sense of political efficacy \cite{verba_participation_1987, highton_political_2001}? Or do the individual-level characteristics of DAO voters reflect those of early crypto adopters more generally—do they tend to be younger, male, more dissatisfied with existing institutions and tech-savvy, for example? Moreover, do DAOs with compulsory voting laws \cite{lijphart_patterns_2012}, unicameralism \cite{jackman_political_1987}, or proportional representation rather than plurality systems \cite{blais_turnout_1998} also tend to have higher levels of voter turnout? What mobilization strategies are most effective in DAOs—for example invoking social pressure by publicizing individuals’ voting behavior \cite{bryan_motivating_2011}, or online variations of get-out-the-vote efforts \cite{green_get_2015}?

\subsection{Real-time experimentation}

\textit{Summary: DAOs are experimenting rapidly with both traditional and novel governance strategies, with online programmability widening the design space for what is possible. The rapid experimentation of DAOs provide political scientists a unique and, to date, mostly unexploited opportunity to study the effect of institutions on political behavior and collective outcomes.}

Unlike political governing bodies which historically take decades or even centuries to change, DAOs are experimenting with the design of governing bodies at exponentially quicker scales (Hall and Smith, 2022). Optimism\footnote{Optimism is a Layer 2 network and can be thought of as a sort of blockchain built on top of another ``Layer 1'' blockchain to reduce fees and transaction speed.}, for example, has designed a bicameral governance structure where governance matters are divided between the Token House and the Citizens’ House. In the Token House, an individual with more tokens has more votes. In the Citizen House, every individual ``citizen'' is granted one non-transferable citizen token which yields precisely one vote per individual citizen. Optimism has also created a system of delegation that allows users to select a trusted third party to make governance decisions on their behalf, though said user retains full ownership of their tokens and can choose at any time whether to vote directly instead of via the delegate. Other DAOs are experimenting with new voting mechanisms including: quadratic voting which allows individuals to express how strongly they care about a given issue, quorum voting which requires a certain threshold of votes in order for a proposal to pass, conviction voting in which proposals accumulate enough votes to pass over time, or versions of liquid democracy which employ elements of both direct and representative democracy. 

These and other real-time applications provide political scientists with a unique opportunity to analyze large amounts of time-stamped, individual-level data on governance participation. This opens up new opportunities for scholars to empirically test centuries of political thought as well as test new bold ideas for voting and governance. This could offer scholars and practitioners the opportunity to apply lessons from DAO governance to potentially improve political processes and coordination in our ever-digitizing society. 

\subsection{Self-governance}

\textit{Summary: Previous scholarship has explored decentralized forms of governance in non-state realms, shedding light on factors that contribute to successful self-governance and coordination which are relevant for understanding the governance of decentralized digital governance. Political science can help address questions about power allocation, participant vetting, community norm enforcement, and the ideal level of gatekeeping in self-governing communities, while considering the coexistence with external regulatory jurisdictions and the importance of informal authority and community building in decentralized organizations.}

Contemporary political science increasingly includes the study of non-state realms of governance, including the governance of firms and other government-recognized legal entities; tribes and other socio-political groups; or gangs, protest movements, and other non-state networks. Scholars have documented how and when decentralized governance and coordination is successful in the absence of strong formal governing institutions, which is relevant for understanding the governance of distributed online networks. 13th-century Mediterranean traders, for example, created a unique institutional system based on norms and practices that promoted trust, cooperation, and reputation-building among traders, enabling traders to overcome the risks associated with long-distance trade and resolve disputes without violence \cite{avner_institutions_2006}. Other scholarship has examined the role of social capital, villager associations, and informal monitoring networks in facilitating the provision of public goods in the absence of democratic or bureaucratic mechanisms of accountability \cite{tsai_accountability_2007}. 

Self-governance can exist in the absence of formal rules. In examining the conditions in which criminal groups exert state-like functions in their spheres of influence, scholars have documented how the internal organization of a Mexican mafia gang functions much like a traditional cooperative with a written constitution \cite{kostelnik_governance_2013}, whereas a prison gang lacks clearly defined control rights and is worker-owned, with each member having a vote and relying on an internal rule structure \cite{skarbek_governance_2011}. To gain membership into either a prison or mafia gang, recruits must make costly sacrifices to improve member quality, limit free riding, and reveal information about the recruit’s abilities and dedication. DAOs face similar questions regarding power allocation, participant vetting to improve quality of deliberation, and community norm enforcement in self-governing organizations. Other explanations for self-governance look beyond state solutions or market-oriented approaches to explain how societies and groups regularly devise rules and enforcement mechanisms to govern the commons \cite{ostrom_governing_1990}. This scholarship reveals that these types of self-governing cooperatives tend to work because they’re rather homogeneous and have ``clearly defined community boundaries'' \cite{rozas_when_2021}; extensions of commons scholarship to online communities have demonstrated the framework’s power for understanding virtual and computational resource governance as well \cite{frey_emergence_2019,kollock_managing_1996}. As DAOs aspire to be flatter and more egalitarian with power distributed across tokenholders who cast votes, this open, participatory, and heterogeneous structure raises questions about how best to scale. In terms of clearly defining community boundaries, DAOs face questions regarding the ideal level of gatekeeping—how open, and how participatory should a values-aligned, mission-driven DAO be, and who should decide this?

Of course, DAOs do not operate in a vacuum and are exploring ways to self-govern while co-existing with evolving external, formal regulatory jurisdictions. Similarly, existing research examines the relationships that arise through interactions between state and informal actors at the local level (e.g., hostility, acquiescence, collusion) and seeks to understand what factors make this type of governance stable over time. Baldwin  \cite{baldwin_paradox_2015}, for example, argues that unelected traditional leaders (chiefs) in Africa improve the responsiveness of democratic governments by helping to hold political leaders accountable, depending on how well local politicians work with chiefs. Just as political scientists focus on many aspects of governance beyond the formal constitutions, there are many aspects of DAO governance beyond questions about hard-coded governance rules, for example online community building and informal authority delegation.

\subsection{Global governance}

\textit{Summary: Similar to nation-states regulating themselves without a central governing body and designing their own mechanisms for enforcement, DAOs operate without a central governing DAO and alongside territorial nation-state laws. The scholarship on non-state actors and international institutions is relevant for understanding inter-DAO operations both between DAOs, and in the context of national regulatory landscapes.}

Studying DAOs is also useful for theorizing about how technological innovations might transform transnational interactions between nation states and non-state actors. Within political science, the subfield of international relations (IR) examines why, and how, states and non-state actors interact. This involves interactions between sovereign states, as well as interactions between non-state actors across international borders. Between states, these interactions include war, trade, and diplomacy. Non-state actors include intergovernmental organizations, multinational corporations, multilateral financial institutions, international non-governmental and humanitarian organizations, transnational diaspora communities, drug cartels, and armed resistance groups. A core assumption of IR is the anarchic nature of the international system - states regulate themselves without a central governing body, or without a ``world police." DAOs similarly face questions about governing in the absence of a pre-existing central governing body. 

With cross-chain interoperability protocols such as Layer Zero or IBC enabling interactions between chains, can we think of chains functioning as self-governing, decentralized platforms themselves, connected by bridges? What determines coalition formation in these ecosystems, and what is the role of inter-chain, or inter-DAO governance? Could lessons from global governance be applied here? Just as the League of Nations was formed to hold nation-states accountable,and for example regional trade commissions, global health bodies, or the international criminal court were designed to pool expert knowledge and enforce common norms, DAOs also grapple with balancing expert knowledge and centralized accountability with individual autonomy and representation. An interesting framework might connect parallels between interactions of states and non-state actors in the international system, with interactions between DAOs and L1 and L2 protocols, forming alliances and coalitions, and facing similar questions about balancing the desire for expert knowledge and centralized accountability with the desire for autonomy and representation.

Moreover, scholars have argued that globalization and the proliferation of non-state actors in recent decades has led to the erosion of power and sovereignty of the traditional Westphalian nation state (Keck and Sikkink, 1998). While some early IR scholarship focuses on the use-case of cross-border payments via cryptocurrencies \cite{nershi_how_2020,chey_cryptocurrencies_2023}, many open questions remain about how DAOs - and the underlying blockchain tech more broadly - might reshape how we think about cooperation, governance, and power in the international system.

The above paragraphs highlight fruitful avenues for connecting research on DAO governance to existing scholarship on formal, informal, or global governance. In addition to both theoretical and empirical advances scholarship on governance, political institutions, and civic behavior, studying DAOs offers an additional important output: Understanding DAO governance could also potentially inform the design of new digital processes for online civic engagement and assembly. While other aspects of society have rapidly digitized, political institutions continue to employ processes designed for pre-digital times. The reason for holding national political leaders accountable only every 2-4 years, for example, is essentially a technology problem and there is arguably little reason to believe existing systems for representation, accountability, civic participation, or even the organization of nation states under the current Westphalian system will be immune to digital disruption \cite{bernholz_digital_2021,srinivasan_network_2022}. Studying DAO governance offers the opportunity for scholars to combine centuries of political thought with lessons from the trial and error of early web3 collectives to build democratic political institutions for the future. 

\subsection{Political philosophy}

\textit{Summary: DAOs present far-reaching  questions related to deliberation, justice, access, fairness, representation, and self-determination in digital governance, necessitating productive dialogue between political philosophers and DAO members to bridge the gap between traditional and techno-normative concepts. Applying lessons from political philosophy will offer insights into the complexities of the digital public sphere, civic foundations of discourse, and legitimacy in group action.}

Concepts, methodologies, and frameworks from political philosophy also speak to contemporary issues raised by the presence of DAOs in digital governance infrastructures. Though to date political theorists have shown little interest in DAO-based governance, DAOs clearly raise questions on the nature of deliberation in the digital public sphere, justice, access to information, fairness and political equality, representation and participation, and self-determination.

The most important immediate work will need to be translational, such that political philosophers and DAO members can speak to each other productively. Currently, political theorists speak in highly technical terminology inherited from a long tradition of normative and analytical thinking, whereas DAO builders and members may describe their work using inventive lexicons native to the blockchain space. A significant contribution to the literature on DAOs would translate between traditional concepts in political philosophy (such as equality, liberty, non-domination, and reciprocity), and what some have called ``techno-normative concepts," cited as desirable by builders and community members \cite{allen_ethics_2023}. This parallels early work on AI ethics and on bioethics.

As an operative medium for social organization and coordination, DAOs raise questions around the emerging role of the digital public sphere. Research in this area might examine recent experiments in efforts in more healthy, democratic digital conversations and decision making, such as Polis. Importantly, it’s worth considering whether this ``democratization" of digital conversations is having a collectivizing effect rather than simply empowering atomized individuals. Research here might also turn to Jürgen Habermas’s formulation of the public sphere \cite{habermas_public_2000} and recent work looking at Habermasian approaches to digital technologies and the digital possibilities (and perils) of healthy, deliberative democracy. Further research in political thought on DAOs might ask whether traditional models of deliberative democracy suffice in capturing the complexities of the digital public sphere and whether DAOs offer new technological means to empower individuals to collaborate as political, economic, and social equals as members of collective decision-making bodies.

Among the key concerns of contemporary and 20th century political philosophy, riding on the coattails of the Rawlsian tradition in \textit{A Theory of Justice} \cite{rawls_theory_1971}, are questions of distributive justice. These concern the fair distribution of benefits and burdens of social cooperation and mutual reciprocity in a society of individuals with competing needs and claims. Building on Ciepley’s work on political theorizing about corporations \cite{ciepley_beyond_2013}, DAOs cannot be satisfactorily assimilated to the dichotomous categories that form the basis for liberal political theorizing, such as public/private, government/market, state/society, privilege/equality, and status/contract. DAOs may require political theorists to carve out a new set of rules and norms of regulation that address ``meso-level" institutions. First off, can we even apply existing theories of distributive justice to questions of DAO governance? If so, what counts as a distributive good in DAOs (e.g. tokens as financial compensation or reputational markers, social bases for self-respect) and according to what criterion of justice should these goods be distributed (e.g. equality, desert, market forces)? There also arises the question of global territorial scope \cite{caney_cosmopolitan_2005,nine_global_2012,stilz_territorial_2019} and of justice in production or ``productive justice" \cite{stanczyk_productive_2012}. Can DAO members be compelled to work to produce primary goods whose principles of distributive justice, if any, governs to be produced? What principles should govern the production of such goods in DAOs and to what degree ought such principles incur obligations (of work) on DAO members? What are the conditions of DAO membership as an analogue of citizenship?

Provided that members of a DAO span various cross-cultural identities that include a plurality of geographies, citizenships, and political affiliations, questions arise on the civic foundations of discourse across boundaries of difference. Some scholars cite the importance of egalitarian pluralism as a starting ground \cite{allen_ethics_2023,jain_algorithmic_2023}. We must ask questions about how DAOs raise both opportunities for individuals to act as decision-making equals in political and economic forums. Perhaps, we must challenge the presumption that token-based participation in DAOs, even in theory, can offer robust conditions of political equality such as those demanded by Danielle Allen’s recent interpretation and critique of Rawls (Allen, 2023).

Another potentially fruitful area for research in political philosophy centers around questions of legitimacy and group action. DAOs provide a unique cross-national medium for the technological constitution of groups. Recent work in political philosophy has suggested that democratic legitimacy is not simply a descriptive, positivistic notion, but is deeply normative \cite{applbaum_legitimacy_2010,lazar_legitimacy_2022}. Much of this work centers around questions about the possibility and definition of group actors and group agency. Philosophers might seek to define more formally the conditions under which decision-making in and by DAOs can be legitimate. Even further, philosophers might study DAOs as potentially novel structures for collective decision-making which shed light on traditional work on group agents.

\section{Major coordination problems}

\begin{quote}
Editors: Helena Rong and Joshua Tan. Contributors: Helena Rong, Joshua Tan, Jonathan Dotan, Ryan L. Thomas, Kevin Owocki, Ariana Fowler, Scott Moore, Benedict Lau.
\end{quote}

DAOs and digitally constituted organizations have been the subject of extraordinary hype; various boosters have advertised them as ``changing the way we work'' or as ``the next generation of organizations''—including this essay’s own abstract. This is a problem not only for colleagues outside the field trying to sift fact from hyperbole but also for researchers and builders whose contributions are held up against unrealistic expectations.

In lieu of hopes and hype, we define a set of clear coordination, collective action, and social problems that DAOs and other social technologies may be able to solve. These problems are well-understood to be hard and without good solutions. Solving them would help not only mark a significant contribution to society but help validate some of the claims of the technologists to those not already convinced. Failing to solve these problems would also mark a significant contribution to knowledge insofar as the failures clearly demonstrate the limitations and relative fit of different types of solutions for different types of problems.

The problems, in short, are:

\begin{enumerate}
	\item Sustaining open-source software
	\item Democratic governance of artificial intelligence
	\item Regenerative finance
\end{enumerate}

Each problem, organized into subsections below, highlights a major coordination, collective action, or social problem and a survey of recent work to address them, including but not limited to work in the DAO space. Each problem is accompanied by a set of clear success criteria modeled on the XPRIZE\footnote{XPRIZE is a non-profit organization that hosts public competitions aimed at encouraging technological development to benefit humanity. See \url{https://www.xprize.org/.}} structure, with the broad caveat that complete success is hard to define in these complex problem domains. In evaluating success, we have sought to choose metrics and indicators that are concrete, persuasive, and sustainable.\footnote{By concrete, we mean that a metric has a definite and measurable answer that is not dependent on subjective judgment. A concrete metric is often but not always numeric or categorical. By persuasive, we mean that the metric should be able to communicate success to the general public; genuine success along a metric should ``pop'' to the public. Finally, by sustainable we mean that a metric (or a slate of metrics, considered holistically) should be not too expensive to collect and, ideally, collectible through automated means.}

While we have curated problems where some form of social coordination or collective action failure is the sticking point—and where technological solutions seem promising—many of these problems admit solutions that have nothing to do with DAOs and even nothing to do with coordination per se (e.g. by sharply reducing the cost of a key process or increasing the availability of a scarce resource). We have tried to write evaluation criteria with this in mind. Sometimes, the best way to solve a coordination problem is to sidestep it altogether.

\subsection{Sustaining open-source software}

\subsubsection{The problem}

Open-source software (OSS) is a multi-billion dollar industry and the backbone of major enterprise-level software products and services \cite{schweik_internet_2012}. Values inherent to open-source, including peer review, open code contribution, and community feedback, are now widely recognized as key catalysts for fostering innovation and ensuring software quality. Despite its role as critical infrastructure for many devices and applications that form an integral part of our daily lives, OSS suffers from chronic underfunding and resource shortages; this has led to a crisis in its maintenance and sustainability \cite{eghbal_working_2020}.

Open-source software projects are largely volunteer-driven initiatives in which developers join, cooperate, or cease contributing of their own volition. They face challenges typical of public goods maintenance, notably the ``free rider'' problem and the ``tragedy of the commons'', where individuals benefit from OSS without contributing to its maintenance, placing an undue burden on those who do \cite{samuelson_diagrammatic_1955}. This volunteer-driven model often leads to the problem of \textit{contributor burnout}, especially among key maintainers, resulting in a disproportionate distribution of resources and effort. For instance, although the top 1000 projects on the popular OSS hosting site GitHub show an average of around 80 contributors \cite{bao_large_2021}, many other projects---including critical digital infrastructure \cite{zuckerman_what_2020}---are maintained by just one person, often the overstretched founder of that project. As OSS becomes increasingly global, its coordination challenges further grow exponentially. These complexities can become particularly pronounced when large companies and states, losing either interest or capacity, fail to continuously support OSS projects. This situation is exacerbated when these entities themselves adopt free rider behavior at an institutional level, similar to individuals. As a result, these public goods face severe coordination failures, despite their ability to generate substantial positive externalities for increasingly large populations. This dichotomy between the immense impact of OSS and the lack of adequate support and coordination mechanisms underscores the urgency of this crisis. The question becomes: what can we do to help sustain this unique class of public goods given the value they continue to create?

\subsubsection{Current progress}
FOSS foundations such as the Linux Foundation, Apache, NumFOCUS, and Python offer critical support and guidance by leveraging their extensive experience and understanding of common challenges faced by OSS projects. They provide a range of services including mentorship, formal incubation processes for new projects, project operations, tooling, training, and even legal and IP management. These services help OSS projects navigate complex issues like governance, asset management, and community growth.

In terms of management tooling, Open Collective provides crucial support for open-source projects by managing the administrative burdens associated with grant-funded work. This includes handling the complexities of grant reporting, tracking success, and maintaining or finding suitable charitable entities to support projects, thereby allowing developers to focus more on their core project activities without the overhead of administrative tasks. An often under-appreciated fact about Open Collective is that the fiscal sponsorship path effectively gives open source projects access to a bank account---similar to the use-case for a multisig or certain DAOs.

Other efforts, from new OSS licenses to new donation systems, offer ways to protect and shape the incentives of OSS contributors. For example, the \href{https://opensource.org/license/cal-1-0}{Cryptographic Autonomy License} (CAL) protects user autonomy by ensuring users have control over their data and the software they use, while the \href{https://www.mongodb.com/legal/licensing/server-side-public-license}{Server Side Public License} (SSPL) introduced by MongoDB is designed to ensure that if a service provider leverages the software to deliver services, they must also share the service's entire source code under the same license, thus promoting transparency and fairness in the use of OSS. With respect to funding, GitHub Sponsors is an example of a platform that facilitates community-driven financial support, allowing individual developers and organizations to receive direct contributions from those who value their open-source work.

In the DAO ecosystem, a key focus has been developing effective ways to reward contributors fairly and promoting more equitable ownership. Many DAOs are experimenting with community currencies \cite{adriano_technological_2021} to deter free-riding and ensure balanced opportunities for voice and exit \cite{berg_exit_2017}. These currencies enable contributors—whether they are coding, supporting, or adding value in other ways—to engage in the governance of these organizations. Some may even gain from the broader adoption and application of the projects they contribute to, thus nurturing a more sustainable and participatory ecosystem.

DAOs like Gitcoin and Optimism have made significant progress towards finding new models that allow any group to curate, evaluate, and consequently allocate capital to developers that do not necessarily have any immediate financial return but have a significant impact on the ecosystems they exist in. Developers are funded based on their impact either as judged by community sentiment through quadratic funding (as in the case of Gitcoin) or by expert decision-makers curated via a web of trust (as in the case of Optimism). Both models have their trade-offs but fundamentally deviate from prior grants-making processes typically led by closed-ended and opaque committees rather than open communities. In the case of Gitcoin, the platform has conducted tests across over one hundred distinct grant pools, collectively distributing over 50M USD. This significant funding milestone was achieved in partnership with prominent organizations, including the Ethereum Foundation and UNICEF. Among the funded projects are notable entities like the Electronic Frontier Foundation, Internet Archive, Uniswap, and WalletConnect.

\subsubsection{Success criteria}

The criteria defined below are intended to capture results over a one-year period for a particular cohort of open-source projects.
We expect all open-source projects included in the evaluation to:

\begin{itemize}
	\item Have a working code.

	\item Have published code under a widely accepted open-source license, e.g., the MIT License or a Creative Commons license.

	\item Have at least 50 stars on GitHub, GitLab, or an equivalent metric as a measure of user adoption.

	\item Have at least one active maintainer who has created or resolved an issue.

\end{itemize}

\paragraph{Proposal criteria}
The following criteria are intended to help teams define and plan their solution. It may also be used by funders considering proposals that promise to address open-source sustainability.

\begin{table}[H]
\begin{adjustbox}{max width=\textwidth}
\begin{tabular}{p{2.91cm}p{4.47cm}p{9.13cm}p{2.91cm}p{4.47cm}p{9.13cm}}
\hline
\multicolumn{1}{|p{2.91cm}}{\raggedright
\textbf{Criteria}} & 
\multicolumn{1}{|p{4.47cm}}{\raggedright
\textbf{Description}} & 
\multicolumn{1}{|p{9.13cm}|}{\raggedright
\textbf{Evaluation Metric}} \\ 
\hline
\multicolumn{1}{|p{2.91cm}}{{\footnotesize Contributor sustainability}} & 
\multicolumn{1}{|p{4.47cm}}{{\footnotesize Quality and participation rate (over time) of contributors}} & 
\multicolumn{1}{|p{9.13cm}|}{{\footnotesize The planned solution must create a path for contributors to take part in the decision-making and direction of the project.\par}} \\ 
\hline
\multicolumn{1}{|p{2.91cm}}{{\footnotesize Maintainer sustainability}} & 
\multicolumn{1}{|p{4.47cm}}{{\footnotesize Quality and participation rate (over time) of maintainers}} & 
\multicolumn{1}{|p{9.13cm}|}{{\footnotesize The planned solution must account for the maintainer’s participation over time, productivity, satisfaction with the solution, and burnout. \par}} \\ 
\hline
\multicolumn{1}{|p{2.91cm}}{{\footnotesize Governance sustainability}} & 
\multicolumn{1}{|p{4.47cm}}{{\footnotesize Contribution of non-code contributions}} & 
\multicolumn{1}{|p{9.13cm}|}{{\footnotesize The planned solution must allow all participants to contribute to the governance of the software. }} \\ 
\hline
\multicolumn{1}{|p{2.91cm}}{{\footnotesize Financial sustainability}} & 
\multicolumn{1}{|p{4.47cm}}{{\footnotesize Monetary incentives and financial support of the project}} & 
\multicolumn{1}{|p{9.13cm}|}{{\footnotesize The planned solution must include a way to reward contributors and maintainers for continued motivation to build value for the project. \par}} \\ 
\hline
\end{tabular}
\end{adjustbox}
\caption{Proposal criteria}
\end{table}

\paragraph{Implementation criteria}

The following implementation criteria cover metrics that need to be evaluated on any implementation of the solution or intervention. Note, we expect proposals to evolve over the course of implementation.

\begin{table}[H]
\begin{adjustbox}{max width=\textwidth}
\begin{tabular}{p{2.88cm}p{3.73cm}p{9.9cm}p{2.88cm}p{3.73cm}p{9.9cm}}
\hline
\multicolumn{1}{|p{2.88cm}}{\textbf{Criteria}} & 
\multicolumn{1}{|p{3.73cm}}{\textbf{Description}} & 
\multicolumn{1}{|p{9.9cm}|}{\textbf{Evaluation Metric}} \\ 
\hline
\multicolumn{1}{|p{2.88cm}}{{\footnotesize User adoption}} & 
\multicolumn{1}{|p{3.73cm}}{{\footnotesize Growing number of users}} & 
\multicolumn{1}{|p{9.9cm}|}{{\footnotesize The user base of the OSS project (across all forks) \textit{does not decline} over the course of implementing the solution.\footnote{ The correlation between raw end-user adoption and project sustainability is considered mild. High adoption rates do not always lead to adequate maintenance of critical digital infrastructure. Often, engagement with primary software does not extend to underlying OSS repositories. Moreover, while growth in centralized platforms may increase resources for maintaining a single OSS repository, this does not necessarily translate to broader support. Measuring end-user adoption effectively remains a challenge but could potentially improve with data from package managers or platforms like GitHub. Overall, project competitiveness and the effects of forks also significantly influence user adoption metrics.}\par}} \\ 
\hline
\multicolumn{1}{|p{2.88cm}}{\multirow{4}{*}{\parbox{2.88cm}{{\footnotesize Contributor sustainability}}}} & 
\multicolumn{1}{|p{3.73cm}}{{\footnotesize Participation rate over time of contributors}} & 
\multicolumn{1}{|p{9.9cm}|}{{\footnotesize For 75$\%$ of projects in the cohort, the average participation rate (measured in submitted PRs, comments, contributions, and all other interactions with the repository) of all active contributors increases by at least 25$\%$.\par}} \\ 
\hhline{~--}
\multicolumn{1}{|p{2.88cm}}{} & 
\multicolumn{1}{|p{3.73cm}}{{\footnotesize New contributors}} & 
\multicolumn{1}{|p{9.9cm}|}{{\footnotesize For 75$\%$ of projects in the cohort, the number of active contributors \textit{does not decline}. An active contributor is a person who actively submits PRs, comments on issues, or contributes to discussions within a 3-month window.\footnote{ Projects may only require attaining a "replacement level" for their contributors. It is not necessary for all projects to expand their pool of contributors in order to achieve sustainability.}\par}} \\ 
\hhline{~--}
\multicolumn{1}{|p{2.88cm}}{} & 
\multicolumn{1}{|p{3.73cm}}{{\footnotesize Quality of commits by new contributors}} & 
\multicolumn{1}{|p{9.9cm}|}{{\footnotesize New contributor pull requests must be of high quality, meaning (1) more than 75$\%$ of PRs of new contributors are eventually merged, and (2) less than 5$\%$ of code in the PR needs to be revised before merging.\par}} \\ 
\hhline{~--}
\multicolumn{1}{|p{2.88cm}}{} & 
\multicolumn{1}{|p{3.73cm}}{{\footnotesize Onboarding speed}} & 
\multicolumn{1}{|p{9.9cm}|}{{\footnotesize The average time between a new contributor’s first interaction with the project (defined as starring, forking, or posting an issue), and their first PR, is less than 30 days.\par}} \\ 
\hline
\multicolumn{1}{|p{2.88cm}}{\multirow{4}{*}{\parbox{2.88cm}{{\footnotesize Maintainer sustainability}}}} & 
\multicolumn{1}{|p{3.73cm}}{{\footnotesize Participation over time}} & 
\multicolumn{1}{|p{9.9cm}|}{{\footnotesize \textbf{Free the maintainer}: Aim for 75$\%$ of the maintainers to report increased flexibility in their engagement with contributors and projects within a 3-month period. \par}
{\footnotesize \textbf{Commit to the commit}: Within 6 months, 75$\%$ of maintainers should have established and communicated clear guidelines for contributor engagement and responsibility acceptance.\par}} \\
\hhline{~--}
\multicolumn{1}{|p{2.88cm}}{} & 
\multicolumn{1}{|p{3.73cm}}{{\footnotesize Productivity}} & 
\multicolumn{1}{|p{9.9cm}|}{{\footnotesize The ratio of issues being opened to issues being closed over a 3-month time span is no greater than 2:1.\par}} \\ 
\hhline{~--}
\multicolumn{1}{|p{2.88cm}}{} & 
\multicolumn{1}{|p{3.73cm}}{{\footnotesize Satisfaction with solution}} & 
\multicolumn{1}{|p{9.9cm}|}{{\footnotesize After 10 months, at least 70$\%$ of maintainers should respond neutrally to positively about their experience with the solution. \par}} \\ 
\hhline{~--}
\multicolumn{1}{|p{2.88cm}}{} & 
\multicolumn{1}{|p{3.73cm}}{{\footnotesize Burnout}} & 
\multicolumn{1}{|p{9.9cm}|}{{\footnotesize After 10 months, at least 70$\%$ of maintainers should rate ``good" or ``excellent" on the survey question about their work-life balance.\par}} \\ 
\hline
\multicolumn{1}{|p{2.88cm}}{{\footnotesize Governance sustainability }} & 
\multicolumn{1}{|p{3.73cm}}{{\footnotesize Incentive structures}} & 
\multicolumn{1}{|p{9.9cm}|}{{\footnotesize \textbf{Raise the value of non-code contributions}: Recognizing and elevating the contributions of all project participants will legitimize their concerns, raise their profile, leverage the same motivations that lead developers to contribute, and create stronger software that is the product of a more diverse range of perspectives and skills.\par}} \\ 
\hline
\multicolumn{1}{|p{2.88cm}}{{\footnotesize Financial sustainability}} & 
\multicolumn{1}{|p{3.73cm}}{{\footnotesize Financial support and finance management}} & 
\multicolumn{1}{|p{9.9cm}|}{{\footnotesize \textbf{Use money as an incentive for open source}: Provide a stable foundation of support while enabling contributors and maintainers to continue to build value within projects.\par}

{\footnotesize \textbf{Recognize, value, and invest in OSS}: After 10 months, at least a 50$\%$ increase in project funding will be conditional on the project's value. \par}

{\footnotesize \textbf{Lower barriers for open source projects to manage finances}: Provide mechanisms to simplify financial management processes for open source projects to enhance accessibility and efficiency.\par}} \\ 
\hline
\end{tabular}
\end{adjustbox}
\caption{Implementation criteria}
\end{table}

\paragraph{Scale criteria}

In addition to the implementation criteria above, a successful solution must be scalable.

\begin{table}[H]
\begin{adjustbox}{max width=\textwidth}
\begin{tabular}{p{2.88cm}p{3.68cm}p{9.95cm}p{2.88cm}p{3.68cm}p{9.95cm}}
\hline
\multicolumn{1}{|p{2.88cm}}{\textbf{Criteria}} & 
\multicolumn{1}{|p{3.68cm}}{\textbf{Description}} & 
\multicolumn{1}{|p{9.95cm}|}{\textbf{Evaluation Metric }} \\ 
\hline
\multicolumn{1}{|p{2.88cm}}{{\footnotesize Scale}} & 
\multicolumn{1}{|p{3.68cm}}{{\footnotesize Teams must scale their solutions to as many OSS projects as possible within 10 months.}} & 
\multicolumn{1}{|p{9.95cm}|}{{\footnotesize The target population size will be at least 1000 contributors across at least 20 OSS projects.\footnote{ This estimation is derived from analyzing GitHub's top 1000 projects, which have an average of 80 contributors each. For a more realistic scope within a one-year timeframe, the base figure was adjusted to 50 contributors per project. Selecting 20 projects, instead of 100, balances realism with the necessity for a scalable technological or methodological solution. This approach pragmatically narrows the focus while maintaining broad applicability across numerous projects.}\par}
{\footnotesize Teams will be evaluated based on the criteria from Round 2.}} \\ 
\hline
\end{tabular}
\end{adjustbox}
\caption{Scale criteria}
\end{table}

\subsubsection{Rationale}

In defining the cohort, we expect that by selecting projects that already demonstrate a foundational level of engagement and quality—evidenced by working code, recognized licensing, a minimum user adoption rate, and active maintenance—we can establish a fertile ground for testing and enhancing sustainability solutions for OSS. The rationale behind these criteria is to focus on projects that, while already valuable and impactful, are at a crucial juncture where strategic interventions can significantly bolster their long-term viability and effectiveness. User adoption is a critical indicator of a project's relevance and potential for broader impact, guiding our efforts toward those initiatives most likely to benefit from and contribute to the overarching goals of open-source sustainability.

\subsection{Democratic governance of artificial intelligence}

\subsubsection{The problem}

Artificial intelligence (AI) is quickly becoming an integral part of the modern world. By 2030, it is estimated that AI could contribute up to \$15.7 trillion to the global economy \cite{pwc_generative_2024}, impacting all industries and societies with new forms of automation and efficiency.

However, as AI’s impact and importance grow, so do the coordination problems related to its creation and governance. Training effective AI algorithms—including but not limited to large language models (LLMs) and other foundation models—requires a collaborative effort among various stakeholders, including researchers, engineers, and entrepreneurs. Ensuring AI is responsible and equitable requires these same experts to work together with end-users and society at large to provide the feedback needed to set guardrails that render AI safe and inclusive.

The problem of governing AI, as we define it here, is thus the problem of specifying and implementing a multi-party, federated, and iterative endeavor. This endeavor is \textit{multi-party} insofar as it spans many entities, roles, and domains of expertise. It is \textit{federated} insofar as it integrates intelligence across ecosystems of different vendors and entities that provide data, code, and infrastructure. And it is \textit{iterative} insofar as it is constantly evolving as it adapts to the ever-changing requirements of creators and end-users.

\paragraph{Additional considerations for democratic AI}

The unprecedented adoption of generative AI has been led by transformative LLMs that consume unprecedented amounts of computing and data. In this opening stage of mainstream AI, the most performant foundational models are being released by either the most valuable corporations in the world, their affiliates and investees, or nation-states. Together they have formed a powerful monopsony over AI infrastructure. The concentration of power in just a handful of entities has created critical vulnerabilities in the safety and effectiveness of the most dominant AI models. Even open-source AI projects released by these companies have not solved the problem, as nearly all do not open-source their training data.

Like prior technology eras, the fear is that a lack of competition can lead to market distortions, more brittle products, and slower innovation. As Tim O’Reilly put it, this ``dark pattern'' ensures that the market no longer picks the winners, but instead delivers ``unequal returns to entrepreneurs, investors, and society'' \cite{oreilly_ai_2024}. A new concern, driven by geopolitics, is that these same large companies could be pushed and subsidized by governments to prioritize competition over greater societal norms, such as protecting privacy and minimizing bias. While AI accountability is taking shape with national-state-level regulations, many new laws are vague and do not prescribe specific actions for compliance or enforcement. Further, it is not clear, even with more precision in regulation, that practically responsible AI can be enforced from the top down.

Further, this logic leaves unquestioned a prevailing myth that only large AI models can and will be effective. We see, instead, smarter AI can follow the logic of ``subsidiarity'' that scholars such as Amy Hasinoff and Nathan Schneider persuasively argue allows local social units to have meaningful autonomy to promote local social goals and push out the gains of AI to its members \cite{hasinoff_scalability_2022}. The responsibility of preventing and addressing harm is therefore accountable to the people who are most directly affected. Context-sensitive AI governance can then scale across larger networks through federalist or polycentric forms of community \cite{thiel_polycentric_2023}. Instead of one distorted market, decentralized AI markets can syndicate and amplify different norms, rules, and objectives. Self-sovereign AI, created in a community is best positioned to represent the perspectives of its constituents and participants.

These considerations motivate us to focus on a specific class of decentralized, democratic, and bottom-up approaches to AI governance in this problem specification. We believe DAOs and other coordination technologies can sustain a new, bottom-up form of AI governance that is more responsive and representative of the stakeholders impacted by AI. Indeed, supporters of bottom-up AI governance have argued that DAOs could easily pool capital and resources to scale and compete with AI giants with more purchasing power \cite{pham_why_2024}. This short-term win might inject new competitive vigor into AI markets to make training and inference more affordable and democratic.

\subsubsection{Current progress}

Social coordination research, as applied to AI governance, shapes how democratic AI might work in practice. For example, constitutional AI approaches, pioneered by Anthropic, align AI algorithms to a set of pre-determined principles that they are benchmarked against \cite{bai_constitutional_2022}. These pre-defined rules give auditors, regulators, and end-users a clear view of which choices were made during model training. Alignment assemblies go one step further in bringing a democratic process to forming constitutions for AI through collective input through conversations about user needs and goals for AI \cite{ganguli_collective_2023}.

DAOs offer a powerful structure to coordinate these activities to create responsible AI by using cryptographic primitives to underwrite a new layer of verifiability across the entire AI lifecycle. By knowing what each party contributed to AI development and tracing that impact from training to inference, new forms of incentives, coalition building, and value creation are unlocked. DAOs can also action business logic on verifiable credentials, zero-knowledge proofs, and other cryptographic artifacts that can bind and audit AI governance frameworks. Finally, by establishing verifiable custody and attribution for contributions to AI models, DAOs could be the most efficient corporate structure to reward the contributors to AI algorithms, thereby arriving at the most accurate value of self-sovereign AI assets in the next generation of AI-focused marketplaces.

Putting some of these ideas into practice, the first AI DAO legally registered is the Endowment of Climate Intelligence (ECI). Formed in March 2024, the ECI governs ClimateGPT: an ensemble of the first task-specific open-source LLMs and benchmarks on the impact of climate change \cite{thulke_climategpt_2024}. The DAO is registered as a DLT Foundation in the Abu Dhabi Global Market free trade zone and uses new cryptographic methods to create new binding and legally enforceable smart contracts. Through successive steps of progressive decentralization \cite{esber_progressive_2023}, the ECI brings together AI engineers, climate change experts, civil society, and enterprises in working groups to refine and govern the model. As a cryptographic coordination layer, the DAO serves three purposes: first, it bootstraps and operates a private ledger that registers machine and human attestations across the AI development lifecycle; second, it coordinates the policies and rules of the Endowment; and third, it enables collective ownership, custody and governance of the Endowment as it seeks to reinvest royalties from paid access to the models.

The ECI’s design as a distributed governance organization is intuitive because of its mission. Climate change is a global problem that will only be solved with cooperation from stakeholders from across the world. As a pioneering AI DAO, the ECI is primed to set the pace for collective AI governance. It does this by first organizing an array of motivated parties to pool the capital to procure compute powered by renewable energy and recruit the engineering talent needed to train the foundational model best. The ECI then seeks to facilitate binding guidance from expert and local stakeholders impacted by climate change to curate training, instruction fine-tuning, and RAG datasets that help ensure the quality and inclusivity of the model. Examples include expert aggregation of high-quality open science research alongside Creative Commons to both meet quality levels and respect fair-use copyright terms. Of equal importance is the inclusivity of the model to build into the inference the views of climate change activists that may have been underrepresented in traditional model training. For instance, the DAO is making a careful effort to include indigenous or inner-city youth’s perspectives in the model that are often overlooked.

Finally, the ECI uses DAOs to manage revenue in a treasury. A percentage of fiat proceeds from paid commercial revenue will then flow back to the ECI to finance subsidies of inference to qualified researchers and further model training. Similarly, cryptographic block rewards from cryptographic utilities, such as Filecoin and forthcoming systems like Gensyn and Fluence, that are used by the system are also brought back into the treasury. These royalties and rewards are reinvested to create new and more valuable versions of the model, which in turn produce more royalties and block rewards. As a new kind of public utility, the models can become a new class of AI regenerative endowments and/or bonds — sustainable in every sense of the term.

The ECI’s regenerative non-profit approach points to possibilities for nascent commercial AI marketplaces. In a community where the users and stakeholders own an AI algorithm, value is determined holistically. Growth gives way to more regenerative priorities such as loyalty, quality, safety, and specialization that come with reinvestment in AI algorithms rather than solely extracting profits. This push towards a more pluralistic notion of AI governance will reflect smarter, agile, safer, and more valuable forms of AI that can serve as commercial and public goods.

DAOs are, therefore, inspiring discussions on how to reimagine valuation methods that can determine contributions to AI and syndicate them in a marketplace. For instance, beyond the canonical Shapley valuation method, Sim et al. propose three incentive conditions for revenue allocation: individual rationality, stability of the grand coalition, and group welfare \cite{sim_collaborative_2020}. This opens up the possibility for more metrics for value, such as loyalty and longevity of coordination. At best DAOs should spur many decentralized markets with different norms, rules, and objectives. This is the true endgame for compliance: the formation of collective governance of AI that can simultaneously be syndicated across various groups and be refined by different stakeholders with different priorities and incentives.

\subsubsection{Success criteria}

The criteria defined below are intended to capture results over a one-year period for a particular cohort of responsible AI projects that intend to be governed as a DAO.

We expect all responsible AI projects included in the evaluation to:

\begin{itemize}
	\item Set up a DAO-type governing infrastructure that allows for transparent coordination of the creation, evaluation, and maintenance of an AI system.
	\item Have at least 5 active members in the DAO.
	\item Have trained, published and deployed a model that can be accessed on an open source platform such as HuggingFace.
	\item Have responsible AI technical documentation for the AI by system, including model card and evidence of a responsible AI framework implementation.
	\item Use cryptographic methods to generate verifiable attestations across the entire AI lifecycle, including all data and source code, governance, and, where possible, compute.
\end{itemize}

\paragraph{Proposal criteria}
The following criteria are intended to help teams define and plan their solution. It may also be used by funders considering proposals that promise to address responsible AI.

\begin{table}[H]
\begin{adjustbox}{max width=\textwidth}
\begin{tabular}{p{2.91cm}p{4.47cm}p{9.13cm}p{2.91cm}p{4.47cm}p{9.13cm}}
\hline
\multicolumn{1}{|p{2.91cm}}{\textbf{Criteria}} & 
\multicolumn{1}{|p{4.47cm}}{\textbf{Description}} & 
\multicolumn{1}{|p{9.13cm}|}{\textbf{Evaluation Metric}} \\ 
\hline
\multicolumn{1}{|p{2.91cm}}{{\footnotesize Responsible AI sustainability}} & 
\multicolumn{1}{|p{4.47cm}}{{\footnotesize Quality of responsible AI implementation}} & 
\multicolumn{1}{|p{9.13cm}|}{{\footnotesize The planned solution must implement responsible AI standards.}} \\ 
\hline
\multicolumn{1}{|p{2.91cm}}{{\footnotesize AI system sustainability}} & 
\multicolumn{1}{|p{4.47cm}}{{\footnotesize Quality and usability of AI system }} & 
\multicolumn{1}{|p{9.13cm}|}{{\footnotesize The planned solution must include a functioning AI model and application.}} \\ 
\hline
\multicolumn{1}{|p{2.91cm}}{{\footnotesize Decentralized governance sustainability}} & 
\multicolumn{1}{|p{4.47cm}}{{\footnotesize Quality of governance infrastructure }} & 
\multicolumn{1}{|p{9.13cm}|}{{\footnotesize The planned solution must implement a decentralized governance structure, enabling the transparent coordination of the creation, evaluation, and maintenance of an AI system \par}} \\ 
\hline
\multicolumn{1}{|p{2.91cm}}{{\footnotesize AI lifecycle transparency sustainability}} & 
\multicolumn{1}{|p{4.47cm}}{{\footnotesize Quality and usability of documentation}} & 
\multicolumn{1}{|p{9.13cm}|}{{\footnotesize The planned solution must provide thorough documentation and cryptographic attestations of the AI lifecycle, enabling third-party audits.\par}} \\ 
\hline
\multicolumn{1}{|p{2.91cm}}{{\footnotesize Contributor sustainability} \newline
} & 
\multicolumn{1}{|p{4.47cm}}{{\footnotesize Quality and participation rate of contributors } \newline
} & 
\multicolumn{1}{|p{9.13cm}|}{{\footnotesize The planned solution must encourage active participation from experts and impacted stakeholders.}} \\ 
\hline
\multicolumn{1}{|p{2.91cm}}{{\footnotesize Representative governance}} & 
\multicolumn{1}{|p{4.47cm}}{{\footnotesize Governance should be representative of end-user concerns, especially stakeholders who are most vulnerable to the model’s impact} \newline
} &
\multicolumn{1}{|p{9.13cm}|}{{\footnotesize The planned solution should include special working groups for representatives of end-users and create mechanisms for feedback and grievances to be registered and addressed. \par}} \\ 
\hline
\end{tabular}
\end{adjustbox}
\caption{Proposal criteria}
\end{table}

\paragraph{Implementation criteria}
The following implementation criteria cover metrics that need to be evaluated on any implementation of the solution or intervention. We expect proposals to evolve over the course of implementation.

\begin{table}[H]
\begin{adjustbox}{max width=\textwidth}
\begin{tabular}{p{2.88cm}p{3.73cm}p{9.9cm}p{2.88cm}p{3.73cm}p{9.9cm}}
\hline
\multicolumn{1}{|p{2.88cm}}{\textbf{Criteria}} & 
\multicolumn{1}{|p{3.73cm}}{\textbf{Description}} & 
\multicolumn{1}{|p{9.9cm}|}{\textbf{Evaluation Metric }} \\ 
\hline
\multicolumn{1}{|p{2.88cm}}{{\footnotesize Responsible AI consideration }} & 
\multicolumn{1}{|p{3.73cm}}{{\footnotesize Responsible AI framework}} & 
\multicolumn{1}{|p{9.9cm}|}{{\footnotesize For 75$\%$ of projects in the cohort, clear implementation of an industry-standard responsible AI framework must be provided.\par}} \\ 
\hline
\multicolumn{1}{|p{2.88cm}}{\multirow{2}{*}{\parbox{2.88cm}{{\footnotesize Functional AI system}}}} & 
\multicolumn{1}{|p{3.73cm}}{{\footnotesize Model}} & 
\multicolumn{1}{|p{9.9cm}|}{{\footnotesize For 75$\%$ of projects in the cohort, an AI model repository must be published on Github and Hugging Face, along with an open-source license.\par}} \\ 
\hhline{~--}
\multicolumn{1}{|p{2.88cm}}{} & 
\multicolumn{1}{|p{3.73cm}}{{\footnotesize Application}} & 
\multicolumn{1}{|p{9.9cm}|}{{\footnotesize For 75$\%$ of projects in the cohort, an AI application using the published model must be deployed and made available to users. \par}} \\ 
\hline
\multicolumn{1}{|p{2.88cm}}{{\footnotesize Decentralized governance infrastructure}} & 
\multicolumn{1}{|p{3.73cm}}{{\footnotesize Governance structure for transparent coordination }} & 
\multicolumn{1}{|p{9.9cm}|}{{\footnotesize For 75$\%$ of projects in the cohort, a DAO-type governing infrastructure must be established and documented. The governance structure must allow for transparent coordination of the creation, evaluation, and maintenance of the proposed AI system.\par}} \\ 
\hline
\multicolumn{1}{|p{2.88cm}}{{\footnotesize AI lifecycle transparency}} & 
\multicolumn{1}{|p{3.73cm}}{{\footnotesize Documentation}} & 
\multicolumn{1}{|p{9.9cm}|}{{\footnotesize For 75$\%$ of projects in the cohort, documentation regarding the AI lifecycle must be provided. This includes, but is not limited to, a model card with substantial evaluation results. Cryptographic methods are encouraged to generate verifiable attestations across the entire lifecycle, including all data and source code, governance, and, where possible, compute.\par}} \\ 
\hline
\multicolumn{1}{|p{2.88cm}}{{\footnotesize User adoption}} & 
\multicolumn{1}{|p{3.73cm}}{{\footnotesize Growing number of users}} & 
\multicolumn{1}{|p{9.9cm}|}{{\footnotesize The user base of the AI model does not decline over the course of implementing the solution.}} \\ 
\hline
\multicolumn{1}{|p{2.88cm}}{\multirow{2}{*}{\parbox{2.88cm}{{\footnotesize Contributor sustainability}}}} & 
\multicolumn{1}{|p{3.73cm}}{{\footnotesize Participation of experts}} & 
\multicolumn{1}{|p{9.9cm}|}{{\footnotesize For 75$\%$ of projects in the cohort, clear working groups are established for expert contributions to the model, and the average participation rate of both technical and domain experts (measured in active membership and active votes on governance proposals) increases by at least 25$\%$.\par}} \\ 
\hhline{~--}
\multicolumn{1}{|p{2.88cm}}{} & 
\multicolumn{1}{|p{3.73cm}}{{\footnotesize Stakeholder feedback}} & 
\multicolumn{1}{|p{9.9cm}|}{{\footnotesize For 75$\%$ of projects in the cohort, clear working groups are established for stakeholders to help fine-tune the model and set automated and human benchmarking (such as reinforcement learning from human feedback). Stakeholder groups should have proper representation for end-user viewpoints and understanding of trust and safety concerns and have active contributions to working groups (measured in active membership and votes) increased by at least 25$\%$. \par}} \\ 
\hline
\end{tabular}
\end{adjustbox}
\caption{Implementation criteria}
\end{table}

\subsection{Regenerative finance: Toward a sustainable and equitable planetary economy}

\subsubsection{The problem}

Traditional finance and investment models are primarily concerned with, and generally legally obligated to ensure maximizing shareholder returns, often at the expense of the environment and marginalized communities. The financial system frequently overlooks the need to invest in realizable activities that regenerate natural capital, such as reforestation and carbon removal, or social capital, such as community development and equitable access to credit. This has resulted in substantial inequalities, environmental degradation, and economic instability. Regenerative Finance (ReFi) is a relatively new concept-turned-movement that seeks to address these inherent flaws in the existing financial system.\footnote{ The underlying concept of this approach traces its roots to "Regenerative Capitalism," a term first coined by economist John Fullerton in 2015 [CITE fullerton\_regenerative\_2015]. Though Fullerton did not originally link his concept to decentralized technologies, his eight principles of a regenerative economy–(1) in right relationship; (2) views wealth holistically; (3) innovative, adaptive, responsive; (4) empowered participation; (5) honors community and place; (6) edge effect abundance; (7) robust circulatory flow; and (8) seeks balance$\#$ have since been adopted by impact-oriented professionals within the Web3 space. These practitioners have used Fullerton's principles to construct systems that promote behaviors that enhance overall systemic health and discourage actions that cause its deterioration.} A key distinction between extractive economic systems and regenerative economic systems is that while the former assigns value to the \textit{consumption} of CPRs, such as natural assets and increases profits for a few individuals, the latter seeks to assign value to the \textit{protection }and \textit{regeneration} of these assets where the profit of their exchange is distributed fairly. 

ReFi can be thought of in terms of ``big problems" and ``small primitives." The ``big problem" ReFi intends to address is to build a more ecologically and socially sustainable alternative in governing global CPRs. In essence, the ReFi movement seeks to encourage ecological restoration, social justice, and economic resilience by aligning financial incentives with the principles of sustainability, equity, and plurality; thus promoting a symbiotic catallaxy between nature and people, galvanized by primitives that support a viable planetarity. ReFi poses itself to address these multifaceted, planetary scale issues through the microeconomic use of fundamental cryptoeconomics tools and coordination mechanisms, which are colloquially called ``primitives." In what ways can we build tools for more positive-sum games that increase resource capacity?  This refers to not only financial resource capacity, but across the eight types of capital, including–natural, financial, material, intellectual, experiential, social, cultural, and spiritual \cite{pecone_eight_2023}. Using small primitives that agglomerate to resolve big problems, ReFi aims to challenge a speculative, extractive, and ultimately harmful status quo at the macroeconomic level. This is to say that to be ``regenerative," protocols, mechanisms, and organizations must themselves address harms generated by traditional finance and investment models in the capitalist system. This concept intentionally works to restore faith in operations of many enterprises that can fundamentally provide technical utility while resolving these externalized harms through systemically positive-sum dynamics. 

In a recent analysis and report by Carbon Copy on \textit{The State of ReFi} in Q1 2024, written in coordination with ReFi DAO, the authors state, ``At a more practical level, ReFi can be described as Web3-powered ecological and social impact. In other words, taking the available Web3 solutions — blockchain, cryptocurrency, smart contracts, and decentralized autonomous organizations (DAOs) — and combining them with other modern technologies to build solutions that address our systemic issues" \cite{carbon_copy_state_2024}. The key assertion in this consensus among many varied definitions of this concept is the addressing of systemic issues, not just providing novel utilities atop of their faults. 

For some within the broader collective of ReFi organizations, developers, and evangelists, the definitions presented are still narrow, albeit broad in their scope. This is because there is a significant bottom-up, grassroots, and often subaltern base from which ReFi can be viewed. Bottom-up in the sense that the primitives being developed are reinventing basic fundamentals of traditional finance that are taken as given. Some may include crowdfunding protocols for allocating capital based on community amplification, or mechanisms for disbursing funds for those who do provable virtuous actions, or providing transparent retroactive certifications of promissory actions to come, among many other novel primitives. Through ReFi as a movement, extractivist goods and transactional services become public goods and services to the planet. The grassroots and subaltern state of many of these efforts is given that several significant projects in this vein are being developed in the Global South, by people whose actions have been usually beyond the tunnel vision of Silicon Valley startup funders and multinational corporate investment firms. ReFi has become a platform for typically unsung innovations in the traditional financial system. This said, there are still areas of concern regarding coordination in the current state of ReFi, beyond those addressed in the aforementioned report. These include succumbing to what we are calling the ``\textit{accelerationist trap}," the risk of virtue signaling turning into \textit{virtue noise}, and a lack of cohesive \textit{standard setting} between TradFi and ReFi.

\begin{itemize}
	\item \textbf{Sustainability bolstering the accelerationist trap\footnote{ ``Sustainability" itself has been a tenuous term for traditional finance, as its most agreed upon definition was established in the 1987 Report of the United Nations World Commission on Environment and Development, titled Our Common Future, but more commonly known as the Brundtland Report \cite{world_commission_on_environment_and_development_our_1987}. The report has defined sustainable development as, ``development that meets the needs of the present without compromising the ability of future generations to meet their own needs" (ibid). This basis established the terms used by the UN which transcended their Millenium Development Goals toward their more current Sustainable Development Goals (SDGs) and yet the fundamental practices of sustainable development used by global governance has not changed in nearly 40 years.}}: As the ReFi movement gains momentum, there's a risk of falling into the "accelerationist trap" where rapid scale and financial profit motives overshadow the core principles of regeneration, and sustainability becomes a basis for attracting venture capital without proof of impact. This could lead to prioritizing short-term gains over long-term ecological and social values, potentially replicating the same dynamics criticized in traditional finance by the ReFi movement.

	\item \textbf{Virtue signaling becoming virtue noise}: In the ReFi space, the prevalence of virtue signaling—where entities proclaim their sustainability efforts more for marketing than impact—can degrade into "virtue noise," where the abundance of unsubstantiated claims drowns out genuine progress. Ensuring accountability and transparency through verifiable metrics, such as ``Proof-of-Virtue," and ``Proof-of-Useful-Work (PoUW)" is essential to prevent this drift \cite{bachman_incentivizing_2023}.

	\item \textbf{Standard setting between TradFi and ReFi}: The ReFi movement faces challenges establishing cohesive standards that align with traditional finance (TradFi) systems. For example, in the realm of the carbon market, the challenges of price discovery, accounting, and the quality of carbon credits are reflective of the broader issue of inconsistent standards \cite{rong_blockchain_2023}. Without unified frameworks, ReFi initiatives struggle with interoperability and recognition in broader financial markets, hampering their effectiveness and scalability. There is a significant need to develop common standards that could bridge the gap between innovative ReFi mechanisms and established financial protocols.

\end{itemize}

Despite these challenges, however, the ongoing exploration of ReFi reflects its potential to bring about a significant transformation in our economic systems, aligning them more closely with environmental and social sustainability objectives. Given the current development of ReFI, we ask: How can we effectively leverage the innovative tools and approaches of ReFi to ensure they truly contribute regeneratively to sustainable development, rather than merely replicating traditional financial systems' shortcomings under a new greenwashed guise?

\subsubsection{Current progress}

Currently, the ReFi movement seeks to achieve sustainable governance of CPRs through digital monitoring, reporting and verification, asset tokenization, and decentralized governance practices \cite{schletz_blockchain_2023}. We identify two levels of interrelated progress in the space: (1) coordinated financing and governance of CPRs; and (2) regeneration of CPRs through impact-oriented real-world projects.\footnote{Currently, there is a notable absence of Fortune 500 examples within the ReFi sector. This is not necessarily indicative of the movement's effectiveness or ambitions. Rather, the focus is on a collective of smaller, cohesive projects working synergistically toward common regenerative goals, which suggests that the strength of ReFi may lie in the aggregate impact of numerous smaller initiatives rather than the involvement of large corporate entities.} 

Several Web3 innovations have emerged in recent years to address the challenge of sustainably financing and governing CPRs. For example, \href{https://www.optimism.io/}{\uline{\textcolor[HTML]{1155CC}{Optimism}}} has played a crucial role in financing public goods through mechanisms like Retroactive Public Goods Funding (RetroPGF). This initiative demonstrates a pioneering approach by distributing tokens to contributors based on the impact and value they bring to the ecosystem, rather than traditional upfront funding models. By doing so, Optimism incentivizes ongoing contributions to the development and maintenance of public goods, aligning financial incentives with long-term ecological and social benefits. Optimism is notable because of the size of the funding it allocates to public goods, but there are other, smaller players, who are also innovating on creating financing for ReFi projects, such as \href{https://giveth.io/}{\uline{\textcolor[HTML]{1155CC}{Giveth}}}, \href{https://clr.fund/#/}{\uline{\textcolor[HTML]{1155CC}{CLRfund}}}, \href{https://www.gitcoin.co/}{\uline{\textcolor[HTML]{1155CC}{Gitcoin}}}, and \href{https://www.artizen.fund/}{\uline{\textcolor[HTML]{1155CC}{Artizen}}}.

Similarly, \href{https://hypercerts.org/}{\uline{\textcolor[HTML]{1155CC}{HyperCerts}}} offers a novel solution for the certification of sustainable impacts. These digital certificates are blockchain-based records that provide verifiable proof of ecological outcomes, such as carbon sequestration or reforestation efforts. By making these certifications transparent and immutable, HyperCerts provides a ``proof-of-virtue" mechanism and ensures that the environmental benefits claimed by projects are indeed credible and traceable. This transparency is crucial for building trust among investors and stakeholders, facilitating more informed decisions in the financing of regeneration projects.

The new flows of finance into ReFi have helped sustain a new generation of innovators who are using Web3 for regenerative ends. Many projects endeavor to bridge the gap between the tangible world and the financial sphere through the creation, verification, and valuation of token-based representations of real-world assets (RWAs). By facilitating the creation and utilization of novel asset types (like natural resources), these projects propel economies toward progression based on environmental guardianship and conscientious resource administration. Projects like \href{https://www.openforestprotocol.org/}{\uline{\textcolor[HTML]{1155CC}{Open Forest Protocol}}} (OFP) and \href{https://app.regen.network/}{\uline{\textcolor[HTML]{1155CC}{Regen Network}}} use blockchain to provide the infrastructure for transparent monitoring and reporting of environmental data in the carbon credit creation process, enhancing the management of CPRs such as forestry projects. For instance, OFP enables various stakeholders, including environmental groups, corporations, and governments, to generate transparent, immutable, and verified proof-of-impact data in real-time. The project has partnered with organizations like \href{https://www.solid.world/}{\uline{\textcolor[HTML]{1155CC}{Solid World}}}, whose operations are on risk assessment and scoring carbon credit futures, to scale its efforts in network expansion, market access, broadening the technological base, and diversifying funding sources with robust financial support mechanisms.

Asset tokenization projects like Toucan, Flowcarbon, and \href{https://www.klimadao.finance/}{\uline{\textcolor[HTML]{1155CC}{KlimaDAO}}} focus on the tokenization of carbon credits, aiming to create a transparent, efficient marketplace for carbon trading, thus increasing market liquidity and providing verifiable tracking of carbon credit transactions. Expanding beyond the traditional carbon credits market, \href{https://www.kolektivo.network/}{\uline{\textcolor[HTML]{1155CC}{Kolektivo}}}, a pilot initiative in Curaçao, aims to tokenize natural assets such as food forests and coral reefs to promote local self-reliance and economic resilience in marginalized communities. Using the ReFi-focused Celo blockchain and the Astral Protocol, Kolektivo creates a Decentralized Exchange Trading System (DETS) that bolsters environmental resilience and addresses interdependent financial and biospheric risks. This approach incentivizes the preservation of natural resources and broadens access to carbon markets.

Despite these efforts, most ReFi projects are limited to tokenizing carbon and nature credits to broaden the financialization of commodified assets that are traded on markets to incentivize companies and individuals to lower and/or offset their carbon emissions. Although it is true that tokenization holds the promise of allowing companies to support public goods while capturing private value, potentially improving their reputation and customer relationships by integrating these tokens into stakeholder interactions, the sole focus on natural asset tokenization risks the perpetuation of an extractive economic model by commodifying nature, encouraging short-term transactions of carbon credits rather than fostering long-term sustainable investments. While tokenization could alter corporate interactions positively, it might also maintain the status quo without truly regenerating the environment. Further research and deeper integration with regenerative economics are necessary to ensure that ReFi contributes effectively to sustainable and equitable economic practices.

As such, we believe that future projects should engage more proactively and thoroughly with the ``regeneration" aspect of regenerative finance and provide solutions that would genuinely promote long-term environmental sustainability and societal well-being.

\subsubsection{Success criteria}

The criteria defined below are intended to capture results over a one-year period for a cohort of ReFi projects.

We expect all ReFi projects included in the evaluation to:

\begin{itemize}
	\item Have a clear definition of a common pool resource (CPR) that is to be governed (whether it is tokenized carbon removal, renewable energy, nature-based solutions, etc.)
	\item Address early problems laid out in the primitives to be designed or utilized. 
\end{itemize}

\paragraph{Proposal criteria}

The following proposal criteria are intended to help teams define and plan their solution. It may also be used by funders considering proposals that promise to address the sustainable governance and regeneration of common resources.

\begin{table}[H]
\begin{adjustbox}{max width=\textwidth}
\begin{tabular}{p{2.91cm}p{4.47cm}p{9.1cm}p{2.91cm}p{4.47cm}p{9.1cm}}
\hline
\multicolumn{1}{|p{2.91cm}}{\textbf{Criteria}} & 
\multicolumn{1}{|p{4.47cm}}{\textbf{Description}} & 
\multicolumn{1}{|p{9.1cm}|}{\textbf{Evaluation Metric}} \\ 
\hline
\multicolumn{1}{|p{2.91cm}}{{\footnotesize Regenerative Intentionality and Relevance}} & 
\multicolumn{1}{|p{4.47cm}}{{\footnotesize Explicit impact-driven goals and what stakeholders deem this impact of capital relevance to them.} \newline
} & 
\multicolumn{1}{|p{9.1cm}|}{{\footnotesize The planned solution must demonstrate regenerative intentionality in its goals and objectives, ensuring alignment with relevant issues and challenges to generate a positive impact. This includes sustainable funding, financial management, and a net positive increase in resource availability, with a focus on impact significance. Beyond economic relevance - Consider 8 Forms of Capital (natural, financial, material, intellectual, experiential, social, cultural, and spiritual).\par}} \\ 
\hline
\multicolumn{1}{|p{2.91cm}}{{\footnotesize Decentralized Governance and Technical Sustainability}} & 
\multicolumn{1}{|p{4.47cm}}{{\footnotesize Building institutional capacity and incentive structures, and enhancing interoperability of systems. Positive-sum participation.}} & 
\multicolumn{1}{|p{9.1cm}|}{{\footnotesize The planned solution must provide tools for relevant stakeholders to participate and engage in governance without technical or institutional barriers. It must also ensure data traceability, transparency, authenticity, and interoperability. Efficiency of the system must not come at a detriment to its baseline performance or those stakeholders participating in it. Ensuring the systems are not more intensive to operate than the impact they provide. (e.g. power consumption at scale)\par}} \\ 
\hline
\multicolumn{1}{|p{2.91cm}}{{\footnotesize Internalized Impact and Materiality}} & 
\multicolumn{1}{|p{4.47cm}}{{\footnotesize Regeneration of resources without externalizing harms.} \newline
{\footnotesize Translation of nominal units of account into CPRs with traceable impact via novel primitives which generate material change to the fundamental problems.\par}} & 
\multicolumn{1}{|p{9.1cm}|}{{\footnotesize The planned solution should address the internalization of externalities in specific industries or anthropogenic practices, particularly utilizing the creation of regenerative primitives. This involves translating traceable impact into CPRs and ensuring that the creation of these leads to material changes in operational decisions initially addressed, rather than solely circulating through superficial financial transactions. The planned solution must track and ensure the net positive increase in resource availability, without externalizing harm. The net benefit to communities must outweigh the net harm.\par}} \\ 
\hline
\end{tabular}
\end{adjustbox}
\caption{Proposal criteria}
\end{table}

\paragraph{Implementation criteria}

The following implementation criteria cover metrics that need to be evaluated on any implementation of the solution or intervention. We expect proposals to evolve over the course of implementation.

\begin{table}[H]
\begin{adjustbox}{max width=\textwidth}
\begin{tabular}{p{2.88cm}p{3.73cm}p{9.9cm}p{2.88cm}p{3.73cm}p{9.9cm}}
\hline
\multicolumn{1}{|p{2.88cm}}{\textbf{Criteria}} & 
\multicolumn{1}{|p{3.73cm}}{\textbf{Description}} & 
\multicolumn{1}{|p{9.9cm}|}{\textbf{Evaluation Metric }} \\ 
\hline
\multicolumn{1}{|p{2.88cm}}{\multirow{2}{*}{\parbox{2.88cm}{{\footnotesize Regenerative Intentionality and Relevance}}}} & 
\multicolumn{1}{|p{3.73cm}}{{\footnotesize Explicit impact-driven goals}} & 
\multicolumn{1}{|p{9.9cm}|}{{\footnotesize At least 75$\%$ of project goals aligned with regenerative impact objectives. } \newline
} \\ 
\hhline{~--}
\multicolumn{1}{|p{2.88cm}}{} & 
\multicolumn{1}{|p{3.73cm}}{{\footnotesize Project funding}} & 
\multicolumn{1}{|p{9.9cm}|}{{\footnotesize The planned solution must include a way to sustainably fund the project and reward contributors without resorting to accelerationist pressures. High ratio (e.g., more than 50$\%$) of sustainable funding sources to total project funding. \par}} \\ 
\hline
\multicolumn{1}{|p{2.88cm}}{\multirow{4}{*}{\parbox{2.88cm}{{\footnotesize Decentralized Governance and Technical Sustainability}}}} & 
\multicolumn{1}{|p{3.73cm}}{{\footnotesize Institutional capacity}} & 
\multicolumn{1}{|p{9.9cm}|}{{\footnotesize Building institutional capacity for decentralized stakeholders and providing incentive structures for participation. \par}} \\ 
\hhline{~--}
\multicolumn{1}{|p{2.88cm}}{} & 
\multicolumn{1}{|p{3.73cm}}{{\footnotesize Interoperability }} & 
\multicolumn{1}{|p{9.9cm}|}{{\footnotesize Solution standards are recognized and adopted by a sizable number of industry stakeholders. Data should be interoperable across multiple platforms and ledgers. The solution should effectively integrate related on-chain and off-chain systems. \par}} \\ 
\hhline{~--}
\multicolumn{1}{|p{2.88cm}}{} & 
\multicolumn{1}{|p{3.73cm}}{{\footnotesize Quality of data }} & 
\multicolumn{1}{|p{9.9cm}|}{{\footnotesize Data provenance and transparency should be accounted for. Transparency score based on quantifiable metrics (e.g., percentage of data publicly available). \par}} \\ 
\hhline{~--}
\multicolumn{1}{|p{2.88cm}}{} & 
\multicolumn{1}{|p{3.73cm}}{{\footnotesize Proof-of-Virtue} \newline
{\footnotesize Proof-of-Useful Work}} & 
\multicolumn{1}{|p{9.9cm}|}{{\footnotesize Verify the actual realization of claimed project benefits. Percentage of benefits audited and validated after 10 months of project implementation. \par}} \\ 
\hline
\multicolumn{1}{|p{2.88cm}}{{\footnotesize Internalized Impact and Materiality}} & 
\multicolumn{1}{|p{3.73cm}}{{\footnotesize Regeneration of resources }} & 
\multicolumn{1}{|p{9.9cm}|}{{\footnotesize Resource availability must see a net positive increase while avoiding externalizing harms. These can be measured in quantitative indicators such as hectares of forest restored and kilowatts of renewable energy generated. There should be a percentage increase in carbon credits or other regenerative assets generated compared to the baseline, yielded by tangible results. \par}} \\ 
\hline
\multicolumn{1}{|p{2.88cm}}{{\footnotesize User adoption }} & 
\multicolumn{1}{|p{3.73cm}}{{\footnotesize Growing number of stakeholders that recognize the solution as a valid standard}} & 
\multicolumn{1}{|p{9.9cm}|}{{\footnotesize There should be a reasonable growth of relevant stakeholders who either adopt or recognize the project’s legitimacy. \par}} \\ 
\hline
\end{tabular}
\end{adjustbox}
\caption{Implementation criteria}
\end{table}

\paragraph{Scale criteria}
Traditionally, CPRs are associated with small local systems such as irrigation systems and fisheries \cite{ostrom_governing_1990}. However, ReFi has the potential to address much larger-scale problems requiring coordination at a global level. As such, a successful solution must be scalable. 

\begin{table}[H]
\begin{adjustbox}{max width=\textwidth}
\begin{tabular}{p{2.88cm}p{3.68cm}p{9.95cm}p{2.88cm}p{3.68cm}p{9.95cm}}
\hline
\multicolumn{1}{|p{2.88cm}}{\textbf{Criteria}} & 
\multicolumn{1}{|p{3.68cm}}{\textbf{Description}} & 
\multicolumn{1}{|p{9.95cm}|}{\textbf{Evaluation Metric }} \\ 
\hline
\multicolumn{1}{|p{2.88cm}}{{\footnotesize Multi-level Scale}} & 
\multicolumn{1}{|p{3.68cm}}{{\footnotesize Cross-scale linkages }} & 
\multicolumn{1}{|p{9.95cm}|}{{\footnotesize The solution must consider interactions between actors of different levels of political or social organizations. \par}} \\ 
\hline
\multicolumn{1}{|p{2.88cm}}{{\footnotesize Macroscopic Shift from Exchange Value to Use Value}} & 
\multicolumn{1}{|p{3.68cm}}{{\footnotesize ReFi assets and primitives that deploy CPRs can be measured by the extent to which their value transitions from being primarily derived from financial exchange to being derived from their intrinsic utility in generating positive environmental or social outcomes.\par}} & 
\multicolumn{1}{|p{9.95cm}|}{{\footnotesize The solution must yield a greater percentage of total value derived from the use value of the underlying regenerative asset compared to the exchange value it was initially assigned upon instantiation.\par}
{\footnotesize Qualitative assessments of stakeholders' perception of the assets' utility and intrinsic value.\par}
{\footnotesize Analysis of market trends indicating increasing demand for CPRs based on their regenerative use value rather than speculative trading. \par}} \\ 
\hline
\multicolumn{1}{|p{2.88cm}}{{\footnotesize Regenerative} \newline
{\footnotesize Market Insights and Analysis}} & 
\multicolumn{1}{|p{3.68cm}}{{\footnotesize Investment portfolios and ratings agencies develop competitive indices to TradFi stock indices to assess viability of transitioning their asset classes  from speculative ones to regenerative ones.\par}} & 
\multicolumn{1}{|p{9.95cm}|}{{\footnotesize Major investment firms declare significant investment into CPRs via investment vehicles traceable not by price speculation but by regenerative impact performance. Investments would go beyond economic corollary and become causal in promoting direct impact.\par}
{\footnotesize Sovereign wealth funds begin to publish their utilization of CPRs as stable asset classes and support infrastructure that promote positive-sum dynamics for their respective citizens and allies.\par}
{\footnotesize ReFi assets are utilized in parity with, or in lieu of, sovereign currency, government bonds, or debt (in debt-for-nature or debt-for-social impact swaps). This parity promotes units of account that maintain sovereign economic solvency while promoting positive-sum impact beyond economic forecasts in TradFi market cycles. At-scale, CPRs may reliably offset the unforeseen socioeconomic consequences of these typical market cycles.\par}} \\ 
\hline
\end{tabular}
\end{adjustbox}
\caption{Scale criteria}
\end{table}

\section*{Acknowledgements}

We are deeply grateful for suggestions, comments, and feedback from 
Divya Siddarth, 
Anna Weichselbraun, 
Eugene Leventhal, 
Hélène Landemore, 
Tony Douglas, 
Connor Spelliscy, 
Sara Horowitz, 
Dawn Song, 
Nathan Schneider, 
Daniel Kronovet, 
Marina Markezic, 
Scott Moore, 
Benedict Lau,
Dennison Bertram, 
Cent Hosten, 
and Hazel Devjani.

\bibliographystyle{abbrv}
\bibliography{references}

\begin{thebibliography}{100}

\bibitem{abadi_blockchain_2018}
J.~Abadi and M.~Brunnermeier.
\newblock Blockchain {Economics}, Dec. 2018.

\bibitem{abar_agent_2017}
S.~Abar, G.~K. Theodoropoulos, P.~Lemarinier, and G.~M.~P. O’Hare.
\newblock Agent {Based} {Modelling} and {Simulation} tools: {A} review of the
  state-of-art software.
\newblock {\em Computer Science Review}, 24:13--33, May 2017.

\bibitem{abidin_private_2020}
C.~Abidin and G.~de~Seta.
\newblock Private messages from the field: {Confessions} on digital ethnography
  and its discomforts.
\newblock {\em Journal of Digital Social Research}, 2020.

\bibitem{adriano_technological_2021}
A.~Adriano.
\newblock Technological innovation is fueling the resurgence of community
  currencies.
\newblock Technical report, International Monetary Fund, Mar. 2021.

\bibitem{al-breiki_trustworthy_2020}
H.~Al-Breiki, M.~H.~U. Rehman, K.~Salah, and D.~Svetinovic.
\newblock Trustworthy {Blockchain} {Oracles}: {Review}, {Comparison}, and
  {Open} {Research} {Challenges}.
\newblock {\em IEEE Access}, 8:85675--85685, 2020.
\newblock Conference Name: IEEE Access.

\bibitem{alchian_uncertainty_1950}
A.~A. Alchian.
\newblock Uncertainty, {Evolution}, and {Economic} {Theory}.
\newblock {\em Journal of Political Economy}, 58(3):211--221, June 1950.
\newblock Publisher: The University of Chicago Press.

\bibitem{allen_ethics_2023}
D.~Allen, G.~Weyl, D.~Siddarth, J.~Simons, and W.~Lim.
\newblock Ethics of {Decentralized} {Social} {Technologies}: Lessons from
  {Web3}, the {Fediverse}, and {Beyond}.
\newblock Technical report, Harvard University or the Justice, Health \&
  Democracy Impact Initiative, Mar. 2023.

\bibitem{allen_cryptodemocracy_2019}
D.~W. Allen, C.~Berg, and A.~Lane.
\newblock {\em Cryptodemocracy: {How} {Blockchain} {Can} {Radically} {Expand}
  {Democratic} {Choice}}.
\newblock 2019.

\bibitem{althoff_reachability_2010}
M.~Althoff.
\newblock {\em Reachability analysis and its application to the safety
  assessment of autonomous cars}.
\newblock Doctoral {Dissertation}, Technische Universität München, 2010.

\bibitem{alvesson_taking_2000}
M.~Alvesson and D.~Kärreman.
\newblock Taking the {Linguistic} {Turn} in {Organizational} {Research}:
  {Challenges}, {Responses}, {Consequences}.
\newblock {\em The Journal of Applied Behavioral Science}, 36(2):136--158, June
  2000.
\newblock Publisher: SAGE Publications Inc.

\bibitem{applbaum_legitimacy_2010}
A.~I. Applbaum.
\newblock Legitimacy without the {Duty} to {Obey}.
\newblock {\em Philosophy \& Public Affairs}, 38(3):215--239, 2010.

\bibitem{aristotle_nicomachean_353bce}
Aristotle.
\newblock {\em Nicomachean {Ethics}}.
\newblock 353 BCE.

\bibitem{arroyo_dao-analyzer_2022}
J.~Arroyo, D.~Davó, E.~Martínez-Vicente, Y.~Faqir-Rhazoui, and S.~Hassan.
\newblock {DAO}-{Analyzer}: {Exploring} {Activity} and {Participation} in
  {Blockchain} {Organizations}.
\newblock In {\em Companion {Publication} of the 2022 {Conference} on
  {Computer} {Supported} {Cooperative} {Work} and {Social} {Computing}},
  {CSCW}'22 {Companion}, pages 193--196, New York, NY, USA, Nov. 2022.
  Association for Computing Machinery.

\bibitem{ast_when_2021}
F.~Ast and B.~Deffains.
\newblock When {Online} {Dispute} {Resolution} {Meets} {Blockchain}: {The}
  {Birth} of {Decentralized} {Justice}.
\newblock {\em Stanford Journal of Blockchain Law \& Policy}, June 2021.

\bibitem{atkinson_ethnography_1998}
P.~Atkinson and M.~Hammersley.
\newblock Ethnography and participant observation.
\newblock In {\em Strategies of {Qualitative} {Inquiry}}, pages 248--261.
  Thousand Oaks: Sage., 1998.

\bibitem{austgen_dao_2023}
J.~Austgen, A.~Fábrega, S.~Allen, K.~Babel, M.~Kelkar, and A.~Juels.
\newblock {DAO} {Decentralization}: {Voting}-{Bloc} {Entropy}, {Bribery}, and
  {Dark} {DAOs}.
\newblock {\em ArXiv}, Nov. 2023.

\bibitem{avner_institutions_2006}
G.~Avner.
\newblock {\em Institutions and the {Path} to the {Modern} {Economy}: {Lessons}
  from {Medieval} {Trade}}.
\newblock Cambridge University Press, 2006.

\bibitem{bachman_incentivizing_2023}
J.~Bachman, S.~Chakravorti, S.~Rane, and K.~Thyagarajan.
\newblock Incentivizing {Gigaton}-{Scale} {Carbon} {Dioxide} {Removal} via a
  {Climate}-{Positive} {Blockchain}, Aug. 2023.
\newblock arXiv:2308.02653 [cs].

\bibitem{bai_constitutional_2022}
Y.~Bai, S.~Kadavath, S.~Kundu, A.~Askell, J.~Kernion, A.~Jones, A.~Chen,
  A.~Goldie, A.~Mirhoseini, C.~McKinnon, C.~Chen, C.~Olsson, C.~Olah,
  D.~Hernandez, D.~Drain, D.~Ganguli, D.~Li, E.~Tran-Johnson, E.~Perez,
  J.~Kerr, J.~Mueller, J.~Ladish, J.~Landau, K.~Ndousse, K.~Lukosuite,
  L.~Lovitt, M.~Sellitto, N.~Elhage, N.~Schiefer, N.~Mercado, N.~DasSarma,
  R.~Lasenby, R.~Larson, S.~Ringer, S.~Johnston, S.~Kravec, S.~E. Showk,
  S.~Fort, T.~Lanham, T.~Telleen-Lawton, T.~Conerly, T.~Henighan, T.~Hume,
  S.~R. Bowman, Z.~Hatfield-Dodds, B.~Mann, D.~Amodei, N.~Joseph,
  S.~McCandlish, T.~Brown, and J.~Kaplan.
\newblock Constitutional {AI}: {Harmlessness} from {AI} {Feedback}, Dec. 2022.
\newblock arXiv:2212.08073 [cs].

\bibitem{baldwin_paradox_2015}
K.~Baldwin.
\newblock {\em The {Paradox} of {Traditional} {Chiefs} in {Democratic}
  {Africa}}.
\newblock Cambridge University Press, 1 edition, Nov. 2015.

\bibitem{baliga_performance_2018}
A.~Baliga, I.~Subhod, P.~Kamat, and S.~Chatterjee.
\newblock Performance {Evaluation} of the {Quorum} {Blockchain} {Platform}.
\newblock {\em ArXiv}, July 2018.

\bibitem{bannermanquist_marshall_2022}
J.~Bannermanquist.
\newblock Marshall {Islands} legally recognizes {DAOs} as domestic limited
  liability companies, Dec. 2022.

\bibitem{bao_large_2021}
L.~Bao, X.~Xia, D.~Lo, and G.~C. Murphy.
\newblock A {Large} {Scale} {Study} of {Long}-{Time} {Contributor} {Prediction}
  for {GitHub} {Projects}.
\newblock {\em IEEE Transactions on Software Engineering}, 47(6):1277--1298,
  June 2021.

\bibitem{baum_eagle_2022}
C.~Baum, J.~Chiang, B.~David, and T.~Frederiksen.
\newblock Eagle: {Eﬃcient} {Privacy} {Preserving} {Smart} {Contracts}.
\newblock 2022.

\bibitem{baum_survival-enhancing_1998}
J.~A.~C. Baum and P.~Ingram.
\newblock Survival-{Enhancing} {Learning} in the {Manhattan} {Hotel}
  {Industry}, 1898–1980.
\newblock {\em Management Science}, 44(7):996--1016, July 1998.

\bibitem{beauchamp_principles_2009}
T.~Beauchamp and J.~F. Childress.
\newblock {\em Principles of biomedical ethics (6th ed.).}
\newblock New York: Oxford University Press., 2009.

\bibitem{becker_unveiling_2021}
R.~Becker, G.~D’Angelo, E.~Delfaraz, and H.~Gilbert.
\newblock Unveiling the {Truth} in {Liquid} {Democracy} with {Misinformed}
  {Voters}.
\newblock In D.~Fotakis and D.~Ríos~Insua, editors, {\em Algorithmic
  {Decision} {Theory}}, Lecture {Notes} in {Computer} {Science}, pages
  132--146, Cham, 2021. Springer International Publishing.

\bibitem{beinhocker_origin_2007}
E.~D. Beinhocker.
\newblock {\em The {Origin} of {Wealth}: {The} {Radical} {Remaking} of
  {Economics} and {What} it {Means} for {Business} and {Society}}.
\newblock Harvard Business Press, Sept. 2007.
\newblock Google-Books-ID: GtRBDwAAQBAJ.

\bibitem{belchior_survey_2021}
R.~Belchior, A.~Vasconcelos, S.~Guerreiro, and M.~Correia.
\newblock A {Survey} on {Blockchain} {Interoperability}: {Past}, {Present}, and
  {Future} {Trends}.
\newblock {\em ACM Computing Surveys}, 54(8):168:1--168:41, Oct. 2021.

\bibitem{benaloh_end--end_2015}
J.~Benaloh, R.~Rivest, P.~Y.~A. Ryan, P.~Stark, V.~Teague, and P.~Vora.
\newblock End-to-end verifiability, Apr. 2015.
\newblock arXiv:1504.03778 [cs].

\bibitem{benkler_peer_2015}
Y.~Benkler, A.~Shaw, and M.~Benjamin.
\newblock Peer production: {A} form of collective intelligence.
\newblock {\em Handbook for Collective Intelligence}, 175, 2015.

\bibitem{berg_exit_2017}
A.~Berg and C.~Berg.
\newblock Exit, {Voice}, and {Forking}, Dec. 2017.

\bibitem{berg_understanding_2019}
C.~Berg, S.~Davidson, and J.~Potts.
\newblock Understanding the {Blockchain} {Economy}, 2019.

\bibitem{berger_social_2011}
P.~L. Berger and T.~Luckmann.
\newblock {\em The {Social} {Construction} of {Reality}: {A} {Treatise} in the
  {Sociology} of {Knowledge}}.
\newblock Open Road Media, Apr. 2011.
\newblock Google-Books-ID: Jcma84waN3AC.

\bibitem{bernholz_digital_2021}
L.~Bernholz, {Helene Landemore}, and R.~Reich, editors.
\newblock {\em Digital {Technology} and {Democratic} {Theory}}.
\newblock The University of Chicago Press, 2021.

\bibitem{bhargavan_formal_2016}
K.~Bhargavan, A.~Delignat-Lavaud, C.~Fournet, A.~Gollamudi, G.~Gonthier,
  N.~Kobeissi, N.~Kulatova, A.~Rastogi, T.~Sibut-Pinote, N.~Swamy, and
  S.~Zanella-Béguelin.
\newblock Formal {Verification} of {Smart} {Contracts}: {Short} {Paper}.
\newblock In {\em Proceedings of the 2016 {ACM} {Workshop} on {Programming}
  {Languages} and {Analysis} for {Security}}, {PLAS} '16, pages 91--96, New
  York, NY, USA, Oct. 2016. Association for Computing Machinery.

\bibitem{bjornsson_corporate_2017}
G.~Björnsson and K.~Hess.
\newblock Corporate {Crocodile} {Tears}? {On} the {Reactive} {Attitudes} of
  {Corporate} {Agents}.
\newblock {\em Philosophy and Phenomenological Research}, 94(2):273--298, Mar.
  2017.

\bibitem{black_rationale_1948}
D.~Black.
\newblock On the {Rationale} of {Group} {Decision}-making.
\newblock {\em Journal of Political Economy}, 56(1):23--34, 1948.
\newblock Publisher: University of Chicago Press.

\bibitem{blais_turnout_1998}
A.~Blais and A.~Dobrzynska.
\newblock Turnout in electoral democracies.
\newblock {\em European Journal of Political Research}, 33(2):239--262, Mar.
  1998.

\bibitem{blockscience_applying_2022}
BlockScience.
\newblock Applying {Lessons} from {Constitutional} {Public} {Finance} to
  {Token} {System} {Design}, Nov. 2022.

\bibitem{boder_how_2022}
R.~Boder.
\newblock How {DAOs} {Can} {Transform} the {Business} {World}, July 2022.

\bibitem{bowe_zexe_2020}
S.~Bowe, A.~Chiesa, M.~Green, I.~Miers, P.~Mishra, and H.~Wu.
\newblock {ZEXE}: {Enabling} {Decentralized} {Private} {Computation}.
\newblock {\em 2020 IEEE Symposium on Security and Privacy (SP)}, pages
  947--964, May 2020.

\bibitem{brey_ethics_2017}
P.~Brey.
\newblock Ethics of {Emerging} {Technology}.
\newblock In S.~Hassan, editor, {\em Methods for the {Ethics} of {Technology}}.
  Rowman and Littlefield International, 2017.

\bibitem{brody_ideologies_2021}
A.~Brody and S.~Couture.
\newblock Ideologies and {Imaginaries} in {Blockchain} {Communities}: {The}
  {Case} of {Ethereum}.
\newblock {\em Canadian Journal of Communication}, 46(3):19--pp, 2021.

\bibitem{brooks_st_2022}
M.~Brooks.
\newblock St {Helena} {Government} {Budget} {Speech} 2022.
\newblock May 2022.

\bibitem{brown_corda_2018}
R.~Brown.
\newblock The {Corda} {Platform} : {An} {Introduction}.
\newblock 2018.

\bibitem{brummer_legal_2022}
C.~Brummer and R.~Seira.
\newblock Legal {Wrappers} and {DAOs}, May 2022.

\bibitem{bryan_motivating_2011}
C.~J. Bryan, G.~M. Walton, T.~Rogers, and C.~S. Dweck.
\newblock Motivating voter turnout by invoking the self.
\newblock {\em Proceedings of the National Academy of Sciences of the United
  States of America}, 108(31):12653--12656, Aug. 2011.

\bibitem{buchanan_domain_1990}
J.~M. Buchanan.
\newblock The domain of constitutional economics.
\newblock {\em Constitutional Political Economy}, 1(1):1--18, Dec. 1990.

\bibitem{buchanan_calculus_1962}
J.~M. Buchanan and G.~Tullock.
\newblock {\em The {Calculus} of {Consent}: {Logical} {Foundations} of
  {Constitutional} {Democracy}}.
\newblock 1962.

\bibitem{bueno_de_mesquita_paying_2022}
E.~Bueno~de Mesquita and A.~Hall.
\newblock Paying {People} to {Participate} in {Governance}.
\newblock {\em a16z Crypto}, 2022.

\bibitem{bunz_bulletproofs_2018}
B.~Bunz, J.~Bootle, D.~Boneh, A.~Poelstra, P.~Wuille, and G.~Maxwell.
\newblock Bulletproofs: {Short} {Proofs} for {Confidential} {Transactions} and
  {More}.
\newblock {\em 2018 IEEE Symposium on Security and Privacy (SP)}, pages
  315--334, May 2018.

\bibitem{burton_github_2017}
R.~M. Burton, D.~D. Håkonsson, J.~Nickerson, P.~Puranam, M.~Workiewicz, and
  T.~Zenger.
\newblock {GitHub}: exploring the space between boss-less and hierarchical
  forms of organizing.
\newblock {\em Journal of Organization Design}, 6(1):10, Dec. 2017.

\bibitem{buterin_daos_2014}
V.~Buterin.
\newblock {DAOs}, {DACs}, {DAs} and {More}: {An} {Incomplete} {Terminology}
  {Guide}, June 2014.

\bibitem{buterin_minimal_2019}
V.~Buterin.
\newblock Minimal anti-collusion infrastructure, May 2019.

\bibitem{buterin_credible_2020}
V.~Buterin.
\newblock Credible {Neutrality} {As} {A} {Guiding} {Principle}, Jan. 2020.

\bibitem{buterin_most_2021}
V.~Buterin.
\newblock The {Most} {Important} {Scarce} {Resource} is {Legitimacy}, Mar.
  2021.

\bibitem{buterin_nathan_2021}
V.~Buterin.
\newblock On {Nathan} {Schneider} on the limits of cryptoeconomics, Sept. 2021.

\bibitem{buterin_what_2023}
V.~Buterin.
\newblock What do {I} think about {Community} {Notes}?, Aug. 2023.

\bibitem{buterin_eip-1559_2019}
V.~Buterin, E.~Conner, R.~Dudley, M.~Slipper, I.~Norden, and A.~Bakhta.
\newblock {EIP}-1559: {Fee} market change for {ETH} 1.0 chain, Apr. 2019.

\bibitem{buterin_flexible_2019}
V.~Buterin, Z.~Hitzig, and E.~G. Weyl.
\newblock A {Flexible} {Design} for {Funding} {Public} {Goods}.
\newblock {\em Management Science}, 65(11):5171--5187, Nov. 2019.
\newblock arXiv:1809.06421 [econ, q-fin].

\bibitem{bonneau_zether_2020}
B.~Bünz, S.~Agrawal, M.~Zamani, and D.~Boneh.
\newblock Zether: {Towards} {Privacy} in a {Smart} {Contract} {World}.
\newblock volume 12059, pages 423--443, Cham, 2020. Springer International
  Publishing.

\bibitem{caney_cosmopolitan_2005}
S.~Caney.
\newblock Cosmopolitan {Justice}, {Responsibility}, and {Global} {Climate}
  {Change}.
\newblock {\em Leiden Journal of International Law}, 18(4):747--775, Dec. 2005.

\bibitem{ilgen_organizational_2000}
K.~M. Carley.
\newblock Organizational adaptation in volatile environments.
\newblock In D.~R. Ilgen and C.~L. Hulin, editors, {\em Computational modeling
  of behavior in organizations: {The} third scientific discipline.}, pages
  241--273. American Psychological Association, Washington, 2000.

\bibitem{casson_nature_1996}
M.~Casson.
\newblock The {Nature} of the {Firm} {Reconsidered}: {Information} {Synthesis}
  and {Entrepreneurial} {Organisation}.
\newblock {\em MIR: Management International Review}, 36:55--94, 1996.

\bibitem{casson_information_1997}
M.~Casson.
\newblock {\em Information and {Organization}: {A} {New} {Perspective} on the
  {Theory} of the {Firm}.}
\newblock Clarendon Press., 1997.

\bibitem{catalini_simple_2020}
C.~Catalini and J.~S. Gans.
\newblock Some simple economics of the blockchain.
\newblock {\em Communications of the ACM}, 63(7):80--90, June 2020.

\bibitem{cetinkaya_verification_2007}
O.~Cetinkaya and D.~Cetinkaya.
\newblock Verification and {Validation} {Issues} in {Electronic} {Voting}.
\newblock {\em Electronic Journal of e-Government},
  5(2):pp117‑126--pp117‑126, Dec. 2007.
\newblock Number: 2.

\bibitem{chamlee-wright_cultural_2002}
E.~Chamlee-Wright.
\newblock The {Cultural} {Foundations} of {Economic} {Development}: {Urban}
  {Female} {Entrepreneurship} in {Ghana}, 2002.

\bibitem{chamlee-wright_structure_2008}
E.~Chamlee-Wright.
\newblock The {Structure} of {Social} {Capital}: {An} {Austrian} {Perspective}
  on its {Nature} and {Development}.
\newblock {\em Review of Political Economy}, 20:41--58, Feb. 2008.

\bibitem{chandler_strategy_1969}
A.~D. Chandler.
\newblock {\em Strategy and {Structure}}.
\newblock The MIT Press, Aug. 1969.

\bibitem{chaum_scantegrity_2008}
D.~Chaum, A.~Essex, R.~Carback, J.~Clark, S.~Popoveniuc, A.~Sherman, and
  P.~Vora.
\newblock Scantegrity: {End}-to-{End} {Voter}-{Verifiable} {Optical}- {Scan}
  {Voting}.
\newblock {\em IEEE Security \& Privacy}, 6(3):40--46, May 2008.

\bibitem{cheng_ekiden_2019}
R.~Cheng, F.~Zhang, J.~Kos, W.~He, N.~Hynes, N.~Johnson, A.~Juels, A.~Miller,
  and D.~Song.
\newblock Ekiden: {A} {Platform} for {Confidentiality}-{Preserving},
  {Trustworthy}, and {Performant} {Smart} {Contracts}.
\newblock {\em 2019 IEEE European Symposium on Security and Privacy
  (EuroS\&P)}, pages 185--200, June 2019.

\bibitem{chey_cryptocurrencies_2023}
H.-k. Chey.
\newblock Cryptocurrencies and the {IPE} of money: an agenda for research.
\newblock {\em Review of International Political Economy}, 30(4):1605--1620,
  July 2023.
\newblock Publisher: Routledge \_eprint:
  https://doi.org/10.1080/09692290.2022.2109188.

\bibitem{chiu_regulating_2021}
I.~H.-Y. Chiu and J.~Linarelli.
\newblock {\em Regulating the {Crypto} {Economy}: {Business} {Transformations}
  and {Financialisation}}.
\newblock Hart Publishing, Oxford ; New York, Nov. 2021.

\bibitem{ciepley_beyond_2013}
D.~A. Ciepley.
\newblock Beyond {Public} and {Private}: {Toward} a {Political} {Theory} of the
  {Corporation}, Feb. 2013.

\bibitem{clarke_situational_2003}
A.~E. Clarke.
\newblock Situational {Analyses}: {Grounded} {Theory} {Mapping} {After} the
  {Postmodern} {Turn}.
\newblock {\em Symbolic Interaction}, 26(4):553--576, Nov. 2003.

\bibitem{clarke__2016}
A.~E. Clarke, R.~Washburn, C.~Friese, A.~E. Clarke, and R.~Washburn, editors.
\newblock {\em : {Mapping} {Research} with {Grounded} {Theory}}.
\newblock Routledge, New York, June 2016.

\bibitem{clarkson_civitas_2008}
M.~R. Clarkson, S.~Chong, and A.~C. Myers.
\newblock Civitas: {Toward} a {Secure} {Voting} {System}.
\newblock In {\em 2008 {IEEE} {Symposium} on {Security} and {Privacy} (sp
  2008)}, pages 354--368, May 2008.
\newblock ISSN: 2375-1207.

\bibitem{coase_nature_1937}
R.~H. Coase.
\newblock The {Nature} of the {Firm}.
\newblock {\em Economica}, 4(16):386--405, Nov. 1937.

\bibitem{cohen_punishing_2019}
M.~A. Cohen.
\newblock Punishing {Corporations}.
\newblock In M.~L. Rorie, editor, {\em The {Handbook} of {White}‐{Collar}
  {Crime}}, pages 314--333. Wiley, 1 edition, Oct. 2019.

\bibitem{cohen_garbage_1972}
M.~D. Cohen, J.~G. March, and J.~P. Olsen.
\newblock A {Garbage} {Can} {Model} of {Organizational} {Choice}.
\newblock {\em Administrative Science Quarterly}, 17(1):1--25, 1972.

\bibitem{uk_law_commission_decentralised_2022}
U.~L. Commission.
\newblock Decentralised {Autonomous} {Organisations} ({DAOs}) - {Call} for
  {Evidence}, Nov. 2022.

\bibitem{cong_blockchain_2019}
L.~W. Cong and Z.~He.
\newblock Blockchain {Disruption} and {Smart} {Contracts}.
\newblock {\em The Review of Financial Studies}, 32(5):1754--1797, May 2019.

\bibitem{conway_blockchain_2022}
K.~Conway.
\newblock Blockchain {Technology}: {Limited} {Liability} {Companies} and the
  {Need} for {North} {Carolina} {Legislation}.
\newblock {\em Campbell Law Review}, 45(1):127, Jan. 2022.

\bibitem{carbon_copy_state_2024}
C.~Copy.
\newblock The {State} of {ReFi}: {A} {Closer} {Look} at {Web3} {Regenerative}
  {Finance}.
\newblock Technical report, 2024.

\bibitem{crawford_grammar_1995}
S.~E.~S. Crawford and E.~Ostrom.
\newblock A {Grammar} of {Institutions}.
\newblock {\em American Political Science Review}, 89(3):582--600, Sept. 1995.

\bibitem{dagdelen_exploring_2021}
C.~F. Dagdelen, J.~Emmett, M.~Hampshire, S.~Voshmgir, and M.~Zargham.
\newblock Exploring {DAO2DAO} {Collaboration} {Mechanisms}, Mar. 2021.

\bibitem{dahl_democracy_1989}
R.~A. Dahl.
\newblock {\em Democracy and {Its} {Critics}}.
\newblock Yale University Press, 1989.
\newblock Google-Books-ID: l1RQngEACAAJ.

\bibitem{daian_-chain_2018}
P.~Daian, T.~Kell, I.~Miers, and A.~Juels.
\newblock On-{Chain} {Vote} {Buying} and the {Rise} of {Dark} {DAOs}, July
  2018.

\bibitem{daostar_wwwdaostarorg_2023}
DAOstar.
\newblock www.daostar.org, Mar. 2023.
\newblock Accessed March 31, 2023.

\bibitem{davidson_blockchains_2018}
S.~Davidson, P.~D. Filippi, and J.~Potts.
\newblock Blockchains and the economic institutions of capitalism.
\newblock {\em Journal of Institutional Economics}, 14(4):639--658, Aug. 2018.
\newblock Publisher: Cambridge University Press.

\bibitem{davidson_corporate_2022}
S.~Davidson and J.~Potts.
\newblock Corporate {Governance} in a {Crypto}-{World}, May 2022.

\bibitem{de_filippi_blockchain_2018}
P.~De~Filippi and S.~Hassan.
\newblock Blockchain {Technology} as a {Regulatory} {Technology}: {From} {Code}
  is {Law} to {Law} is {Code}, Jan. 2018.
\newblock arXiv:1801.02507 [cs].

\bibitem{de_filippi_report_2022}
P.~De~Filippi, M.~Mannan, J.~Henderson, T.~Merk, S.~Cossar, and K.~Nabben.
\newblock Report on {Blockchain} {Technology} \& {Legitimacy}, Dec. 2022.

\bibitem{de_filippi_alegality_2022}
P.~De~Filippi, M.~Mannan, and W.~Reijers.
\newblock The alegality of blockchain technology.
\newblock {\em Policy and Society}, 41(3):358--372, 2022.

\bibitem{denardis_protocol_2009}
L.~DeNardis.
\newblock {\em Protocol {Politics}: {The} {Globalization} of {Internet}
  {Governance}}.
\newblock The MIT Press, Cambridge, Mass, first edition, 1st printing edition,
  July 2009.

\bibitem{dimaggio_iron_1983}
P.~J. DiMaggio and W.~W. Powell.
\newblock The {Iron} {Cage} {Revisited}: {Institutional} {Isomorphism} and
  {Collective} {Rationality} in {Organizational} {Fields}.
\newblock {\em American Sociological Review}, 48(2):147--160, 1983.
\newblock Publisher: [American Sociological Association, Sage Publications,
  Inc.].

\bibitem{dix_human-computer_2003}
A.~Dix, J.~E. Finlay, G.~D. Abowd, and R.~Beale.
\newblock {\em Human-{Computer} {Interaction}}.
\newblock Pearson, Harlow, England ; New York, 3rd edition edition, Sept. 2003.

\bibitem{djankov_new_2003}
S.~Djankov, E.~Glaeser, R.~LaPorta, F.~Lopez-de Silanes, and A.~Shleifer.
\newblock The {New} {Comparative} {Economics}.
\newblock {\em Journal of Comparative Economics}, 31(4):595--619, 2003.

\bibitem{douglass_chapter_2016}
B.~P. Douglass.
\newblock Chapter 1 - {What} {Is} {Model}-{Based} {Systems} {Engineering}?
\newblock In B.~P. Douglass, editor, {\em Agile {Systems} {Engineering}}, pages
  1--39. Morgan Kaufmann, Boston, Jan. 2016.

\bibitem{driver_moral_2022}
J.~Driver.
\newblock Moral {Theory}.
\newblock June 2022.

\bibitem{dupont_experiments_2017}
Q.~DuPont.
\newblock Experiments in algorithmic governance: {A} history and ethnography of
  “{The} {DAO},” a failed decentralized autonomous organization.
\newblock In {\em Bitcoin and {Beyond}}. Routledge, 2017.
\newblock Num Pages: 21.

\bibitem{eghbal_working_2020}
N.~Eghbal.
\newblock {\em Working in public: the making and maintenance of open source
  software}.
\newblock Stripe Press, San Francisco, first edition edition, 2020.

\bibitem{elsden_making_2018}
C.~Elsden, A.~Manohar, J.~Briggs, M.~Harding, C.~Speed, and J.~Vines.
\newblock Making {Sense} of {Blockchain} {Applications}: {A} {Typology} for
  {HCI}.
\newblock In {\em Proceedings of the 2018 {CHI} {Conference} on {Human}
  {Factors} in {Computing} {Systems}}, {CHI} '18, pages 1--14, New York, NY,
  USA, Apr. 2018. Association for Computing Machinery.

\bibitem{esber_progressive_2023}
J.~Esber and S.~D. Kominers.
\newblock Progressive {Decentralization}: {A} {High}-level {Framework}, Jan.
  2023.

\bibitem{evans_rise_2016}
P.~C. Evans and A.~Gawer.
\newblock The {Rise} of the platform {Enterprise}: {A} {Global} {Survey}.
\newblock Technical report, The Center for Global Enterprise, 2016.

\bibitem{ezcurra_ancient_2022}
E.~Ezcurra, P.~Ezcurra, and B.~Meissner.
\newblock Ancient inhabitants of the {Basin} of {Mexico} kept an accurate
  agricultural calendar using sunrise observatories and mountain alignments.
\newblock {\em Proceedings of the National Academy of Sciences},
  119(51):e2215615119, Dec. 2022.
\newblock Publisher: Proceedings of the National Academy of Sciences.

\bibitem{fama_common_1993}
E.~F. Fama and K.~R. French.
\newblock Common risk factors in the returns on stocks and bonds.
\newblock {\em Journal of Financial Economics}, 33(1):3--56, Feb. 1993.

\bibitem{fama_separation_1983}
E.~F. Fama and M.~C. Jensen.
\newblock Separation of {Ownership} and {Control}.
\newblock {\em The Journal of Law \& Economics}, 26(2):301--325, 1983.
\newblock Publisher: [University of Chicago Press, Booth School of Business,
  University of Chicago, University of Chicago Law School].

\bibitem{fan_digital_2020}
J.~Fan and A.~X. Zhang.
\newblock Digital {Juries}: {A} {Civics}-{Oriented} {Approach} to {Platform}
  {Governance}.
\newblock In {\em Proceedings of the 2020 {CHI} {Conference} on {Human}
  {Factors} in {Computing} {Systems}}, {CHI} '20, pages 1--14, New York, NY,
  USA, Apr. 2020. Association for Computing Machinery.

\bibitem{fannizadeh_lawmakers_2023}
F.~Fannizadeh.
\newblock Lawmakers {In} {New} {Hampshire} {And} {Utah} {Recognize} {DAOs} {As}
  {Legal} {Persons}, July 2023.

\bibitem{farmer_daos_2022}
J.~S. Farmer, J.~Cahill, J.~S. Farmer, and J.~Cahill.
\newblock {DAOs}: {A} game changer in need of new rules.
\newblock {\em Reuters}, Oct. 2022.

\bibitem{faustino_myths_2022}
S.~Faustino, I.~Faria, and R.~Marques.
\newblock The myths and legends of king {Satoshi} and the knights of
  blockchain.
\newblock {\em Journal of Cultural Economy}, 15(1):67--80, Jan. 2022.

\bibitem{feichtinger_hidden_2023}
R.~Feichtinger, R.~Fritsch, Y.~Vonlanthen, and R.~Wattenhofer.
\newblock The {Hidden} {Shortcomings} of ({D}){AOs} -- {An} {Empirical} {Study}
  of {On}-{Chain} {Governance}, Feb. 2023.
\newblock arXiv:2302.12125 [cs].

\bibitem{finke_churching_2005}
R.~Finke and R.~Stark.
\newblock {\em The {Churching} of {America}, 1776-2005: {Winners} and {Losers}
  in {Our} {Religious} {Economy}}.
\newblock Rutgers University Press, New Brunswick, N.J, revised edition
  edition, Mar. 2005.

\bibitem{frantz_institutional_2021}
C.~K. Frantz and S.~Siddiki.
\newblock Institutional {Grammar} 2.0: {A} specification for encoding and
  analyzing institutional design.
\newblock {\em Public Administration}, 99(2):222--247, 2021.
\newblock \_eprint: https://onlinelibrary.wiley.com/doi/pdf/10.1111/padm.12719.

\bibitem{frey_composing_2023}
S.~Frey, J.~Hedges, J.~Tan, and P.~Zahn.
\newblock Composing games into complex institutions.
\newblock {\em PLOS ONE}, 18(3):e0283361, Mar. 2023.
\newblock Publisher: Public Library of Science.

\bibitem{frey_this_2019}
S.~Frey, P.~M. Krafft, and B.~C. Keegan.
\newblock "{This} {Place} {Does} {What} {It} {Was} {Built} {For}": {Designing}
  {Digital} {Institutions} for {Participatory} {Change}.
\newblock {\em Proceedings of the ACM on Human-Computer Interaction},
  3(CSCW):1--31, Nov. 2019.

\bibitem{frey_emergence_2019}
S.~Frey and R.~W. Sumner.
\newblock Emergence of integrated institutions in a large population of
  self-governing communities.
\newblock {\em PLOS ONE}, 14(7):e0216335, July 2019.

\bibitem{friedland_bringing_1991}
R.~Friedland and R.~Alford.
\newblock Bringing {Society} {Back} {In}: {Symbols}, {Practices}, and
  {Institutional} {Contradictions}.
\newblock Jan. 1991.

\bibitem{ganado_mapping_2020}
M.~Ganado, J.~Ellul, G.~Pace, S.~Tendon, and B.~Wilson.
\newblock Mapping the {Future} of {Legal} {Personality}.
\newblock {\em MIT Computational Law Report}, Nov. 2020.

\bibitem{gandhi_political_2008}
J.~Gandhi.
\newblock {\em Political {Institutions} under {Dictatorship}}.
\newblock Cambridge University Press, 2008.

\bibitem{ganguli_collective_2023}
D.~Ganguli, S.~Huang, L.~Lovitt, and D.~Siddarth.
\newblock Collective {Constitutional} {AI}: {Aligning} a {Language} {Model}
  with {Public} {Input}, Oct. 2023.

\bibitem{geertz_interpretation_1973}
C.~Geertz.
\newblock {\em The {Interpretation} {Of} {Cultures}}.
\newblock Basic Books, 1973.
\newblock Google-Books-ID: m3Y4DgAAQBAJ.

\bibitem{ghani_compositional_2018}
N.~Ghani, J.~Hedges, V.~Winschel, and P.~Zahn.
\newblock Compositional {Game} {Theory}.
\newblock In {\em Proceedings of the 33rd {Annual} {ACM}/{IEEE} {Symposium} on
  {Logic} in {Computer} {Science}}, {LICS} '18, pages 472--481, New York, NY,
  USA, July 2018. Association for Computing Machinery.

\bibitem{ghavi_primer_2022}
A.~Ghavi, A.~Qureshi, G.~Weinstein, J.~Schwartz, and S.~Lofchie.
\newblock A {Primer} on {DAOs}, Sept. 2022.

\bibitem{gilbert_counter-exploit_2023}
A.~Gilbert and S.~Haig.
\newblock '{Counter}-{Exploit}' {Claws} {Back} \${202M} {From} {Hacker}, Feb.
  2023.

\bibitem{glaveski_how_2022}
S.~Glaveski.
\newblock How {DAOs} {Could} {Change} the {Way} {We} {Work}.
\newblock {\em Harvard Business Review}, Apr. 2022.

\bibitem{goll_relationships_2005}
I.~Goll and A.~A. Rasheed.
\newblock The {Relationships} between {Top} {Management} {Demographic}
  {Characteristics}, {Rational} {Decision} {Making}, {Environmental}
  {Munificence}, and {Firm} {Performance}.
\newblock {\em Organization Studies}, 26(7):999--1023, July 2005.

\bibitem{green_get_2015}
D.~P. Green and A.~S. Gerber.
\newblock {\em Get {Out} the {Vote}: {How} to {Increase} {Voter} {Turnout}}.
\newblock Brookings Institution Press, 3 edition, 2015.

\bibitem{green_rhetorical_2004}
S.~E. Green.
\newblock A {Rhetorical} {Theory} of {Diffusion}.
\newblock {\em The Academy of Management Review}, 29(4):653--669, 2004.
\newblock Publisher: Academy of Management.

\bibitem{grube_culture_2015}
L.~E. Grube and Storr.
\newblock Culture and {Economic} {Action}, 2015.

\bibitem{habermas_public_2000}
J.~Habermas.
\newblock The {Public} {Sphere}.
\newblock In {\em Media {Studies}: {A} {Reader}}. NYU Press, Mar. 2000.
\newblock Google-Books-ID: 86kZKhuAjlAC.

\bibitem{haig_number_2020}
S.~Haig.
\newblock The number of active {DAOs} is up 660\% since 2019.
\newblock {\em Cointelegraph}, Sept. 2020.

\bibitem{hall_political_1996}
P.~A. Hall and R.~C.~R. Taylor.
\newblock Political {Science} and the {Three} {New} {Institutionalisms}.
\newblock {\em Political Studies}, 1996.

\bibitem{hannan_inertia_1996}
M.~T. Hannan, M.~D. Burton, and J.~N. Baron.
\newblock Inertia and {Change} in the {Early} {Years}: {Employment} {Relations}
  in {Young}, {High} {Technology} {Firms}.
\newblock Jan. 1996.
\newblock Accepted: 2020-11-17T17:23:32Z.

\bibitem{harris-braun_holochain_2017}
E.~Harris-Braun, N.~Luck, and A.~Brock.
\newblock Holochain: scalable agent-centric distributed computing.
\newblock Oct. 2017.

\bibitem{hartshorne_thousand_2019}
J.~K. Hartshorne, J.~R. de~Leeuw, N.~D. Goodman, M.~Jennings, and T.~J.
  O'Donnell.
\newblock A thousand studies for the price of one: {Accelerating} psychological
  science with {Pushkin}.
\newblock {\em Behavior Research Methods}, 51(4):1782--1803, Aug. 2019.

\bibitem{hasinoff_scalability_2022}
A.~A. Hasinoff and N.~Schneider.
\newblock From {Scalability} to {Subsidiarity} in {Addressing} {Online} {Harm}.
\newblock {\em Social Media + Society}, 8(3):205630512211260, July 2022.

\bibitem{hassan_decentralized_2021}
S.~Hassan and P.~De~Filippi.
\newblock Decentralized {Autonomous} {Organization}.
\newblock {\em Internet Policy Review}, 10(2), Apr. 2021.

\bibitem{hasu_new_2021}
Hasu and monetsupply.
\newblock A {New} {Mental} {Model} for {Defi} {Treasuries}, Oct. 2021.

\bibitem{hayek_kinds_1964}
F.~Hayek.
\newblock Kinds of {Order} in {Society}.
\newblock {\em New Individualist Review}, Vol. 1(No. 3):457--466, 1964.

\bibitem{highton_political_2001}
B.~Highton and R.~E. Wolfinger.
\newblock The {Political} {Implications} of {Higher} {Turnout}.
\newblock {\em British Journal of Political Science}, 31(1):179--223, 2001.

\bibitem{hine_ethnography_2015}
C.~Hine.
\newblock {\em Ethnography for the {Internet}: embedded, embodied and
  everyday}.
\newblock Bloomsbury Academic, An imprint of Bloomsbury Publishing Plc, London
  ; New York, 2015.

\bibitem{hitchens_decentralised_2023}
B.~Hitchens and O.~Roberts.
\newblock Decentralised {Autonomous} {Organisations} ({DAOs}): {What} are they?
  {And} can they be parties to a claim?, Jan. 2023.

\bibitem{hubbard_beyond_2022}
S.~Hubbard.
\newblock Beyond the {Buzzwords}: {Web3}, {DAOs}, and the {Future} of {Human}
  {Coordination}, Apr. 2022.

\bibitem{ilyushina_decentralised_2022}
N.~Ilyushina and T.~Macdonald.
\newblock Decentralised {Autonomous} {Organisations}: {A} {New} {Research}
  {Agenda} for {Labour} {Economics}.
\newblock {\em The Journal of The British Blockchain Association}, Apr. 2022.
\newblock Publisher: The British Blockchain Association.

\bibitem{jackman_political_1987}
R.~W. Jackman.
\newblock Political {Institutions} and {Voter} {Turnout} in the {Industrial}
  {Democracies}.
\newblock {\em American Political Science Review}, 81(2):405--423, 1987.

\bibitem{jain_algorithmic_2023}
S.~Jain, V.~Suriyakumar, K.~Creel, and A.~Wilson.
\newblock Algorithmic {Pluralism}: {A} {Structural} {Approach} {To} {Equal}
  {Opportunity}.
\newblock 2023.

\bibitem{jenkinson_remote_2022}
G.~Jenkinson.
\newblock Remote work triggers move to {DAOs} in the post-pandemic world:
  {Survey}.
\newblock {\em Cointelegraph}, Dec. 2022.

\bibitem{jennings_how_2022}
M.~Jennings and D.~Kerr.
\newblock How to pick a {DAO} legal entity, June 2022.

\bibitem{jensen_theory_1976}
M.~C. Jensen and W.~H. Meckling.
\newblock Theory of the firm: {Managerial} behavior, agency costs and ownership
  structure.
\newblock {\em Journal of Financial Economics}, 3(4):305--360, Oct. 1976.

\bibitem{johnson_what_2007}
V.~Johnson.
\newblock What {Is} {Organizational} {Imprinting}? {Cultural}
  {Entrepreneurship} in the {Founding} of the {Paris} {Opera}.
\newblock {\em American Journal of Sociology}, 113(1):97--127, 2007.
\newblock Publisher: The University of Chicago Press.

\bibitem{johnson_backstage_2009}
V.~Johnson.
\newblock {\em Backstage at the {Revolution}: {How} the {Royal} {Paris} {Opera}
  {Survived} the {End} of the {Old} {Regime}}.
\newblock University of Chicago Press, Chicago, IL, Feb. 2009.

\bibitem{juels_coercion-resistant_2005}
A.~Juels, D.~Catalano, and M.~Jakobsson.
\newblock Coercion-resistant electronic elections.
\newblock In {\em Proceedings of the 2005 {ACM} workshop on {Privacy} in the
  electronic society}, {WPES} '05, pages 61--70, New York, NY, USA, Nov. 2005.
  Association for Computing Machinery.

\bibitem{kalodner_arbitrum_2018}
H.~Kalodner, S.~Goldfeder, X.~Chen, S.~M. Weinberg, and E.~W. Felten.
\newblock Arbitrum: {Scalable}, private smart contracts.
\newblock pages 1353--1370, 2018.

\bibitem{kaptchuk_giving_2019}
G.~Kaptchuk, M.~Green, and I.~Miers.
\newblock Giving {State} to the {Stateless}: {Augmenting} {Trustworthy}
  {Computation} with {Ledgers}.
\newblock {\em Proceedings 2019 Network and Distributed System Security
  Symposium}, 2019.

\bibitem{kaptein_corporations_2017}
M.~Kaptein and M.~Constantinescu.
\newblock Corporations as {Moral} {Entities}.
\newblock In D.~C. Poff and A.~C. Michalos, editors, {\em Encyclopedia of
  {Business} and {Professional} {Ethics}}, pages 1--4. Springer International
  Publishing, Cham, 2017.

\bibitem{karakostas_dimitris_sok_2022}
{Karakostas, Dimitris}, {Kiayias, Aggelos}, and {Ovezik, Christina}.
\newblock {SoK}: {A} {Stratified} {Approach} to {Blockchain}
  {Decentralization}.
\newblock {\em ArXiv}, 2022.

\bibitem{karp_understanding_2011}
R.~M. Karp.
\newblock Understanding {Science} {Through} the {Computational} {Lens}.
\newblock {\em Journal of Computer Science and Technology}, 26(4):569--577,
  July 2011.

\bibitem{kelkar_complete_2023}
M.~Kelkar, K.~Babel, P.~Daian, J.~Austgen, V.~Buterin, and A.~Juels.
\newblock Complete {Knowledge}: {Preventing} {Encumbrance} of {Cryptographic}
  {Secrets}, 2023.
\newblock Publication info: Preprint.

\bibitem{kerber_kachina_2021}
T.~Kerber, A.~Kiayias, and M.~Kohlweiss.
\newblock Kachina - {Foundations} of {Private} {Smart} {Contracts}.
\newblock {\em 2021 IEEE 34th Computer Security Foundations Symposium (CSF)},
  pages 1--16, June 2021.

\bibitem{kirzner_competition_1973}
I.~M. Kirzner.
\newblock {\em Competition and {Entrepreneurship}}.
\newblock University of Chicago Press, Chicago, IL, 1973.

\bibitem{koh_sense_2003}
J.~Koh and Y.-G. Kim.
\newblock Sense of {Virtual} {Community}: {A} {Conceptual} {Framework} and
  {Empirical} {Validation}.
\newblock {\em International Journal of Electronic Commerce}, 8(2):75--93,
  2003.
\newblock Publisher: Taylor \& Francis, Ltd.

\bibitem{kollock_managing_1996}
P.~Kollock and M.~Smith.
\newblock Managing the {Virtual} {Commons}: {Cooperation} and conflict in
  computer communities.
\newblock {\em Computer-Mediated Communication: Linguistic, Social, and
  Cross-Cultural Perspectives, edited bySusan Herring. Amsterdam: John
  Benjamins.}, pages Pp. 109--128, Jan. 1996.

\bibitem{korpas_governance_2023}
L.~Korpas and J.~Z. Tan.
\newblock Governance {Surfaces} of {DAOs}, 2023.

\bibitem{korpas_political_2023}
L.~M. Korpas, S.~Frey, and J.~Tan.
\newblock Political, economic, and governance attitudes of blockchain users.
\newblock {\em Frontiers in Blockchain}, 6, 2023.

\bibitem{kosba_hawk_2016}
A.~Kosba, A.~Miller, E.~Shi, Z.~Wen, and C.~Papamanthou.
\newblock Hawk: {The} {Blockchain} {Model} of {Cryptography} and
  {Privacy}-{Preserving} {Smart} {Contracts}.
\newblock {\em 2016 IEEE Symposium on Security and Privacy (SP)}, pages
  839--858, May 2016.

\bibitem{kostelnik_governance_2013}
J.~Kostelnik and D.~Skarbek.
\newblock The governance institutions of a drug trafficking organization.
\newblock {\em Public Choice}, 156(1/2):95--103, 2013.
\newblock Publisher: Springer.

\bibitem{kreutler_prehistory_2021}
K.~Kreutler.
\newblock A {Prehistory} of {DAOs}, 2021.

\bibitem{kwon_cosmos_2019}
J.~Kwon and E.~Buchman.
\newblock Cosmos {Whitepaper}: {A} {Network} of {Distributed} {Ledgers}.
\newblock 2019.

\bibitem{snapshot_labs_wwwsnapshotorg_nodate}
S.~Labs.
\newblock www.snapshot.org.
\newblock Accessed October 27, 2023.

\bibitem{lalley_quadratic_2018}
S.~P. Lalley and E.~G. Weyl.
\newblock Quadratic {Voting}: {How} {Mechanism} {Design} {Can} {Radicalize}
  {Democracy}.
\newblock {\em AEA Papers and Proceedings}, 108:33--37, May 2018.

\bibitem{landemore_democratic_2017}
H.~Landemore.
\newblock {\em Democratic {Reason}: {Politics}, {Collective} {Intelligence},
  and the {Rule} of the {Many}.}
\newblock Princeton University Press, 2017.

\bibitem{larimer_bitcoin_2013}
S.~Larimer.
\newblock Bitcoin and the {Three} {Laws} of {Robotics}, Sept. 2013.

\bibitem{latour_reassembling_2007}
B.~Latour.
\newblock {\em Reassembling the social: {An} introduction to
  actor-network-theory}.
\newblock Oup Oxford, 2007.

\bibitem{lazar_legitimacy_2022}
S.~Lazar.
\newblock Legitimacy, {Authority}, and {Democratic} {Duties} of {Explanation}.
\newblock 2022.

\bibitem{lebar_corporations_2019}
M.~LeBar.
\newblock Corporations, {Moral} {Agency}, and {Reactive} {Attitudes}.
\newblock {\em Georgetown Journal of Law \& Public Policy}, 17:811, 2019.

\bibitem{lehdonvirta_virtual_2014}
V.~Lehdonvirta and E.~Castronova.
\newblock {\em Virtual {Economies}: {Design} and {Analysis}}.
\newblock The MIT Press, 2014.

\bibitem{leiblein_empirical_2003}
M.~J. Leiblein and D.~J. Miller.
\newblock An empirical examination of transaction- and firm-level influences on
  the vertical boundaries of the firm.
\newblock {\em Strategic Management Journal}, 24(9):839--859, Sept. 2003.

\bibitem{li_liquid_2023}
C.~Li, R.~Xu, and L.~Duan.
\newblock Liquid {Democracy} in {DPoS} {Blockchains}, Sept. 2023.
\newblock arXiv:2309.01090 [cs].

\bibitem{lijphart_patterns_2012}
A.~Lijphart.
\newblock {\em Patterns of {Democracy}: {Government} {Forms} and {Performance}
  in {Thirty}-{Six} {Countries}}.
\newblock Yale University Press, 2012.

\bibitem{locke_two_1690}
J.~Locke.
\newblock {\em Two {Treatises} of {Government}}.
\newblock 1690.

\bibitem{lorenz_kumpan_1935}
K.~Lorenz.
\newblock Der {Kumpan} in der {Umwelt} des {Vogels}.
\newblock {\em Journal für Ornithologie}, 83(2):137--213, Apr. 1935.

\bibitem{lounsbury_cultural_2001}
M.~Lounsbury and M.~A. Glynn.
\newblock Cultural entrepreneurship: stories, legitimacy, and the acquisition
  of resources.
\newblock {\em Strategic Management Journal}, 22(6-7):545--564, 2001.
\newblock \_eprint: https://onlinelibrary.wiley.com/doi/pdf/10.1002/smj.188.

\bibitem{low_company_2022}
K.~F.~K. Low, E.~Schuster, and W.~Y. Wan.
\newblock The {Company} and {Blockchain} {Technology}, Oct. 2022.

\bibitem{lumineau_blockchain_2021}
F.~Lumineau, W.~Wang, and O.~Schilke.
\newblock Blockchain {Governance}—{A} {New} {Way} of {Organizing}
  {Collaborations}?
\newblock {\em Organization Science}, 32(2):500--521, Mar. 2021.
\newblock Publisher: INFORMS.

\bibitem{lustig_intersecting_2019}
C.~Lustig.
\newblock Intersecting {Imaginaries}: {Visions} of {Decentralized} {Autonomous}
  {Systems}.
\newblock {\em Proceedings of the ACM on Human-Computer Interaction},
  3(CSCW):210:1--210:27, Nov. 2019.

\bibitem{luther_cryptocurrencies_2016}
W.~J. Luther.
\newblock {CRYPTOCURRENCIES}, {NETWORK} {EFFECTS}, {AND} {SWITCHING} {COSTS}.
\newblock {\em Contemporary Economic Policy}, 34(3):553--571, July 2016.

\bibitem{maddox_netnography_2020}
A.~Maddox.
\newblock Netnography to {Uncover} {Cryptomarkets}.
\newblock In {\em Netnography {Unlimited}}. Routledge, 2020.

\bibitem{maddox_constructive_2016}
A.~Maddox, M.~J. Barratt, M.~Allen, and S.~Lenton.
\newblock Constructive activism in the dark web: cryptomarkets and illicit
  drugs in the digital ‘demimonde’.
\newblock {\em Information, Communication \& Society}, 19(1):111--126, Jan.
  2016.

\bibitem{magazzeni_validation_2017}
D.~Magazzeni, P.~McBurney, and W.~Nash.
\newblock Validation and {Verification} of {Smart} {Contracts}: {A} {Research}
  {Agenda}.
\newblock {\em Computer}, 50(9):50--57, 2017.
\newblock Conference Name: Computer.

\bibitem{magnuson_blockchain_2020}
W.~Magnuson.
\newblock {\em Blockchain {Democracy}: {Technology}, {Law} and the {Rule} of
  the {Crowd}}.
\newblock Cambridge University Press, Cambridge, 2020.

\bibitem{mannan_fostering_2018}
M.~Mannan.
\newblock Fostering {Worker} {Cooperatives} with {Blockchain} {Technology}:
  {Lessons} from the {Colony} {Project}, Dec. 2018.

\bibitem{march_learning_1991}
J.~G. March, L.~S. Sproull, and M.~Tamuz.
\newblock Learning from {Samples} of {One} or {Fewer}.
\newblock {\em Organization Science}, 2(1):1--13, Feb. 1991.

\bibitem{mckinney_pr008_2023}
J.~McKinney.
\newblock {PR008} – {DAO} {Regulation}., Mar. 2023.

\bibitem{messias_johnnatan_understanding_2023}
{Messias, Johnnatan}, {Pahari, Vabuk}, {Chandrasekaran, Balakrishnan},
  {Guummadi, Krishna P.}, and {Loiseau, Patrick}.
\newblock Understanding {Blockchain} {Governance}: {Analyzing} {Decentralized}
  {Voting} to {Amend} {DeFi} {Smart} {Contracts}.
\newblock {\em ArXiv}, 2023.

\bibitem{metisdao_new_2022}
MetisDAO.
\newblock New {Survey} {Indicates} {The} {Rapid} {Growth} of {Decentralized}
  {Organizations} is {Coming}, Dec. 2022.

\bibitem{metjahic_deconstructing_2017}
L.~Metjahic.
\newblock Deconstructing the {DAO}: {The} need for legal recognition and the
  application of securities laws to decentralized organizations.
\newblock {\em Cardozo L. Rev.}, 39:1533, 2017.

\bibitem{meyden_specification_2019}
R.~v.~d. Meyden.
\newblock On the specification and verification of atomic swap smart contracts
  (extended abstract).
\newblock In {\em 2019 {IEEE} {International} {Conference} on {Blockchain} and
  {Cryptocurrency} ({ICBC})}, pages 176--179, May 2019.

\bibitem{midao_midao_2022}
{MIDAO}.
\newblock {MIDAO} awarded {Facilitation} of {DAO} {Registry} {Process} by
  government of {Marshall} {Islands}, Dec. 2022.

\bibitem{miu_innovation_2018}
E.~Miu, N.~Gulley, K.~N. Laland, and L.~Rendell.
\newblock Innovation and cumulative culture through tweaks and leaps in online
  programming contests.
\newblock {\em Nature Communications}, 9(1):2321, June 2018.
\newblock Number: 1 Publisher: Nature Publishing Group.

\bibitem{montesquieu_spirit_1748}
Montesquieu.
\newblock {\em The {Spirit} of the {Laws}}.
\newblock Hafner Publishing Company, 1748.

\bibitem{moore_what_2023}
G.~Moore.
\newblock What {Is} {Quadratic} {Voting} and {Why} {Don}’t {More} {Projects}
  {Use} {It}?, Jan. 2023.

\bibitem{murray_contracting_2019}
A.~Murray, S.~Kuban, M.~Josefy, and J.~Anderson.
\newblock Contracting in the {Smart} {Era}: {The} {Implications} of
  {Blockchain} and {Decentralized} {Autonomous} {Organizations} for
  {Contracting} and {Corporate} {Governance}.
\newblock Apr. 2019.

\bibitem{murray_contracting_2021}
A.~Murray, S.~Kuban, M.~Josefy, and J.~Anderson.
\newblock Contracting in the {Smart} {Era}: {The} {Implications} of
  {Blockchain} and {Decentralized} {Autonomous} {Organizations} for
  {Contracting} and {Corporate} {Governance}.
\newblock {\em Academy of Management Perspectives}, 35(4):622--641, Nov. 2021.

\bibitem{nabben_is_2021}
K.~Nabben.
\newblock Is a "{Decentralized} {Autonomous} {Organization}" a {Panopticon}?:
  {Algorithmic} governance as creating and mitigating vulnerabilities in
  {DAOs}.
\newblock In {\em Proceedings of the {Interdisciplinary} {Workshop} on (de)
  {Centralization} in the {Internet}}, pages 18--25, Virtual Event Germany,
  Dec. 2021. ACM.

\bibitem{nabben_web3_2023}
K.~Nabben.
\newblock Web3 as ‘self-infrastructuring’: {The} challenge is how.
\newblock {\em Big Data \& Society}, 10(1):205395172311590, Jan. 2023.

\bibitem{nershi_how_2020}
K.~Nershi.
\newblock How {Strong} are {International} {Standards} in {Practice}?
  {Evidence} from {Cryptocurrency} {Transactions}.
\newblock {\em Under Review}, 2020.

\bibitem{nguyen_transparency_2022}
C.~T. Nguyen.
\newblock Transparency is {Surveillance}.
\newblock {\em Philosophy and Phenomenological Research}, 105(2):331--361,
  2022.
\newblock \_eprint: https://onlinelibrary.wiley.com/doi/pdf/10.1111/phpr.12823.

\bibitem{nine_global_2012}
C.~Nine.
\newblock {\em Global {Justice} and {Territory}}.
\newblock OUP Oxford, May 2012.
\newblock Google-Books-ID: VCvFb85XvlsC.

\bibitem{north_institutions_1990}
D.~North.
\newblock {\em Institutions, {Institutional} {Change} and {Economic}
  {Performance}}.
\newblock Cambridge University Press, 1990.

\bibitem{senate_of_the_parliament_of_australia_select_2021}
S.~of~the Parliament~of Australia.
\newblock Select {Committee} on {Australia} as a {Technology} and {Financial}
  {Centre}, Oct. 2021.

\bibitem{us_department_of_the_treasury_us_2022}
U.~D. of~the Treasury.
\newblock U.{S}. {Treasury} {Sanctions} {Notorious} {Virtual} {Currency}
  {Mixer} {Tornado} {Cash}, Aug. 2022.

\bibitem{state_of_wyoming_legislature_wyoming_2021}
S.~of~Wyoming~Legislature.
\newblock Wyoming {Decentralized} {Autonomous} {Organization} {Supplement},
  Apr. 2021.

\bibitem{world_commission_on_environment_and_development_our_1987}
W.~C. on~Environment {and}~Development.
\newblock {\em Our {Common} {Future}}.
\newblock Oxford Paper Backs. Oxford University Press., Oxford; New York, 1987.

\bibitem{oreilly_ai_2024}
T.~O'Reilly.
\newblock {AI} {Has} an {Uber} {Problem}: {How} {Silicon} {Valley}’s {Race}
  for {Monopoly} {Inhibits} {Product}-{Market} {Fit}, Apr. 2024.

\bibitem{ostrom_governing_1990}
E.~Ostrom.
\newblock {\em Governing the {Commons}.}
\newblock Cambridge University Press, 1990.

\bibitem{ostrom_designing_1995}
E.~Ostrom.
\newblock Designing complexity to govern complexity.
\newblock In {\em Property {Rights} and the {Environment}: {Social} and
  {Ecological} {Issues}}, pages 33--45. World Bank Publications, 1995.
\newblock Google-Books-ID: O9iBJ\_f7RZ8C.

\bibitem{ostrom_beyond_2009}
E.~Ostrom.
\newblock Beyond {Markets} and {States}: {Polycentric} {Governance} of
  {Complex} {Economic} {Systems}.
\newblock {\em Nobel Prize in Economics documents}, Dec. 2009.

\bibitem{parthemore_what_2013}
J.~Parthemore and B.~Whitby.
\newblock {WHAT} {MAKES} {ANY} {AGENT} {A} {MORAL} {AGENT}? {REFLECTIONS} {ON}
  {MACHINE} {CONSCIOUSNESS} {AND} {MORAL} {AGENCY}.
\newblock {\em International Journal of Machine Consciousness},
  05(02):105--129, Dec. 2013.

\bibitem{patel_meme_2021}
N.~Patel.
\newblock From a meme to \$47 million: {ConstitutionDAO}, crypto, and the
  future of crowdfunding, Dec. 2021.

\bibitem{patka_exploiting_2022}
I.~Patka.
\newblock Exploiting {Inattention} \& {Optimism} in {DAOs}, Oct. 2022.

\bibitem{pecone_eight_2023}
N.~Pecone.
\newblock The {Eight} {Forms} of {Capital}, 2023.

\bibitem{peterson_match_2021}
M.~Peterson.
\newblock A {Match} {Made} in {DeFi}: {Rari} {Capital} and {Fei} {Protocol}
  {Merge} to ‘{FeiRari}’, Dec. 2021.

\bibitem{pham_why_2024}
B.~Pham.
\newblock Why {DAOs} {Should} {Govern} {AI}, Jan. 2024.

\bibitem{pink_digital_2015}
S.~Pink, H.~Horst, J.~Postill, L.~Hjorth, T.~Lewis, and J.~Tacchi.
\newblock {\em Digital ethnography: {Principles} and practice}.
\newblock sage, 2015.

\bibitem{plato_republic_1892}
Plato.
\newblock "{The} {Republic}" in {The} {Dialogues} of {Plato} translated into
  {English} with {Analyses} and {Introductions} by {B}. {Jowett}, {M}.{A}. in
  {Five} {Volumes}.
\newblock Oxford University Press, 3rd edition revised and corrected edition,
  1892.

\bibitem{pol_anti-money_2020}
R.~F. Pol.
\newblock Anti-money laundering: {The} world's least effective policy
  experiment? {Together}, we can fix it.
\newblock {\em Policy Design and Practice}, 3(1):73--94, Jan. 2020.

\bibitem{potts_economic_2021}
J.~Potts, D.~W.~E. Allen, C.~Berg, S.~Davidson, and T.~MacDonald.
\newblock An economic theory of blockchain foundations, May 2021.

\bibitem{przeworski_democracy_2012}
A.~Przeworski.
\newblock {\em Democracy and {Development}: {Political} {Institutions} and
  {Well}-{Being} in the {World}, 1950-1990}.
\newblock Cambridge University Press, 2012.

\bibitem{puranam_whats_2014}
P.~Puranam, O.~Alexy, and M.~Reitzig.
\newblock What's “{New}” {About} {New} {Forms} of {Organizing}?
\newblock {\em Academy of Management Review}, 39(2):162--180, Apr. 2014.

\bibitem{putnam_making_1993}
R.~Putnam.
\newblock {\em Making {Democracy} {Work}.}
\newblock Princeton University Press, 1993.

\bibitem{pwc_generative_2024}
PWC.
\newblock Generative {AI}: {Transform} the future of business and lead with
  trust.
\newblock Technical report, 2024.

\bibitem{rawls_theory_1971}
J.~Rawls.
\newblock {\em A {Theory} of {Justice}: {Original} {Edition}}.
\newblock Harvard University Press, 1971.

\bibitem{rennie_toward_2022}
E.~Rennie, M.~Zargham, J.~Tan, L.~Miller, J.~Abbott, K.~Nabben, and
  P.~De~Filippi.
\newblock Toward a {Participatory} {Digital} {Ethnography} of {Blockchain}
  {Governance}.
\newblock {\em Qualitative Inquiry}, 28(7):837--847, Sept. 2022.

\bibitem{riker_federalism_1964}
W.~Riker.
\newblock {\em Federalism: {Origin}, {Operation}, {Significance}}.
\newblock Little, Brown, and Company, 1964.

\bibitem{rockmore_cultural_2018}
D.~N. Rockmore, C.~Fang, N.~J. Foti, T.~Ginsburg, and D.~C. Krakauer.
\newblock The cultural evolution of national constitutions.
\newblock {\em Journal of the Association for Information Science and
  Technology}, 69(3):483--494, 2018.
\newblock \_eprint: https://onlinelibrary.wiley.com/doi/pdf/10.1002/asi.23971.

\bibitem{ronfeldt_tribes_1996}
D.~Ronfeldt.
\newblock Tribes, {Institutions}, {Markets}, {Networks}: {A} {Framework}
  {About} {Societal} {Evolution}.
\newblock Technical report, RAND Corporation, Jan. 1996.

\bibitem{rong_blockchain_2023}
H.~Rong, E.~Peris, and E.~Jin.
\newblock Blockchain for {Impact} {Workshop}: {Perspectives} on {International}
  {Development}, {Public} {Goods}, and {Regenerative} {Economy}, May 2023.

\bibitem{roth_shared_2017}
A.~S. Roth.
\newblock Shared {Agency}.
\newblock In E.~N. Zalta, editor, {\em The {Stanford} {Encyclopedia} of
  {Philosophy}}. Metaphysics Research Lab, Stanford University, summer 2017
  edition, 2017.

\bibitem{rousseau_social_1762}
J.-J. Rousseau.
\newblock {\em The {Social} {Contract}}.
\newblock 1762.

\bibitem{rozas_when_2021}
D.~Rozas, A.~Tenorio-Fornés, S.~Díaz-Molina, and S.~Hassan.
\newblock When {Ostrom} {Meets} {Blockchain}: {Exploring} the {Potentials} of
  {Blockchain} for {Commons} {Governance}.
\newblock {\em SAGE Open}, 11(1):21582440211002526, Jan. 2021.
\newblock Publisher: SAGE Publications.

\bibitem{ruane_what_2022}
J.~Ruane and A.~McAfee.
\newblock What a {DAO} {Can} — and {Can}’t — {Do}.
\newblock {\em Harvard Business Review}, May 2022.

\bibitem{russo_fair_2020}
C.~Russo.
\newblock "{Fair} {Launch} is a {New} {Way} for {Founders} to {Express}
  {Themselves}:" {Gavin} {McDermott} of {IDEO} {CoLab} - {The} {Defiant}, Oct.
  2020.

\bibitem{ryan_pret_2009}
P.~Ryan, D.~Bismark, J.~Heather, S.~Schneider, and {Zhe Xia}.
\newblock \textit{{PrÊt} À {Voter}:} a {Voter}-{Verifiable} {Voting}
  {System}.
\newblock {\em IEEE Transactions on Information Forensics and Security},
  4(4):662--673, Dec. 2009.

\bibitem{samuelson_diagrammatic_1955}
P.~A. Samuelson.
\newblock Diagrammatic {Exposition} of a {Theory} of {Public} {Expenditure}.
\newblock {\em The Review of Economics and Statistics}, 37(4):350, Nov. 1955.

\bibitem{santana_blockchain_2022}
C.~Santana and L.~Albareda.
\newblock Blockchain and the emergence of {Decentralized} {Autonomous}
  {Organizations} ({DAOs}): {An} integrative model and research agenda.
\newblock {\em Technological Forecasting and Social Change}, 182:121806, Sept.
  2022.

\bibitem{schillig_decentralized_2022}
M.~Schillig.
\newblock Decentralized {Autonomous} {Organizations} ({DAOs}) under {English}
  {Law}, Sept. 2022.

\bibitem{schletz_blockchain_2023}
M.~Schletz, A.~Constant, A.~Hsu, S.~Schillebeeckx, R.~Beck, and M.~Wainstein.
\newblock Blockchain and regenerative finance: charting a path toward
  regeneration.
\newblock {\em Frontiers in Blockchain}, 6:1165133, July 2023.

\bibitem{schneider_cryptoeconomics_2021}
N.~Schneider.
\newblock Cryptoeconomics as a {Limitation} on {Governance}.
\newblock 2021.

\bibitem{scholz_ours_2016}
T.~Scholz and N.~Schneider, editors.
\newblock {\em Ours to {Hack} and to {Own}: {The} {Rise} of {Platform}
  {Cooperativism}, {A} {New} {Vision} for the {Future} of {Work} and a {Fairer}
  {Internet}}.
\newblock OR Books, 2016.

\bibitem{schumpeter_capitalism_1942}
J.~A. Schumpeter.
\newblock {\em Capitalism, {Socialism}, and {Democracy}}.
\newblock Harper \& Brothers, first edition edition, 1942.

\bibitem{schwartz_collective_2011}
J.~Schwartz.
\newblock Collective {Choice}, 2011.

\bibitem{schweik_internet_2012}
C.~M. Schweik and R.~C. English.
\newblock {\em Internet {Success}: {A} {Study} of {Open}-{Source} {Software}
  {Commons}}.
\newblock The MIT Press, June 2012.

\bibitem{scott_institutions_2014}
R.~W. Scott.
\newblock Institutions and {Organizations}. {Ideas}, {Interests} and
  {Identities}.{Paperback}: 360 pages {Publisher}: {Sage} (1995) {Language}:
  {English} {ISBN}: 978-142242224.
\newblock {\em M@n@gement}, 17(2):136--140, 2014.

\bibitem{scott_institutional_1994}
W.~R. Scott and J.~W. Meyer.
\newblock {\em Institutional {Environments} and {Organizations}: {Structural}
  {Complexity} and {Individualism}}.
\newblock SAGE Publications, Inc, Thousand Oaks, Calif, 1st edition edition,
  Apr. 1994.

\bibitem{seaver_algorithms_2017}
N.~Seaver.
\newblock Algorithms as {Culture}: {Some} {Tactics} for the {Ethnography} of
  {Algorithmic} {Systems}.
\newblock {\em Big Data and Society}, 4(2), 2017.

\bibitem{selgin_bitcoin_2022}
G.~Selgin.
\newblock Bitcoin: {Problems} and {Prospects}.
\newblock Apr. 2022.
\newblock Publisher: Cato Institute.

\bibitem{senge_fifth_2006}
P.~M. Senge.
\newblock {\em The fifth discipline: the art and practice of the learning
  organization}.
\newblock Doubleday/Currency, New York, rev. and updated edition, 2006.
\newblock OCLC: ocm65166960.

\bibitem{sharif_ethics_2022}
M.~M. Sharif and F.~Ghodoosi.
\newblock The {Ethics} of {Blockchain} in {Organizations}.
\newblock {\em Journal of Business Ethics}, 178(4):1009--1025, July 2022.

\bibitem{sharma_unpacking_2023}
T.~Sharma, Y.~Kwon, K.~Pongmala, H.~Wang, A.~Miller, D.~Song, and Y.~Wang.
\newblock Unpacking {How} {Decentralized} {Autonomous} {Organizations} ({DAOs})
  {Work} in {Practice}, Apr. 2023.
\newblock arXiv:2304.09822 [cs].

\bibitem{shepherd_are_2015}
J.~Shepherd.
\newblock Are {Corporations} {Moral} {Agents}? {\textbar} {Practical} {Ethics},
  Dec. 2015.

\bibitem{shleifer_survey_1997}
A.~Shleifer and R.~Vishny.
\newblock A {Survey} of {Corporate} {Governance}.
\newblock {\em Journal of Finance}, 52(2):737--783, 1997.

\bibitem{shorin_uniswap_2021}
T.~Shorin, J.~Pope, L.~Lotti, A.~Z. Lewis, and M.~Gomez.
\newblock Uniswap {Research} {Report}: {Discord}, {Governance}, {Community},
  Aug. 2021.

\bibitem{sim_collaborative_2020}
R.~H.~L. Sim, Y.~Zhang, M.~C. Chan, and B.~K.~H. Low.
\newblock Collaborative {Machine} {Learning} with {Incentive}-{Aware} {Model}
  {Rewards}.
\newblock 2020.
\newblock Publisher: [object Object] Version Number: 1.

\bibitem{simon_sciences_2019}
H.~A. Simon.
\newblock {\em The sciences of the artificial}.
\newblock The MIT Press, Cambridge, Massachusetts, third edition [2019 edition]
  edition, 2019.

\bibitem{skarbek_governance_2011}
D.~Skarbek.
\newblock Governance and {Prison} {Gangs}.
\newblock {\em American Political Science Review}, 2011.

\bibitem{smith_keeping_2020}
C.~E. Smith, B.~Yu, A.~Srivastava, A.~Halfaker, L.~Terveen, and H.~Zhu.
\newblock Keeping {Community} in the {Loop}: {Understanding} {Wikipedia}
  {Stakeholder} {Values} for {Machine} {Learning}-{Based} {Systems}.
\newblock In {\em Proceedings of the 2020 {CHI} {Conference} on {Human}
  {Factors} in {Computing} {Systems}}, pages 1--14, Honolulu HI USA, Apr. 2020.
  ACM.

\bibitem{sochat_experiment_2016}
V.~V. Sochat, I.~W. Eisenberg, A.~Z. Enkavi, J.~Li, P.~G. Bissett, and R.~A.
  Poldrack.
\newblock The {Experiment} {Factory}: {Standardizing} {Behavioral}
  {Experiments}.
\newblock {\em Frontiers in Psychology}, 7, 2016.

\bibitem{srinivasan_network_2022}
B.~Srinivasan.
\newblock {\em The {Network} {State}: {How} to {Start} a {New} {Country}}.
\newblock 2022.

\bibitem{stanczyk_productive_2012}
L.~Stanczyk.
\newblock Productive {Justice}.
\newblock {\em Philosophy and Public Affairs}, 40(2):144--164, 2012.

\bibitem{steffen_zkay_2019}
S.~Steffen, B.~Bichsel, M.~Gersbach, N.~Melchior, P.~Tsankov, and M.~Vechev.
\newblock zkay: {Specifying} and {Enforcing} {Data} {Privacy} in {Smart}
  {Contracts}.
\newblock {\em Proceedings of the 2019 ACM SIGSAC Conference on Computer and
  Communications Security}, pages 1759--1776, Nov. 2019.

\bibitem{stilz_territorial_2019}
A.~Stilz.
\newblock Territorial boundaries and history.
\newblock {\em Politics, Philosophy \& Economics}, 18(4):374--385, Nov. 2019.

\bibitem{stinchcombe_social_2000}
A.~L. Stinchcombe.
\newblock Social structure and organizations.
\newblock In J.~A.C.~Baum and F.~Dobbin, editors, {\em Economics {Meets}
  {Sociology} in {Strategic} {Management}}, volume~17 of {\em Advances in
  {Strategic} {Management}}, pages 229--259. Emerald Group Publishing Limited,
  Jan. 2000.

\bibitem{storr_understanding_2013}
V.~Storr.
\newblock Understanding the {Culture} of {Markets}, 2013.

\bibitem{suchman_managing_1995}
M.~C. Suchman.
\newblock Managing legitimacy: {Strategic} and institutional approaches.
\newblock {\em The Academy of Management Review}, 20:571--610, 1995.
\newblock Place: US Publisher: Academy of Management.

\bibitem{suchow_dallinger_2023}
J.~Suchow, T.~Morgan, and T.~Griffiths.
\newblock Dallinger, Aug. 2023.

\bibitem{sulkowski_tao_2019}
A.~J. Sulkowski.
\newblock The {Tao} of {Dao}: {Hardcoding} {Business} {Ethics} on {Blockchain}.
\newblock {\em Business \& Finance Law Review}, 3:146, 2019.

\bibitem{sun_voter_2023}
X.~Sun, X.~Chen, C.~Stasinakis, and G.~Sermpinis.
\newblock Voter {Coalitions} and democracy in {Decentralized} {Finance}:
  {Evidence} from {MakerDAO}, June 2023.
\newblock arXiv:2210.11203 [cs, q-fin].

\bibitem{swartz_theorizing_2022}
L.~Swartz.
\newblock Theorizing the 2017 blockchain {ICO} bubble as a network scam.
\newblock {\em New Media \& Society}, 24(7):1695--1713, July 2022.

\bibitem{tan_tracing_2018}
C.~Tan.
\newblock Tracing {Community} {Genealogy}: {How} {New} {Communities} {Emerge}
  from the {Old}.
\newblock {\em Proceedings of the International AAAI Conference on Web and
  Social Media}, 12(1), June 2018.
\newblock Number: 1.

\bibitem{tan_constitutions_2022}
J.~Z. Tan, M.~Langenkamp, A.~Weichselbraun, A.~Brody, and L.~Korpas.
\newblock The {Constitutions} of {Web3}.
\newblock 2022.

\bibitem{tan_erc-4824_2022}
J.~Z. Tan, I.~Patka, I.~Gershtein, E.~Eithcowich, M.~Zargham, and S.~Furter.
\newblock {ERC}-4824: {Common} {Interfaces} for {DAOs} [{DRAFT}].
\newblock {\em Ethereum Improvement Proposals}, (4824), Feb. 2022.

\bibitem{teblunthuis_identifying_2022}
N.~TeBlunthuis and B.~M. Hill.
\newblock Identifying {Competition} and {Mutualism} between {Online} {Groups}.
\newblock {\em Proceedings of the International AAAI Conference on Web and
  Social Media}, 16:993--1004, May 2022.

\bibitem{thiel_polycentric_2023}
A.~Thiel.
\newblock Polycentric {Governing} and {Polycentric} {Governance}.
\newblock In F.~Gadinger and J.~A. Scholte, editors, {\em Polycentrism: {How}
  {Governing} {Works} {Today}}, page~0. Oxford University Press, May 2023.

\bibitem{thulke_climategpt_2024}
D.~Thulke, Y.~Gao, P.~Pelser, R.~Brune, R.~Jalota, F.~Fok, M.~Ramos, I.~van
  Wyk, A.~Nasir, H.~Goldstein, T.~Tragemann, K.~Nguyen, A.~Fowler, A.~Stanco,
  J.~Gabriel, J.~Taylor, D.~Moro, E.~Tsymbalov, J.~de~Waal, E.~Matusov,
  M.~Yaghi, M.~Shihadah, H.~Ney, C.~Dugast, J.~Dotan, and D.~Erasmus.
\newblock {ClimateGPT}: {Towards} {AI} {Synthesizing} {Interdisciplinary}
  {Research} on {Climate} {Change}.
\newblock 2024.
\newblock Publisher: [object Object] Version Number: 1.

\bibitem{thurman_curve_2021}
A.~Thurman.
\newblock ‘{Curve} {Wars}’ {Heat} {Up}: {Emergency} {DAO} {Invoked} {After}
  ‘{Clear} {Governance} {Attack}’.
\newblock {\em CoinDesk}, Nov. 2021.
\newblock Section: Finance.

\bibitem{tsai_accountability_2007}
L.~L. Tsai.
\newblock {\em Accountability without {Democracy}: {Solidary} {Groups} and
  {Public} {Goods} {Provision} in {Rural} {China}.}
\newblock Cambridge University Press, 2007.

\bibitem{tsebelis_veto_2002}
G.~Tsebelis.
\newblock {\em Veto {Players}: {How} {Political} {Institutions} {Work}}.
\newblock Princeton University Press, 2002.

\bibitem{vaitiekunas_dialektos_2022}
T.~Vaitiekunas.
\newblock Dialektos: {Privacy}-preserving {Smart} {Contracts}.
\newblock {\em IACR Cryptol. ePrint Arch.}, 2022.

\bibitem{deepdao_ventures_wwwdeepdaoio_nodate}
D.~Ventures.
\newblock www.deepdao.io.
\newblock Accessed October 26, 2022.

\bibitem{verba_participation_1987}
S.~Verba and N.~H. Nie.
\newblock {\em Participation in {America}: {Political} {Democracy} and {Social}
  {Equality}}.
\newblock University of Chicago Press, 1987.

\bibitem{volberda_contingency_2012}
H.~W. Volberda, N.~Van Der~Weerdt, E.~Verwaal, M.~Stienstra, and A.~J. Verdu.
\newblock Contingency {Fit}, {Institutional} {Fit}, and {Firm} {Performance}:
  {A} {Metafit} {Approach} to {Organization}–{Environment} {Relationships}.
\newblock {\em Organization Science}, 23(4):1040--1054, Aug. 2012.

\bibitem{voshmgir_foundations_2019}
S.~Voshmgir and M.~Zargham.
\newblock Foundations of {Cryptoeconomic} {Systems}.
\newblock Technical report, Research Institute for Cryptoeconomics, Vienna,
  2019.

\bibitem{wang_decentralized_2019}
S.~Wang, W.~Ding, J.~Li, Y.~Yuan, L.~Ouyang, and F.-Y. Wang.
\newblock Decentralized {Autonomous} {Organizations}: {Concept}, {Model}, and
  {Applications}.
\newblock {\em IEEE Transactions on Computational Social Systems},
  6(5):870--878, Oct. 2019.

\bibitem{weingast_economic_1995}
B.~R. Weingast.
\newblock The {Economic} {Role} of {Political} {Institutions}:
  {Market}-{Preserving} {Federalism} and {Economic} {Development}.
\newblock {\em Journal of Law, Economics, \& Organization}, 1995.

\bibitem{werhane_corporate_2016}
P.~H. Werhane.
\newblock Corporate {Moral} {Agency} and the {Responsibility} to {Respect}
  {Human} {Rights} in the {UN} {Guiding} {Principles}: {Do} {Corporations}
  {Have} {Moral} {Rights}?
\newblock {\em Business and Human Rights Journal}, 1(1):5--20, Jan. 2016.

\bibitem{white_market_2015}
L.~White.
\newblock The {Market} for {Cryptocurrencies}.
\newblock {\em Cato Journal}, 35(2):383--402, 2015.
\newblock Publisher: Cato Journal, Cato Institute.

\bibitem{wigderson_mathematics_2019}
A.~Wigderson.
\newblock {\em Mathematics and {Computation}: {A} {Theory} {Revolutionizing}
  {Technology} and {Science}}.
\newblock Princeton University Press, Princeton, NJ, Oct. 2019.

\bibitem{wigginton_cooperatives_2023}
J.~Wigginton, M.~Janoff, J.~Perkins, T.~Morrey, S.~Burton, J.~Volz, A.~Truths,
  D.~Antoni, and S.~Heiser.
\newblock Cooperatives: {An} {Ownership} {Model} for {Digital} {Networks}.
\newblock Technical report, National Society of Accountants for Cooperatives,
  Jan. 2023.

\bibitem{williamson_markets_1975}
O.~E. Williamson.
\newblock Markets and {Hierarchies}: {Analysis} and {Antitrust} {Implications}:
  {A} {Study} in the {Economics} of {Internal} {Organization}, 1975.

\bibitem{woetzel_secret_2022}
C.~Woetzel.
\newblock Secret {Network}: {A} {Privacy}-{Preserving} {Secret} {Contract} \&
  {Decentralized} {Application} {Platform}.
\newblock 2022.

\bibitem{wood_polkadot_2016}
G.~Wood.
\newblock Polkadot: {Vision} for a heterogenous multi-chain framework.
\newblock Nov. 2016.

\bibitem{wright_rise_2021}
A.~Wright and C.~P. o. L. a. B. N. C. S.~o. Law.
\newblock The {Rise} of {Decentralized} {Autonomous} {Organizations}:
  {Opportunities} and {Challenges}.
\newblock {\em Stanford Journal of Blockchain Law \& Policy}, June 2021.

\bibitem{wright_measuring_2021}
S.~A. Wright.
\newblock Measuring {DAO} {Autonomy}: {Lessons} {From} {Other} {Autonomous}
  {Systems}.
\newblock {\em IEEE Transactions on Technology and Society}, 2(1):43--53, Mar.
  2021.

\bibitem{xie_zkbridge_2022}
T.~Xie, J.~Zhang, Z.~Cheng, F.~Zhang, Y.~Zhang, Y.~Jia, D.~Boneh, and D.~Song.
\newblock {zkBridge}: {Trustless} {Cross}-chain {Bridges} {Made} {Practical}.
\newblock In {\em Proceedings of the 2022 {ACM} {SIGSAC} {Conference} on
  {Computer} and {Communications} {Security}}, {CCS} '22, pages 3003--3017, New
  York, NY, USA, Nov. 2022. Association for Computing Machinery.

\bibitem{yermack_corporate_2017}
D.~Yermack.
\newblock Corporate {Governance} and {Blockchains}*.
\newblock {\em Review of Finance}, 21(1):7--31, Mar. 2017.

\bibitem{zargham_aligning_2022}
M.~Zargham and K.~Nabben.
\newblock Aligning ‘{Decentralized} {Autonomous} {Organization}’ to
  {Precedents} in {Cybernetics}, Apr. 2022.

\bibitem{zargham_economic_2020}
M.~Zargham, K.~Paruch, and J.~Shorish.
\newblock Economic {Games} as {Estimators}.
\newblock In P.~Pardalos, I.~Kotsireas, Y.~Guo, and W.~Knottenbelt, editors,
  {\em Mathematical {Research} for {Blockchain} {Economy}}, Springer
  {Proceedings} in {Business} and {Economics}, pages 125--142, Cham, 2020.
  Springer International Publishing.

\bibitem{zargham_generalized_2022}
M.~Zargham and J.~Shorish.
\newblock Generalized {Dynamical} {Systems} {Part} {I}: {Foundations}.
\newblock Technical report, Institute for Cryptoeconomics, Interdisciplinary
  Research, WU Vienna University of Economics and Business, Vienna, 2022.

\bibitem{zargham_curved_2020}
M.~Zargham, J.~Shorish, and K.~Paruch.
\newblock From {Curved} {Bonding} to {Configuration} {Spaces}.
\newblock In {\em 2020 {IEEE} {International} {Conference} on {Blockchain} and
  {Cryptocurrency} ({ICBC})}, pages 1--3, May 2020.

\bibitem{zhang_luyao_sok_2023}
{Zhang, Luyao}, {Ma, Xinshi}, and {Liu, Yulin}.
\newblock {SoK}: {Blockchain} {Decentralization}.
\newblock {\em ArXiv}, 2023.

\bibitem{zhong_institutional_2022}
Q.~Zhong and S.~Frey.
\newblock Institutional similarity drives cultural similarity among online
  communities.
\newblock {\em Scientific Reports}, 12(1):18982, Nov. 2022.

\bibitem{zhong_quantifying_2022}
Q.~Zhong, S.~Frey, and M.~Hilbert.
\newblock Quantifying the {Selective}, {Stochastic}, and {Complementary}
  {Drivers} of {Institutional} {Evolution} in {Online} {Communities}.
\newblock {\em Entropy}, 24(9):1185, Sept. 2022.

\bibitem{zilber_institutionalization_2002}
T.~B. Zilber.
\newblock Institutionalization as an {Interplay} between {Actions}, {Meanings},
  and {Actors}: {The} {Case} of a {Rape} {Crisis} {Center} in {Israel}.
\newblock {\em The Academy of Management Journal}, 45(1):234--254, 2002.
\newblock Publisher: Academy of Management.

\bibitem{zucker_role_1977}
L.~G. Zucker.
\newblock The {Role} of {Institutionalization} in {Cultural} {Persistence}.
\newblock {\em American Sociological Review}, 42(5):726--743, 1977.
\newblock Publisher: [American Sociological Association, Sage Publications,
  Inc.].

\bibitem{zuckerman_what_2020}
E.~Zuckerman.
\newblock What is {Digital} {Public} {Infrastructure}?
\newblock Technical report, Center for Journalism \& Liberty, 2020.

\bibitem{zyskind_enigma_2015}
G.~Zyskind, O.~Nathan, and A.~Pentland.
\newblock Enigma: {Decentralized} {Computation} {Platform} with {Guaranteed}
  {Privacy}, June 2015.
\newblock arXiv:1506.03471 [cs].

\end{thebibliography}

\end{document}